\documentclass[aps,reprint]{revtex4-2}
\usepackage{etoolbox} % for \appto
\usepackage{blindtext}
\usepackage{graphicx}
\usepackage{subcaption}
\usepackage{multirow}%
\usepackage{amssymb}
\usepackage{xcolor}
\usepackage{graphicx}
\usepackage{floatrow}
\usepackage[noend]{algpseudocode}
\usepackage[fleqn]{amsmath}
\usepackage[font=small,skip=3pt]{caption}
\usepackage{makecell}
\usepackage{textcomp}%
\usepackage{hyperref} 
\usepackage{cleveref}
\usepackage{url}
\usepackage{tikz}
\usepackage{siunitx}
\DeclareSIUnit\MWh{MWh}
\DeclareSIUnit\dollar{\mathdollar}
\usepackage[version=4]{mhchem} % added by JAS for typesetting isotopes
\usepackage{booktabs} % added by JAS for tables
\usepackage{bm}
\let\mrm\mathrm
\makeatletter
\appto{\appendix}{%
  \@ifstar{\def\theequation@prefix{A.}}%
          {}%!TEX encoding = UTF-8 UnicodeX
}
\makeatother

\newcolumntype{B}{>{\color{red}}c}

\begin{document}
\title{Isotope Production in Fusion Systems}
\author{J. F. Parisi$^{1}$}
\email{jason@marathonfusion.com}
\author{J. A. Schwartz$^{1}$, S. E. Wurzel$^2$, A. Rutkowski$^{1}$, J. Harter$^{1}$}
\affiliation{$^1$Marathon Fusion, 150 Mississippi, San Francisco, 94107, CA, USA}
\affiliation{$^2$Fusion Energy Base, New York, NY 10003, USA}

\begin{abstract}
Fusion systems producing isotopes via neutron-driven transmutation can achieve economic viability well before reaching energy breakeven. Incorporating carefully selected feedstock materials in a blanket allows fusion systems to generate both electrical power and high-value isotopes, expanding the space of viable concepts, significantly enhancing the economic value of fusion energy, and supporting an accelerated path to adoption. We calculate the value of this co-generation and derive a new economic breakeven condition based on net present value. At lower plasma gain, $Q_{\mathrm{plas}}\lesssim 1$, high-value transmutation, such as medical radioisotopes, enables pure transmuter fusion systems operating at only watts to megawatts of fusion power: for example, a 3 megawatt system transmuting ${}^{102}\mathrm{Ru}\rightarrow{}^{99}\mathrm{Mo}$ could fulfill global ${}^{99}\mathrm{Mo}$ demand with $Q_{\mathrm{plas}} \ll 1$. At higher gain $Q_{\mathrm{plas}}\gtrsim 3$, it becomes viable to generate electricity in addition to isotopes. For example, co-production of electricity and gold, transmuted from mercury in a fusion blanket, can reduce the required plasma gain for economic viability from $Q_{\mathrm{plas}}\sim10$–$100$ to $Q_{\mathrm{plas}}\sim3-5$. We further highlight techniques to enhance transmutation with asymmetric neutron wall loading. Fusion neutron-driven transmutation therefore offers a revenue-positive pathway for deploying fusion energy at terawatt-scale, starting from smaller watt-to-megawatt-scale machines for radioisotope production and then scaling up to co-producing electricity and gold in larger fusion power plants.
%combining power generation with critical-material and radioisotope production, fundamentally improving fusion's economic and societal value.
\end{abstract}

\maketitle

\section{Introduction}

For fusion systems to achieve widespread deployment they must generate valuable products. To date, the dominant focus has been selling electricity from fusion power plants (FPPs) \cite{sheffield2001study}. Although the electricity market is vast (9TW total capacity in 2023 and higher today \cite{EIA2025ElectricityCapacity}), electricity is highly fungible with many low-cost competitors, resulting in the value per fusion neutron for electricity generation being relatively low at $ \sim \$10^{-20}$ / neutron \cite{parisi2026valuecostfusionneutrons}. However, electricity need not be the only product of a fusion power plant. Fusion systems can increase the value per neutron by generating additional outputs, most notably through neutron-driven transmutation.
%However, fusion systems can increase the value per neutron by generating other outputs.
%However, fusion systems can also produce valuable isotopes because deuterium-tritium (D-T) fusion reactions
In particular, deuterium–tritium (D–T) fusion reactions
\begin{equation}
  \mathrm{d} + \mathrm{t}
  \longrightarrow{}^{4}_{2}\mathrm{He}
  + \mathrm{n},
  \label{eq:DTreaction}
\end{equation}
%have a unique feature: high-energy neutrons - with kinetic energy $E_\mathrm{n} = \qty{14.1}{MeV}$ at high flux - can drive transmutation on a feedstock material to make valuable materials such as radioisotopes and precious metals 
produce $E_\mathrm{n} = \qty{14.1}{MeV}$ neutrons at high flux, enabling the conversion of feedstock materials into valuable radioisotopes \cite{engholm1986radioisotope,Bourque1988FAME,Leung2018_CompactNG,Handley2021EarlyFusionMarkets,Honney2023FusionNeutrons,shine_neutron_imaging_2025,evitts2025theoretical,parisi2025j} and precious metals \cite{Bourque1988FAME,rutkowski2025scalable} while co-producing electricity \cite{rutkowski2025scalable}. The economic benefits of selling these outputs could significantly accelerate fusion's deployment.

%and $\alpha$ particles carrying $E_{\alpha}=\qty{3.5}{MeV}$. 
 %Among proposed fusion-neutron-driven transmutation pathways, 
 
To date, one fusion-neutron-driven transmutation product has been found to have both a sufficiently large market and value per neutron to support terawatt-scale fusion deployment: producing gold from mercury via $(\mathrm{n},\mathrm{2n})$ reactions \cite{rutkowski2025scalable},
\begin{small}
\begin{align}
&{}^{198}_{80}\mathrm{Hg}(\mathrm{n,2n}){}^{197}_{80}\mathrm{Hg}
   \;\xrightarrow[\;T_{1/2}=64.1\ \text{h}\;]{\varepsilon}\;
   {}^{197}_{79}\mathrm{Au}\;+\;\nu_{e}, \\
& \mathrm{^{198}_{80}Hg}(\mathrm{n,2n}){}^{197\mathrm{m}}_{80}\mathrm{Hg}
   \xrightarrow[\;T_{1/2}=23.8\ \text{h}\;]{%
\substack{94.7\%\,\mathrm{IT}\\5.3\%\,\varepsilon}}
   \begin{cases}
      \mathrm{^{197}_{80}Hg}\;+\;\gamma \; (\mathrm{IT})\\[2pt]
      \mathrm{^{197}_{79}Au}\;+\;\nu_{e} \; (\varepsilon).
   \end{cases}
\label{eq:alchemy_pathway}
\end{align}
\end{small}

\begin{figure}[bt!]
    \centering
    \begin{subfigure}[t]{0.92\textwidth}
    \centering
    \includegraphics[width=1.0\textwidth]{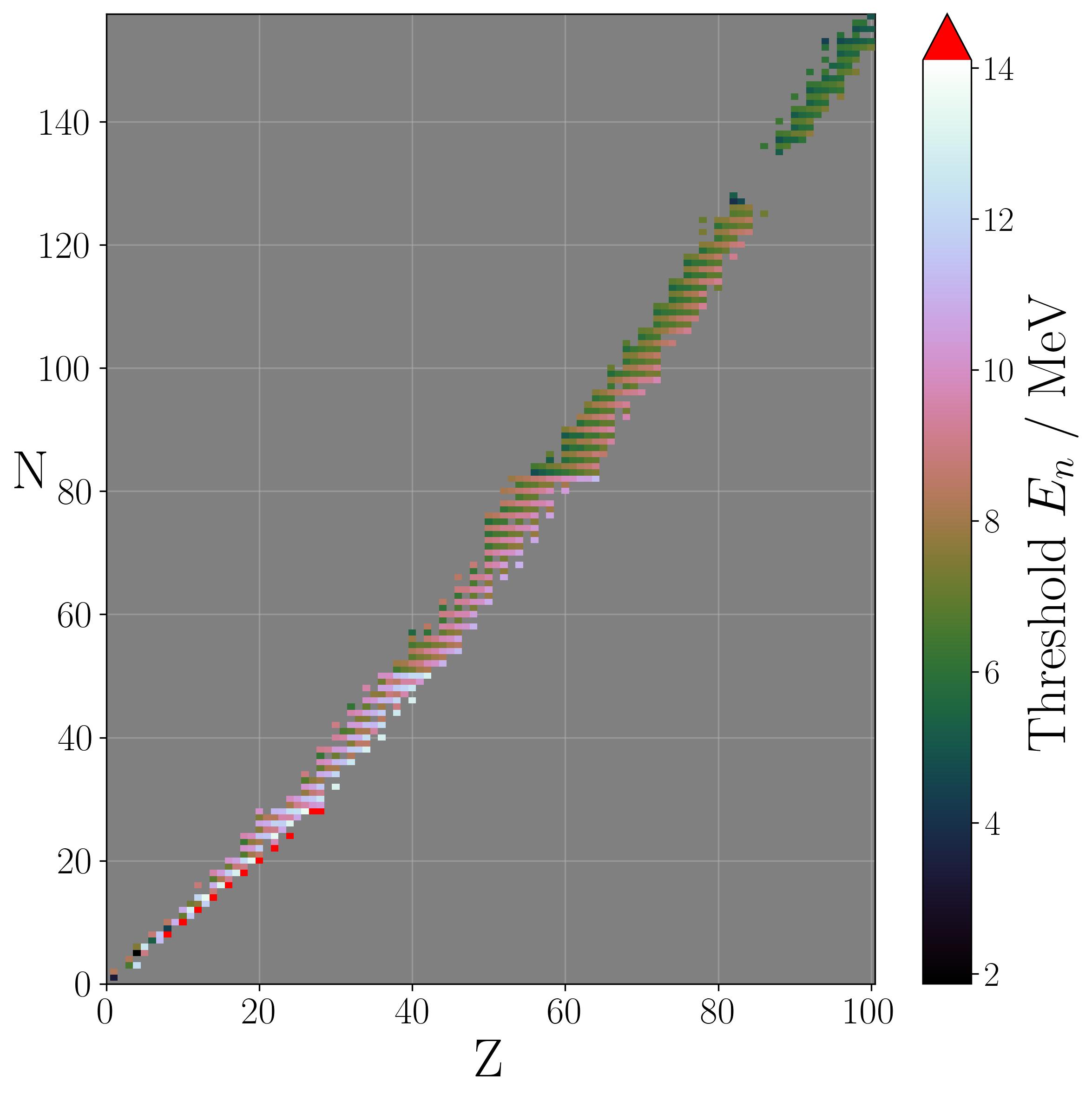}
    \end{subfigure}
    \caption{Threshold energy for $(\mathrm{n},\mathrm{2n})$ reactions versus proton number (Z) and neutron number (N). Red coloring above \qty{14.1}{MeV} indicates the $(\mathrm{n},\mathrm{2n})$ reaction is inaccessible at \qty{14.1}{MeV}. Data from \cite{Brown20181}.}
    \label{fig:n2n_thresholdenergy}
\end{figure}

\begin{figure*}[tb!]
    \centering
    \begin{subfigure}[t]{0.92\textwidth}
    \centering
    \includegraphics[width=1.0\textwidth]{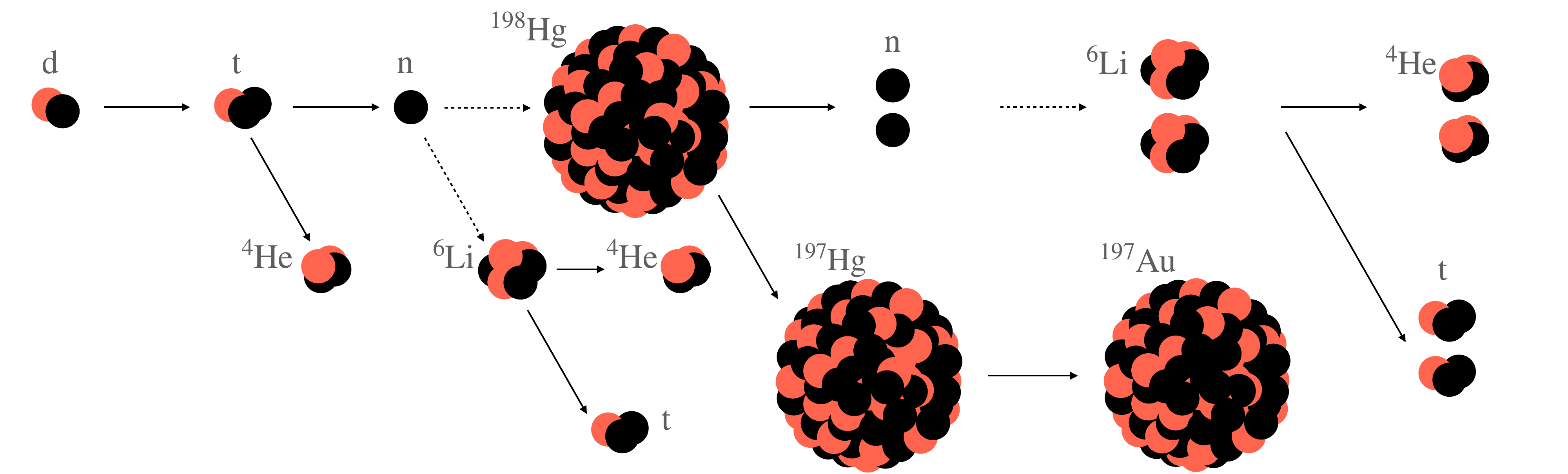}
    \end{subfigure}
    \caption{Driving $(\mathrm{n},\mathrm{2n})$ reactions on ${}^{198}\mathrm{Hg}$ to produce stable gold ${}^{197}\mathrm{Au}$ using D-T neutrons \cite{rutkowski2025scalable}, along with subsequent (n,t) reactions on ${}^{6}\mathrm{Li}$. Dashed lines indicate that fewer than 100\% of the incoming neutrons will drive the next reaction.}
    \label{fig:blanket_schematic}
\end{figure*}

\noindent In this scheme, enriched mercury-198 feedstock is placed in the fusion blanket, where a fraction of the D-T neutrons induce $(\mathrm{n,2n})$ reactions. The remainder are either captured parasitically - via $(\mathrm{n},\gamma)$ - or breed tritium via $\ce{^6Li}(\ce{n},\ce{t})\alpha$ reactions. At present (2025) gold and electricity prices, revenue from gold production is estimated to exceed the revenue from electricity sales \cite{rutkowski2025scalable}, more than doubling the economic value of outputs of a fusion system.

While the Hg $\rightarrow$ Au pathway supports large-scale FPP deployment, other transmutation products with higher value per neutron \cite{pietropaolo2021sorgentina,li2023feasibility,evitts2025theoretical,parisi2025j,parisi2025k} - albeit with much smaller overall market size - could enable an economically viable pathway to scaling fusion energy starting from smaller machines that only produce isotopes \cite{parisi2025j,parisi2025k}. 

In this work, we quantify how revenue from transmutation products relaxes plasma performance requirements, allowing near-term deployment of machines with lower plasma performance but high economic value, producing isotopes with very high value per D-T neutron. Once a fusion system has sufficiently high performance, it becomes economically viable to co-produce electricity and gold - while the value per neutron will be much lower than smaller fusion systems \cite{parisi2026valuecostfusionneutrons}, the market sizes for gold and electricity are many orders of magnitude higher, supporting a much larger deployment of fusion energy capacity.

This paper is organized as follows. In \Cref{sec:fusion_neutron_transm} we introduce fusion-neutron-driven transmutation. In \Cref{sec:governing} we introduce economic models for fusion systems producing isotopes and/or electricity. In \Cref{sec:neutron_wall_asymmetry} we examine how asymmetric neutron wall loading can boost feedstock burn rate. In \Cref{sec:market_size} we discuss the relation between market size, value per neutron, and fusion gain. In \Cref{sec:medicalradioisotopes} we present a short case study for a fusion medical radioisotope facility. We conclude in \Cref{sec:discuss}. Transmutation scalings for tokamaks and magnetic mirrors are presented in Appendices \ref{sec:trans_only} and \ref{sec:trans_only_mirror}.

Throughout this work we use default parameter values for our analysis (unless mentioned otherwise), listed in Appendix \ref{app:default_params}. The key quantities used in this work are described in \Cref{tab:tab0}.

\section{Fusion-Neutron-Driven Transmutation} \label{sec:fusion_neutron_transm}

%In this section we introduce transmutation driven by fusion neutrons.

D-T is the leading candidate fuel mix for creating high-energy neutrons at high flux due to its high reaction cross section at experimentally realizable temperatures. Importantly, the 14.1 MeV neutrons from this reaction are well above the binding energy of the least-bound neutron in most isotopes. At these energies, D–T neutrons can efficiently drive nuclear reactions that change the target nucleus mass number. For example, neutron multiplying (n,2n) reactions -- where a nucleus absorbs one neutron and subsequently emits two neutrons -- are all endothermic, but their threshold energies are usually below the D-T neutron birth energy: the lowest is $\sim$\qty{1.7}{MeV} for ${}^9\mathrm{Be}$ \cite{fischer1957cross} while the majority fall between 5--10\,MeV. Shown in \Cref{fig:n2n_thresholdenergy}, nearly all known $(\mathrm{n,2n})$ channels are accessible for \qty{14.1}{MeV} fusion neutrons.

The $(\mathrm{n,2n})$ reaction plays an important role in D-T fusion \cite{ihli2008review,gilbert2014comparative} because tritium is bred in the blanket through lithium reactions such as $^6\mrm{Li}(\mrm{n,t})\alpha$. Without in-situ tritium breeding, a D-T fusion system would depend on external supplies of tritium - which could be impractical above a certain fusion power. The transmutation chain for mercury-to-gold transmutation - including subsequent tritium breeding - is illustrated in \Cref{fig:blanket_schematic}.

Other threshold reactions such as (n,p) and (n,$\alpha$) are attractive candidates for radioisotope production \cite{li2023feasibility,evitts2025theoretical,parisi2025j} because they produce isotopes with different proton number to to the feedstock, allowing chemical isotope extraction. At fast neutron energy, because (n,p) and (n,$\alpha$) cross sections are typically at least one-to-three orders of magnitude lower than (n,2n) reactions, they are compatible with tritium self-sufficient systems as long as the feedstock has a significant (n,2n) cross section, low (n,$\gamma$) cross section, and meets the relevant requirements for chemical properties.

One example is the $\ce{^102Ru(n,\alpha)^99Mo}$ pathway \cite{gascoine2021towards}, which also has a (n,2n) cross section comparable to blanket multiplier candidates. Furthermore, blanket materials with (n,p) and (n,$\alpha$) pathways may not even need to be tritium self-sufficient. This is mainly due to two reasons. The first is that the value per neutron in selected (n,p) and (n,$\alpha$) pathways can be orders of magnitude higher than the value of neutrons used to breed tritium, and therefore funding tritium externally is economically viable, especially because machines using these pathways are likely to be small. Second, the (n,p) and (n,$\alpha$) cross sections are typically orders of magnitude smaller than (n,2n) \cite{parisi2025j}, and therefore are barely parasitic on tritium breeding (although their (n,2n) cross sections may also be smaller, resulting in lower neutron multiplication).

\begin{figure}[bt!]
    \centering
    \begin{subfigure}[]{\textwidth}
    \centering
    \includegraphics[width=\textwidth]{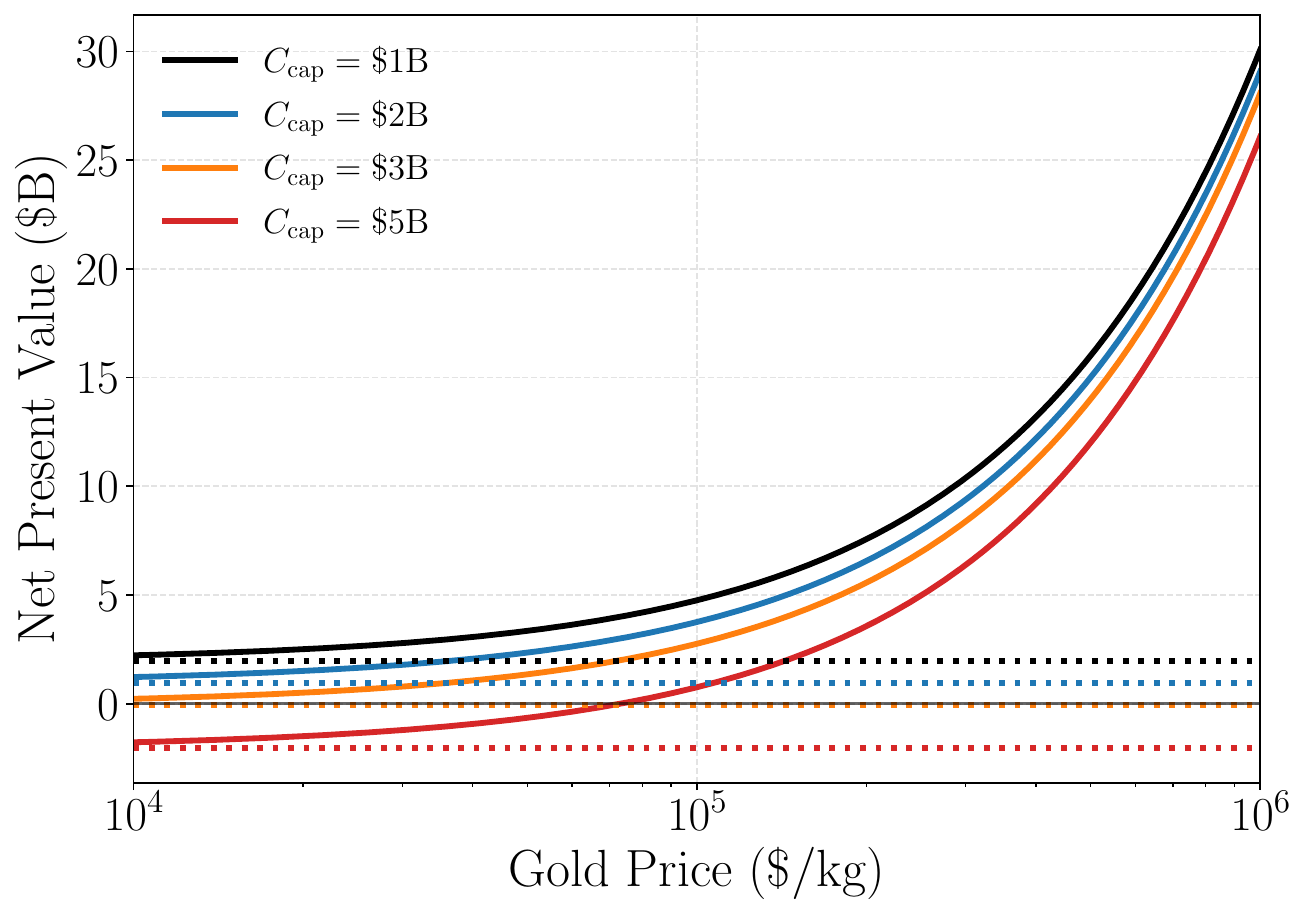}
    \end{subfigure}
    \caption{Net present value versus product price for a fusion power plant with 1GW$_\mrm{th}$ power, $Q_\mrm{plas} = 80$, and electricity price \$50/MWh. Solid curves: co-production of electricity and gold; dotted lines are electricity-only.}
    \label{fig:NPV_simpler}
\end{figure}

As a brief example of the economic utility of transmutation, we show how a fusion system co-producing electricity and gold has significantly improved economics: in \Cref{fig:NPV_simpler} we plot net present value (NPV) versus gold price for a 1GW$_\mrm{th}$ FPP with a plasma gain $Q_\mrm{plas} = 80$ and electricity valued at \$50/MWh. For the electricity-only option, the NPV is between \$$2$B and negative \$$2$B. However, co-producing gold alongside electricity significantly improves NPV by around \$$3$B at present (2026) gold prices (\$160k/kg): the present gold price of $\sim 1.6 \cdot 10^5$ \$/kg gives NPV$>0$ for all considered capital costs. Throughout this work, we will show how neutron-driven-transmutation significantly improves NPV in fusion systems.

\subsection{Neutron and Transmutation Rates}

\begin{figure}[tb]
    \centering
    \begin{subfigure}[tb]{0.79\textwidth}
    \centering
    \includegraphics[width=1.0\textwidth]{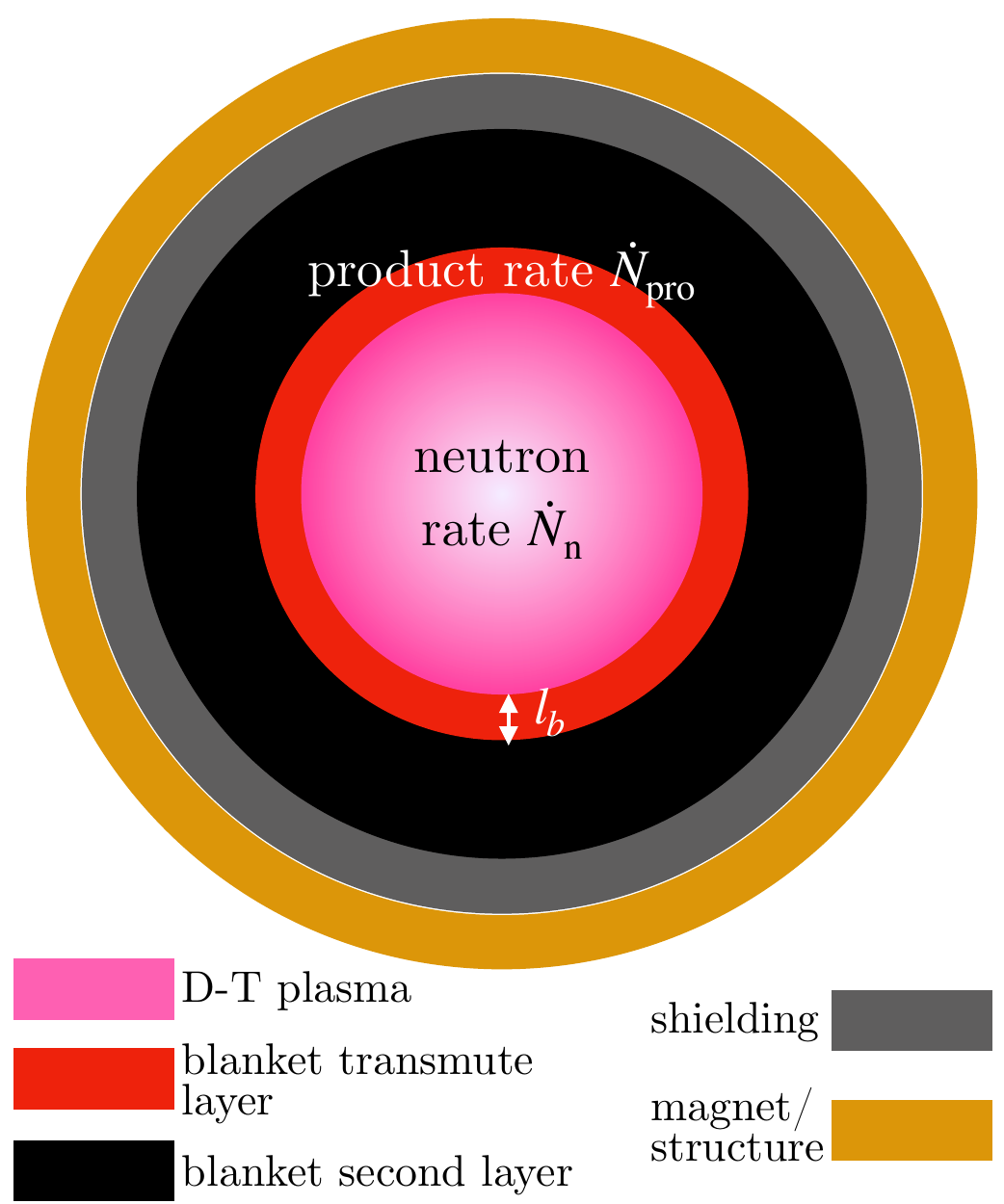}
    \end{subfigure}
    \caption{Simplified blanket model layout.}
    \label{fig:blanket_schematic_layout}
\end{figure}

A D-T system with fusion power $P_\mrm{fus}$ has neutron birth rate
\begin{equation}
    \dot{N}_\mathrm{n} = P_\mrm{fus} / E_\mrm{fus},
    \label{eq:Ndotn_intro}
\end{equation}
where $E_\mrm{fus}$ is the total energy released from a D-T reaction,
\begin{equation}
    E_\mrm{fus} = E_\mathrm{n} + E_\alpha = \qty{17.6}{MeV},
    \label{eq:Efus}
\end{equation}
and $\alpha$ particles have $E_{\alpha}=\qty{3.5}{MeV}$. In a fusion blanket surrounding the neutron source, shown schematically in \Cref{fig:blanket_schematic_layout}, the transmutation rate of a given isotope - the number of isotope atoms produced per second - is $\dot{N}_\mrm{pro}$. The neutron birth and transmutation rates are related by the neutron transmutation fraction,
\begin{equation}
    \eta_\mrm{pro} \equiv \frac{\dot{N}_\mrm{pro}}{\dot{N}_\mrm{n}},
    \label{eq:eta_prod}
\end{equation}
which is the fraction of all neutrons produced by fusion reactions that drive transmutation in the desired pathway.

In a simple model where the transmutation reaction driving $\dot{N}_\mrm{pro}$ dominates the total non-scattering cross section, the fraction of incoming neutrons that drive transmutation is
\begin{equation}
\eta_\mathrm{pro}  = 1 - \exp \left(- \tau \right), 
\label{eq:etapro_exponential_main}
\end{equation}
where 
\begin{equation}
    \tau \equiv \Sigma \, l_b.
\end{equation}
Here, the macroscopic cross section $\Sigma$ of the feedstock with a given number density $n_\mathrm{feed}$ is
\begin{equation}
    \Sigma \equiv \sigma \, n_\mathrm{feed},
    \label{eq:Sigma_macro}
\end{equation}
where $\sigma$ is the transmutation reaction cross section of interest and $l_b$ is the thickness of the blanket transmutation layer. Therefore, in our simple model the neutron transmutation fraction depends only on $l_b$ and $\Sigma$. 
%See \Cref{app:linear_model} for more details of the approximations made in this section. 
The initial feedstock inventory is
\begin{equation}
N_\mrm{feed,0} = n_\mathrm{feed,0} \, V_b
\end{equation}
where $n_\mathrm{feed,0}$ is the initial feedstock density and $V_b$ is the transmutation layer blanket volume. In the thin blanket limit - where the blanket thickness $l_b$ is thin relative to the distance from the neutron source - as shown in \Cref{fig:blanket_schematic_layout}, the initial feedstock is
\begin{equation}
N_\mrm{feed,0} \approx n_\mathrm{feed,0} \, l_b \, A_b,
\label{eq:Nfeed0_requirement}
\end{equation}
where $A_b$ is the surface area facing the D-T neutron source and the blanket thickness is,
\begin{equation}
l_b = \frac{| \ln ( 1 - \eta_\mathrm{pro} ) |}{\Sigma }.
\label{eq:blanket_thick}
\end{equation}
Therefore the transmutation rate is approximately
\begin{equation}
\dot{N}_\mathrm{pro} = \left[ 1 - \exp \left( - l_b \, \sigma \, n_\mathrm{feed,0}  \right)   \right] \, \dot{N}_\mathrm{n},
\end{equation}
which for fixed feedstock density, cross section, and neutron birth rate, can only increase with a thicker blanket. Motivated by algebraic simplicity, in this work we proceed with the thin blanket limit, but we caution that the approximation should always be checked.

\subsection{Feedstock burn rate and Inventory} \label{sec:FBR}

It may not always be practical to increase blanket thickness in order to increase $\dot{N}_\mathrm{pro}$. Two practical limitations could be feedstock scarcity and/or cost. For example, in mercury-to-gold transmutation, enriched ${}^{198}\mathrm{Hg}$ may be expensive or difficult to acquire, and the initial blanket loading could represent a large capital cost. Efficient feedstock utilization, quantified by the \emph{feedstock burn rate} (FBR), therefore becomes important: the faster feedstock is burned, the more efficiently it is used as a resource.

FBR provides a metric connecting physics, blanket design, and economics, as it directly determines how much product value is extracted per unit of initial feedstock. In this section, we develop a simple model for calculating the FBR and estimating the required feedstock inventory in fusion transmutation systems.

We consider the FBR over one-year periods, which we call the annual FBR,
\begin{equation}
  \mathrm{FBR}_a \equiv \frac{\int_\mrm{1 \;yr} \dot{N}_\mrm{pro} \, dt}{N_\mrm{feed,0}},
  \label{eq:Befficiency}
\end{equation}
where $t$ is time. Recent work has shown that the annual FBR is several fractions of a percent in a tokamak power plant \cite{rutkowski2025scalable}, although as we show in this work, the FBR can become significantly larger with various optimizations and in fusion other concepts with higher neutron flux on the feedstock.

\begin{figure}[bt!]
    \centering
    \begin{subfigure}[t]{0.92\textwidth}
    \centering \includegraphics[width=1.0\textwidth]{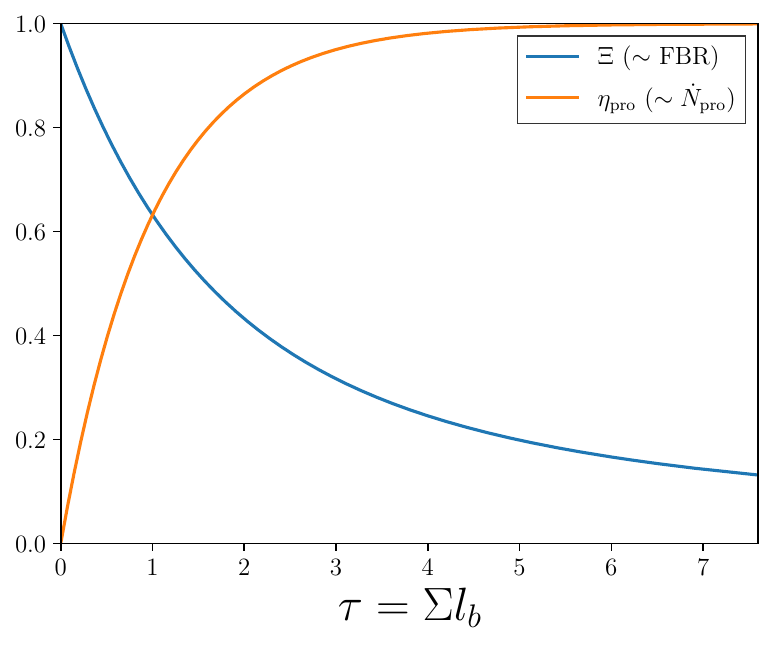}
    \caption{}
    \end{subfigure}
    \centering
    \begin{subfigure}[t]{0.92\textwidth}
    \centering \includegraphics[width=1.0\textwidth]{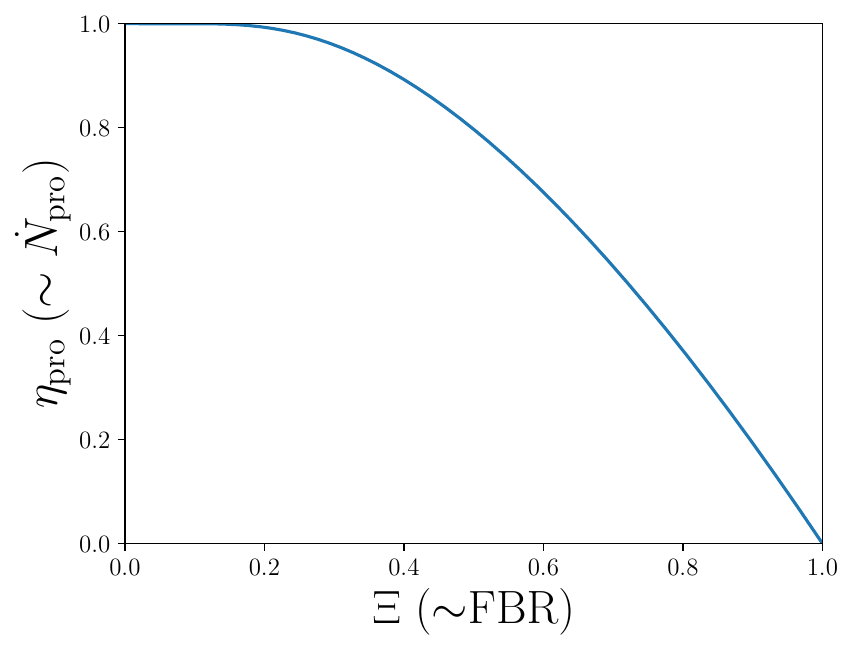}
    \caption{}
    \end{subfigure}
    \caption{(a) $\Xi$ (\Cref{eq:Xiprefac}) and $\eta_\mathrm{pro}$ (\Cref{eq:eta_prod,eq:eta_prod}) versus $\Sigma l_b$ and (b) $\eta_\mathrm{pro}$ versus $\Xi$.}
    \label{fig:FBR_general}
\end{figure}

%The FBR is an important quantity because it couples the neutron rate from a D-T source $\dot{N}_\mathrm{n}$ with the blanket transmutation rate $\dot{N}_\mrm{pro}$ and the initial feedstock inventory $N_\mrm{feed,0}$.

Fusion systems must also manage lithium inventories, since ${}^{6}\mathrm{Li}$ - comprising only 7.5\%at of natural lithium - has a high $(\ce{n,t})$ cross section at thermal energies -- and enriched ${}^{6}\mathrm{Li}$ is widely assumed in breeding designs \cite{hartley1978potential,fasel2005availability,de2008lead,de2021lithium}. Therefore any tritium-breeding fusion system using lithium must also consider the lithium FBR, as lithium inventory can represent a major cost \cite{ward2025lithium}.

\subsection{Constant Transmutation Limit}

We make the simplifying assumption that $\dot{N}_\mathrm{pro}$ is constant over a year timescale, giving the annual number of product nuclei
\begin{equation}
    N_\mathrm{pro,a} = \int_\mrm{1 \;yr} \dot{N}_\mrm{pro} \, dt \approx \dot{N}_\mrm{pro} \, T_\mrm{year}.
    \label{eq:Npro_annual}
\end{equation}
The constant $\dot{N}_\mrm{pro}$ limit simplifies the expression for $\mathrm{FBR}_a$ in \Cref{eq:Befficiency},
\begin{equation}
\mathrm{FBR}_a \simeq \eta_\mathrm{pro}  \frac{\dot{N}_\mathrm{n}}{N_\mathrm{feed,0}} T_\mathrm{year} = \Xi \frac{\sigma}{A_b} \frac{P_\mathrm{fus} }{E_\mathrm{fus}} T_\mathrm{year},
\label{eq:FBR_heuristic}
\end{equation}
where $\Xi$ is a prefactor depending on $\eta_\mathrm{pro}$,
\begin{equation}
\Xi \equiv \frac{\eta_\mathrm{pro}}{ | \ln ( 1 - \eta_\mathrm{pro} ) |},
\label{eq:Xiprefac}
\end{equation}
and $T_\mathrm{year}$ is the number of seconds in a year.

\begin{figure}[bt!]
    \centering
    \begin{subfigure}[t]{0.92\textwidth}
    \centering
    \includegraphics[width=1.0\textwidth]{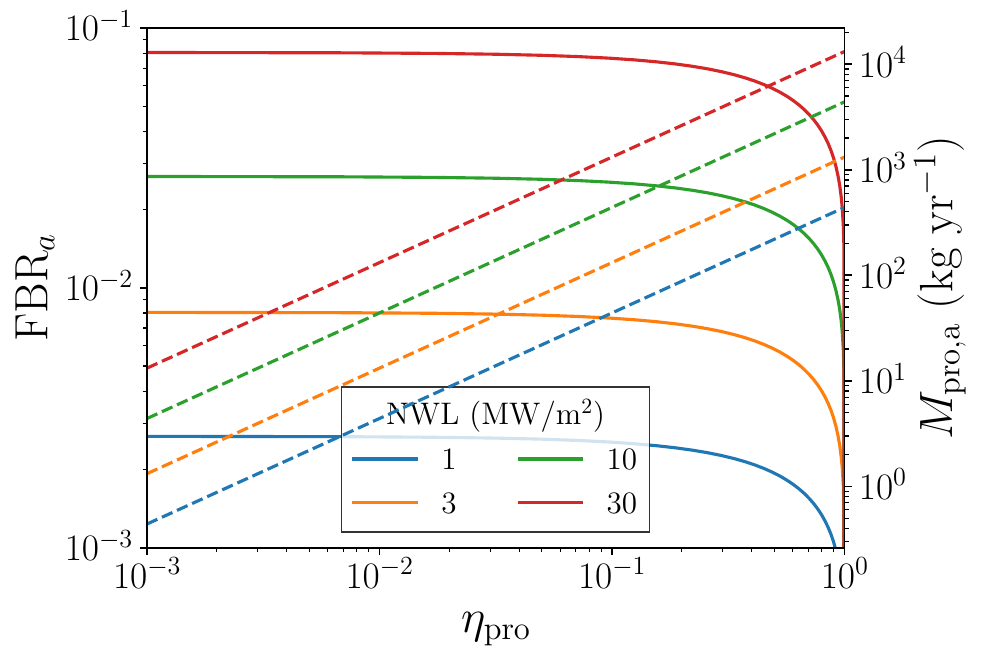}
    \caption{}
    \end{subfigure}
    \centering
    \begin{subfigure}[t]{0.92\textwidth}
    \centering
    \includegraphics[width=1.0\textwidth]{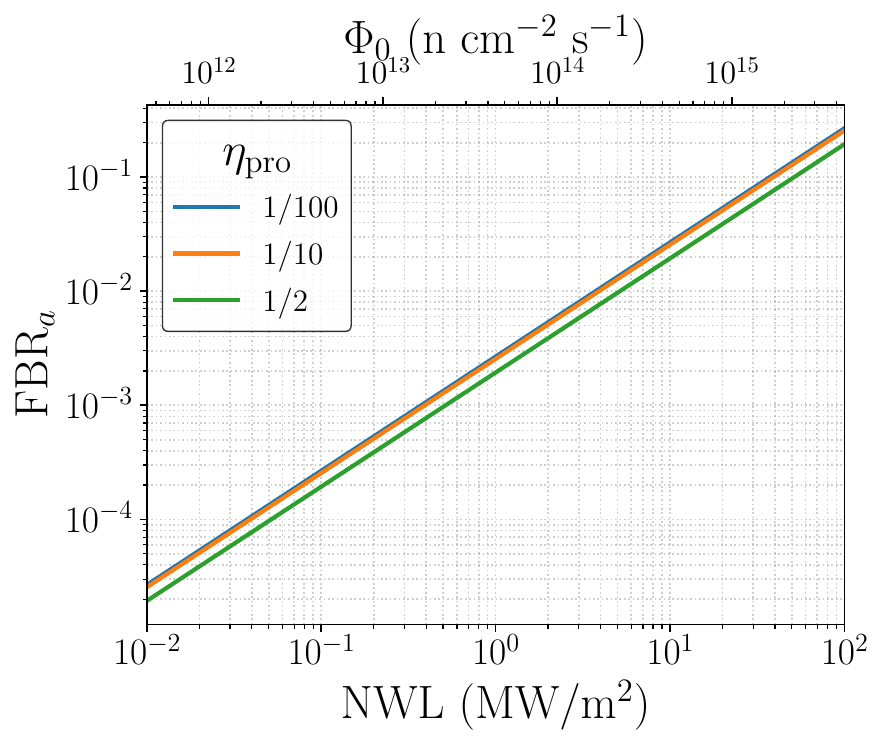}
    \caption{}
    \end{subfigure}
    \caption{$\mrm{FBR}_a$ (solid lines, left y-axis) and product in kg per year (dashed lines, right y-axis) versus (a) neutron capture efficiency ($\eta_\mathrm{pro}$) and (b) neutron wall loading (NWL). We assume $\sigma$ = 2b and production of $\ce{^197Au}$ from $\ce{^198Hg}$.}
    \label{fig:FBR_examples}
\end{figure}

\noindent In \Cref{fig:FBR_general}(a) we plot $\Xi$ versus $\Sigma l_b$, showing how $\Xi$ becomes small for $\Sigma l_b \gg 1$. Because FBR is proportional to $\Xi$, thinner blankets (lower $l_b$) achieve much higher FBR. However, the tradeoff is thinner blankets also produce less total product (i.e. $\dot{N}_\mrm{pro}$ is lower) because a smaller fraction of neutrons $\eta_\mrm{pro}$ are absorbed --- this is shown by $\eta_\mrm{pro}$ versus $\Sigma l_b$ in \Cref{fig:FBR_general}(a). Thus there is a tradeoff between total product $\sim \dot{N}_\mathrm{pro}$ and FBR, shown by $\eta_\mathrm{pro}$ versus $\Xi$ in \Cref{fig:FBR_general}(b). Both total product and FBR are important, and the optimal point between them depends on the transmutation pathway and overall machine design. \Cref{fig:FBR_general}(b) suggests optima for the tradeoff between $\eta_\mrm{pro}$ and $\Xi$: very low values of $\Xi$ unnecessarily decrease FBR without much gain in $\eta_\mrm{pro}$. However, for $\Xi \approx 0.5$, $\eta_\mrm{pro}$ only decreases from 1.0 to approximately 0.8, indicating a favorable design region. 

We now use plausible power plant values to determine the tradeoff between FBR and total product mass per year,
\begin{equation}
    M_\mathrm{pro,a} = N_\mathrm{pro,a} \, m_\mathrm{pro} = \eta_\mathrm{pro} P_\mathrm{fus} \frac{m_\mathrm{pro} T_\mrm{year}}{E_\mrm{fus}},
    \label{eq:Mpro_annual}
\end{equation}
where $m_\mathrm{pro}$ is the mass of a single product atom. It is also helpful to write the FBR in terms of the average neutron wall loading (NWL),
\begin{equation}
\mathrm{NWL}  \equiv \frac{P_\mathrm{n} }{A_b},
\label{eq:p_NWL}
\end{equation}
where $P_\mathrm{n} = (4/5) P_\mathrm{fus}$ is the neutron power. \Cref{eq:FBR_heuristic} therefore becomes
\begin{equation}
\mathrm{FBR}_a \simeq \frac{5}{4} \, \Xi \, \sigma \, \mathrm{NWL} \frac{T_\mathrm{year} }{E_\mathrm{fus}}.
\label{eq:FBR_heuristic_nwl}
\end{equation}

% in kilograms. 
%To obtain the final relation in \Cref{eq:Mpro_annual}, we assumed the annual number of atoms produced is

\begin{figure}[tb]
    \centering
    \begin{subfigure}[tb]{0.99\textwidth}
    \centering
    \includegraphics[width=1.0\textwidth]{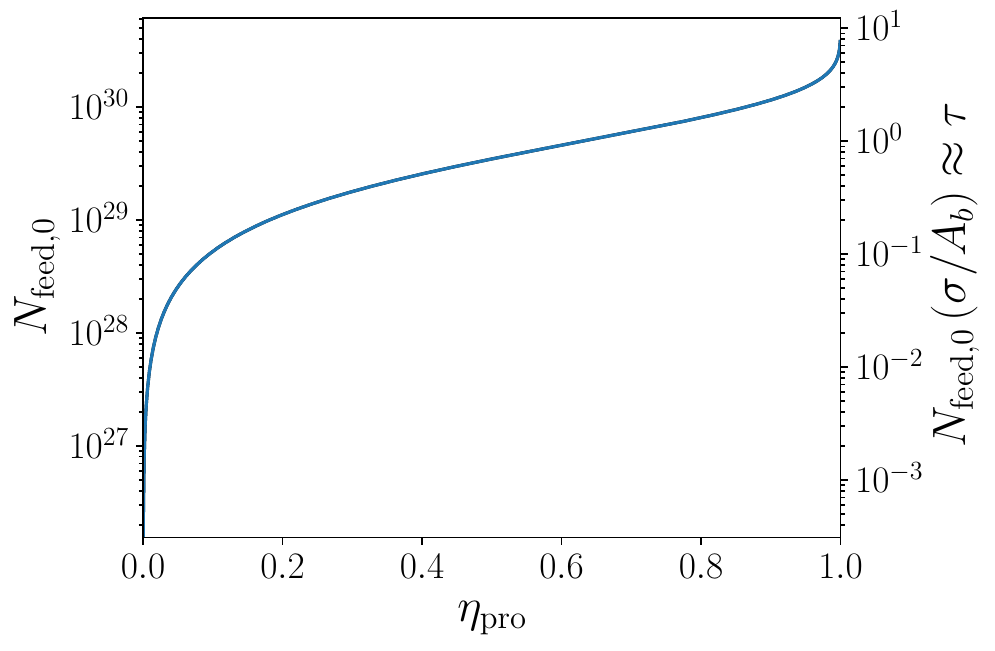}
    \end{subfigure}
    \caption{Left y-axis: inventory $N_\mrm{feed,0}$ versus $\eta_\mathrm{pro}$ (\Cref{eq:Nfeed0scaling}) for ${}^{198}\mathrm{Hg}$. Right y-axis $N_\mrm{feed,0} \sigma/A_b \approx \tau$ (\Cref{eq:Nfeed0normalizedscaling}).}
    \label{fig:Nfeed0}
\end{figure}

In \Cref{fig:FBR_examples}(a) we plot $\mrm{FBR}_a$ and $M_\mathrm{pro,a}$ versus $\eta_\mathrm{pro}$ in a \ce{^198Hg} blanket for four NWL values. Increasing the NWL always increases FBR and total gold production. However, in realistic systems the FBR cannot increase indefinitely through higher power density because of power limits on the plant walls from alpha particle heating of the plasma, and associated constraints around heat rejection from the system. \Cref{fig:FBR_examples}(b) shows $\mrm{FBR}_a$ versus NWL. The top x-axis of \Cref{fig:FBR_examples}(b) also shows the average first-wall neutron flux,
\begin{equation}
    \Phi_0 \equiv \frac{\dot{N}_\mathrm{n}}{A_b} = \frac{5}{4} \frac{\mathrm{NWL}}{E_\mathrm{fus}}.
\end{equation}
The tradeoff between blanket inventory and $\eta_\mathrm{pro}$ is
\begin{equation}
N_\mrm{feed,0} \approx \frac{A_b}{\sigma} | \ln ( 1 - \eta_\mathrm{pro} ) | = \frac{A_b}{\sigma} \tau.
\label{eq:Nfeed0scaling}
\end{equation}
We plot solutions to \Cref{eq:Nfeed0scaling} in \Cref{fig:Nfeed0}. The left y-axis shows $N_\mrm{feed,0}$ for \ce{^{198}Hg}, assuming the $(\mathrm{n},\mathrm{2n})$ reaction has a cross section $\sigma = 2\,$b. The right y-axis shows the normalized quantity
\begin{equation}
N_\mrm{feed,0} \frac {\sigma}{A_b} \approx \tau.
\label{eq:Nfeed0normalizedscaling}
\end{equation}
There are many refinements that can be made to the above model, such as time-dependent depletion, a thicker blanket, and multi-species neutron reactions. However, while the model is simple, it is in decent agreement with the Monte Carlo simulations in \cite{rutkowski2025scalable} where a $P_\mrm{fus} = \qty{1.5}{GW}$ FPP with $\mrm{FBR}_a = 0.005$ produces $M_\mrm{pro,a} \approx \qty{3000}{kg/yr}$; our simple model in \Cref{fig:FBR_examples}(b) shows the same FPP produces $M_\mrm{pro,a} \approx \qty{3500}{kg/yr}$. Monte Carlo neutronics simulations contain additional effects so very close agreement with our simple linear model is not expected. Many valuable transmutation products are radioisotopes that must be extracted much faster than their characteristic decay times -- we discuss this in Appendix \ref{app:radioisotopes}.

\subsection{Power Density Constraints} \label{subsec:powerdensity}

The performance of a fusion transmutation system is constrained not only by nuclear physics in the blanket but also by the tolerance of materials facing the plasma. Constraints on wall heat flux \cite{ueda2017baseline}, erosion \cite{kirschner2023erosion}, and neutron loading \cite{fischer2015neutronics} restrict the maximum achievable power density in the blanket and thus the attainable transmutation rate. In this section, we examine how wall heat-flux limits constrain the relationship between power density, plasma performance, and transmutation efficiency. The local value of $p_\mrm{wall}$ is determined by material constraints on power handling and typically cannot exceed $\sim$\qty{10}{MW/m^2} for solid walls \cite{kotschenreuther2007heat} as an absolute upper value - because  $p_\mrm{wall}$ typically varies spatially across the wall, the maximum achievable $\langle p_\mrm{wall} \rangle$ will typically be lower than $\sim$\qty{10}{MW/m^2}.

\begin{figure}[tb]
    \centering
    \begin{subfigure}[tb]{0.95\textwidth}
    \centering
\includegraphics[width=1.0\textwidth]{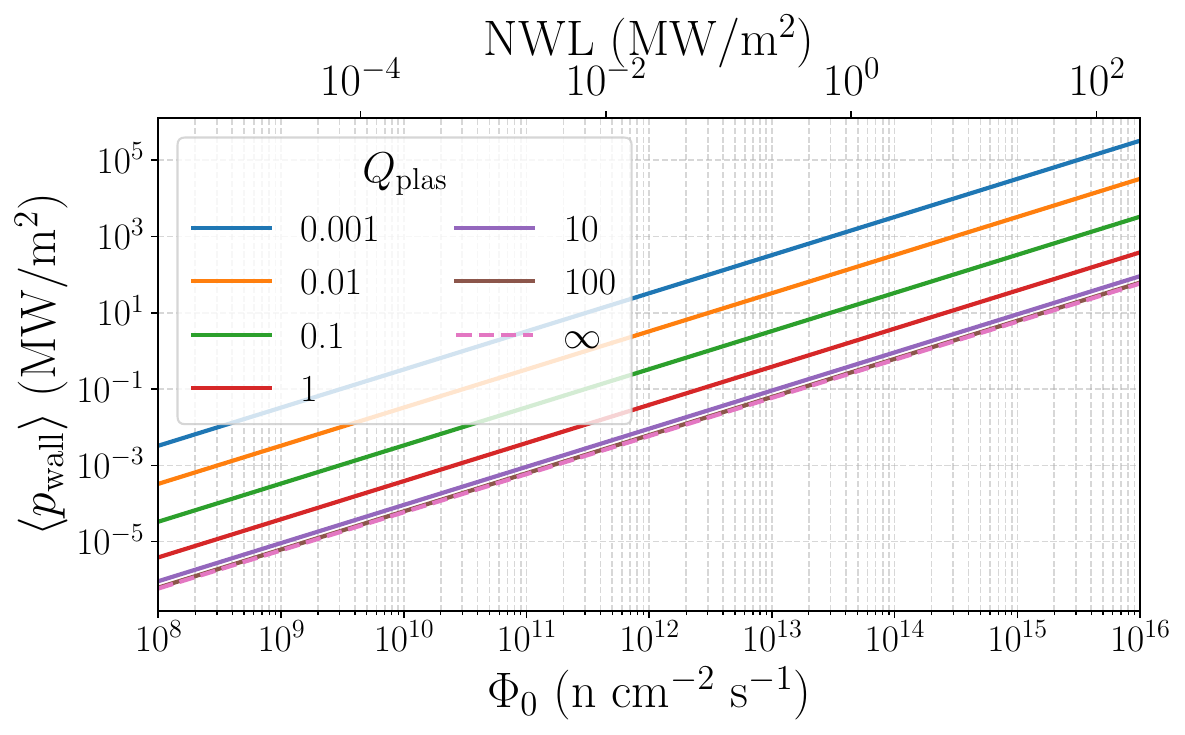}
    \caption{}
    \end{subfigure}
    \centering
    \begin{subfigure}[tb]{0.95\textwidth}
    \centering
\includegraphics[width=1.0\textwidth]{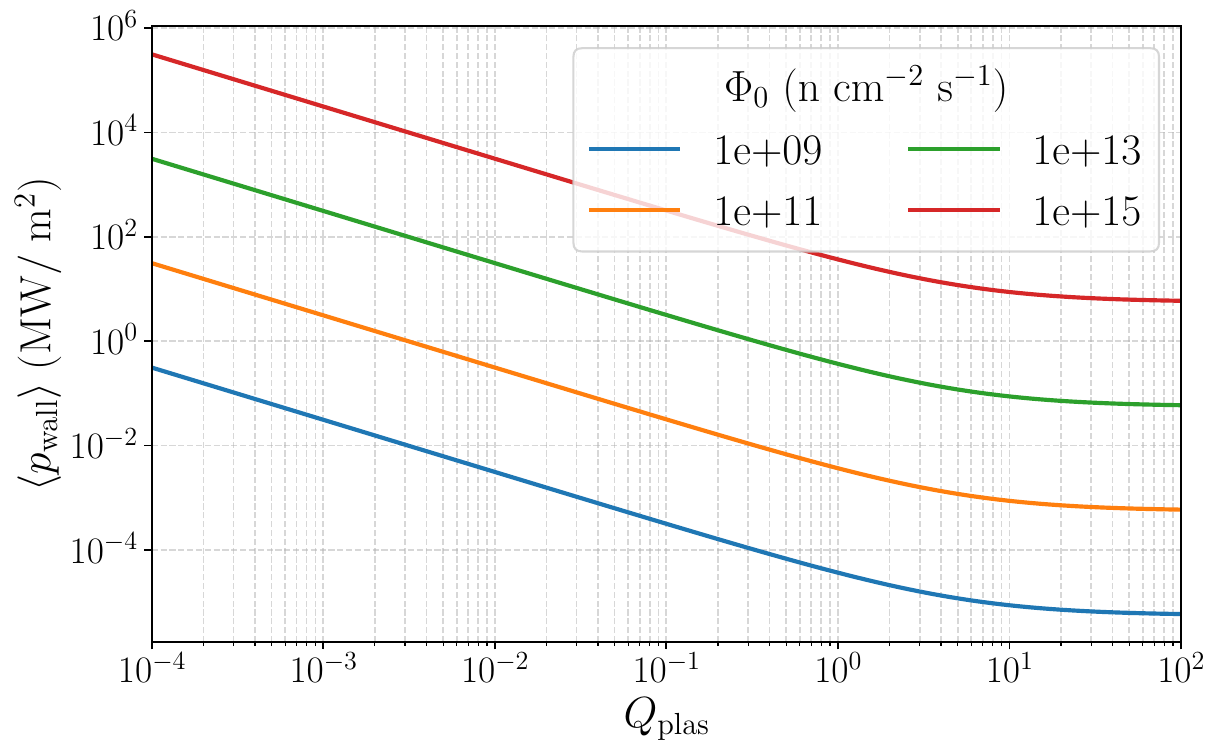}
    \caption{Fixed fusion power.}
    \end{subfigure}
    \centering
    \begin{subfigure}[tb]{0.95\textwidth}
    \centering
\includegraphics[width=1.0\textwidth]{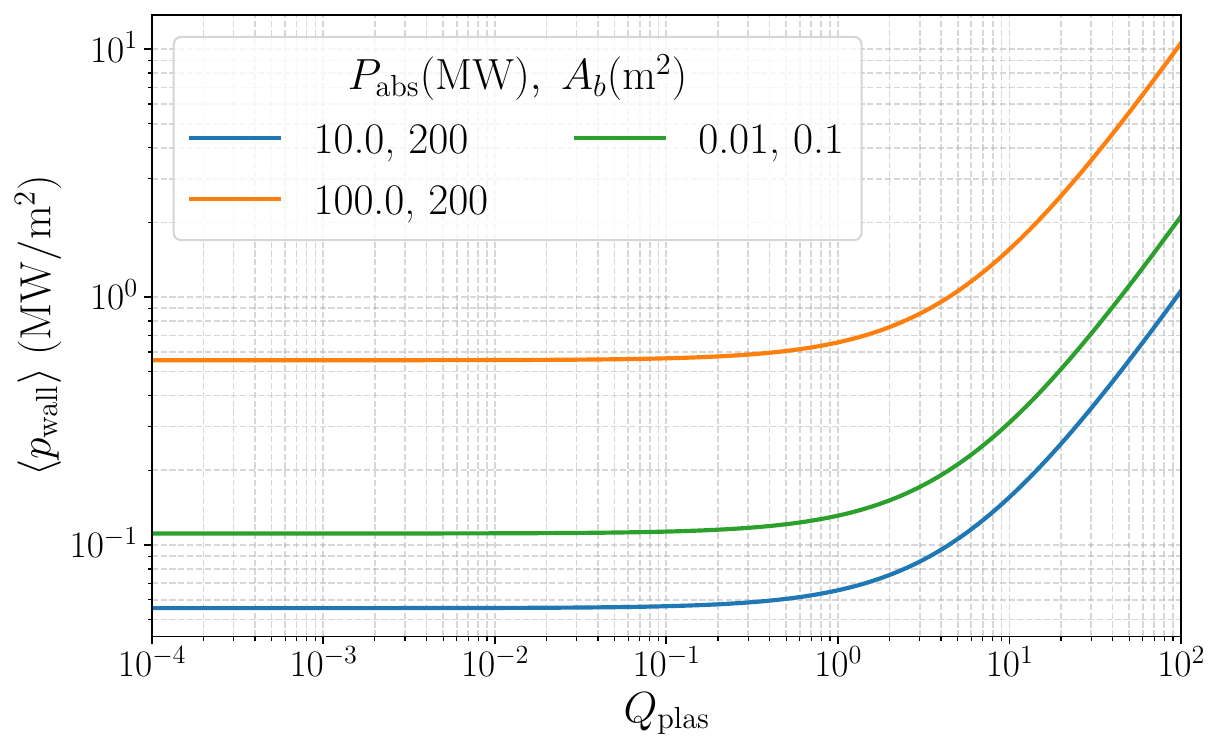}
    \caption{Each curve has fixed absorbed plasma heating and fixed blanket area.}
    \end{subfigure}
    \caption{(a) Averaged wall power per unit area $\langle p_\mathrm{wall} \rangle $ versus neutron flux for different $Q_\mrm{plas}$ values. (b) $\langle p_\mathrm{wall} \rangle $ versus $Q_\mrm{plas}$ at fixed values of $\Phi_0$. (c) $\langle p_\mathrm{wall} \rangle $ versus $Q_\mrm{plas}$ at fixed values of $P_\mathrm{heat} $ and $A_b$ (see \Cref{eq:pwall_fixed_heating}).}
    \label{fig:pwall_neutron_flux}
\end{figure}

\begin{figure*}[tb]
    \centering
    \begin{subfigure}[tb]{0.48\textwidth}
    \centering
    \includegraphics[width=1.0\textwidth]{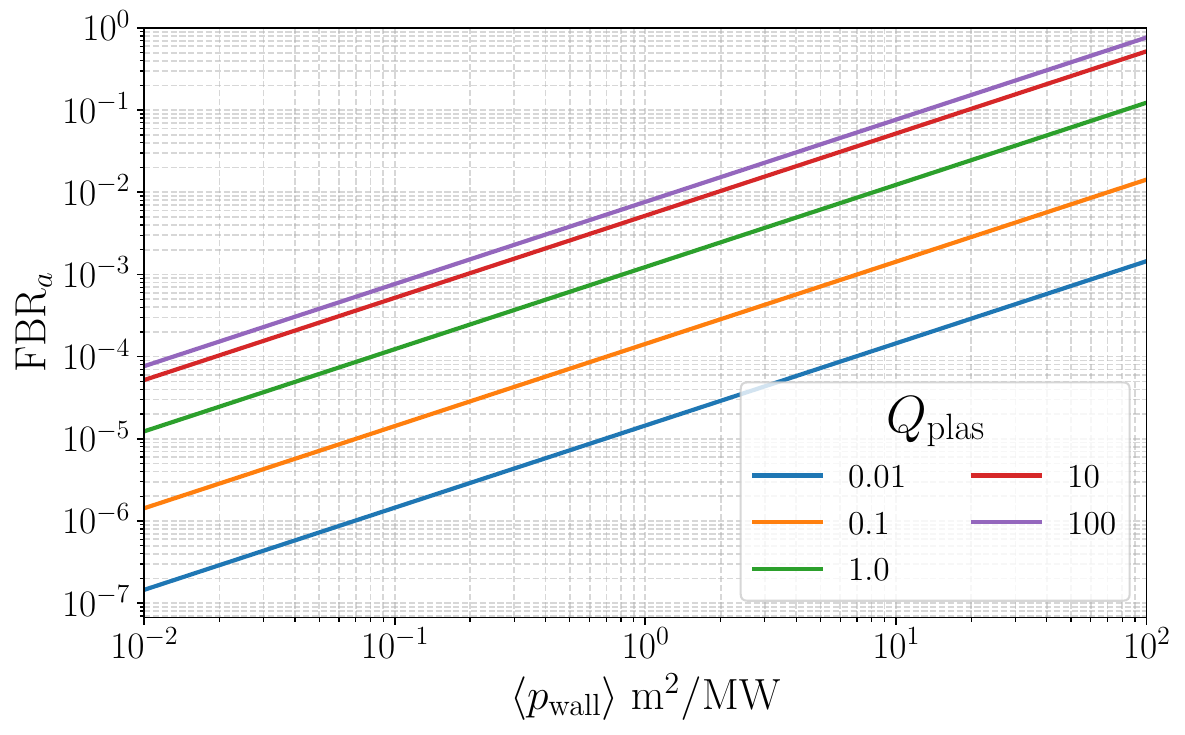}
    \caption{}
    \end{subfigure}    
    \begin{subfigure}[tb]{0.48\textwidth}
    \centering
    \includegraphics[width=1.0\textwidth]{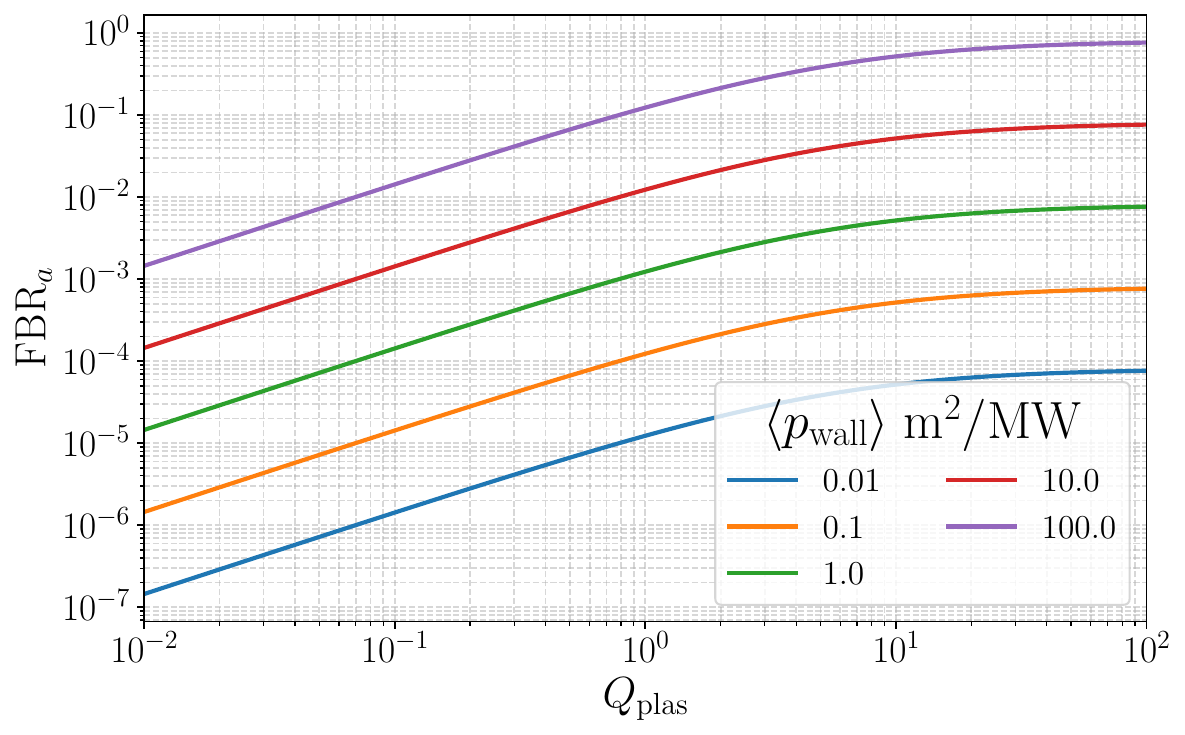}
    \caption{}
    \end{subfigure}    
    \caption{$\mrm{FBR}_a$ versus (a) $\langle p_\mathrm{wall} \rangle$ and (b) plasma gain $Q_\mathrm{plas}$. We use $\sigma = 2.0$ b, $\eta_\mathrm{pro} =0.5$.}
    \label{fig:FBR_etapro_pwall}
\end{figure*}

The total non-neutron power to the first wall surface of a fusion machine is
\begin{equation}
    P_\mrm{wall} = P_\mrm{\alpha} + P_\mrm{heat},
\end{equation}
where $P_\alpha$ is the alpha particle heating power and $P_\mrm{heat}$ is the external heating power injected into the plasma. The average power per unit area to the plasma-facing wall is given by the fusion alpha power plus the heating power,
\begin{equation}
    \langle p_\mrm{wall} \rangle \equiv \frac{P_\mrm{wall}}{A_b} = E_\mrm{fus} \, \Phi_0 \, \left(  f_\alpha  + \frac{1}{ \eta_\mrm{abs} Q_\mrm{plas}} \right),
    \label{eq:pwall_initial}
\end{equation}
where $f_\alpha = 1/5$ is the fusion power fraction carried by alpha particles, $\eta_\mrm{abs}$ is the plasma heating absorption efficiency,
\begin{equation}
\eta_\mrm{abs} \equiv \frac{P_\mrm{abs}}{P_\mrm{heat}}
\label{eq:etaabs}
\end{equation}
where $P_\mrm{abs}$ is the total absorbed heating power in a plasma, and $Q_\mrm{plas}$ is the plasma gain
\begin{equation}
    Q_\mrm{plas} \equiv \frac{P_\mrm{fus}}{ P_\mrm{abs}} = \frac{P_\mrm{fus}}{\eta_\mrm{abs} P_\mrm{heat}}.
    \label{eq:Qplas}
\end{equation}
In \Cref{fig:pwall_neutron_flux}(a) we plot $\langle p_\mrm{wall} \rangle$ versus $\Phi_0$ for different $Q_\mrm{plas}$ values assuming $\eta_\mrm{abs} = 0.9$ - this is a fairly general result that shows increasing $Q_\mrm{plas}$ is beneficial because it decreases the heat flux per neutron. We emphasize this in \Cref{fig:pwall_neutron_flux}(b), showing the importance of increasing $Q_\mrm{plas}$ for lowering $\langle p_\mrm{wall} \rangle$ at sufficiently low values of $\eta_\mathrm{abs} Q_\mrm{plas}$. When $Q_\mathrm{plas}$ satisfies
\begin{equation}
Q_\mrm{plas}  \ll \frac{1}{ \eta_\mrm{abs} f_\alpha}, 
\end{equation}
the wall heat loading scales as
\begin{equation}
\langle p_\mrm{wall} \rangle \approx \frac{ E_\mrm{fus} \Phi_0}{ \eta_\mathrm{abs} Q_\mathrm{plas}} \sim Q_\mathrm{plas}^{-1}.
\label{eq:pwall_fixed_heating}
\end{equation} 
Given that typically $\eta_\mathrm{abs} f_\alpha \gtrsim 0.18$, increasing fusion performance by increasing $Q_\mathrm{plas}$ when $Q_\mathrm{plas} \lesssim 1$ has a significant benefit for reducing $\langle p_\mrm{wall} \rangle$. Note that each curve in \Cref{fig:pwall_neutron_flux}(b) has fixed fusion power. If $Q_\mathrm{plas}$ changes with fusion power, the wall heat flux becomes
\begin{equation}
    \langle p_\mrm{wall} \rangle = \frac{Q_\mrm{plas} P_\mathrm{abs} }{A_b}  \left(  f_\alpha  + \frac{1}{ \eta_\mrm{abs} Q_\mrm{plas}} \right).
    \label{eq:pwall_later}
\end{equation}

We plot \Cref{eq:pwall_later} in \Cref{fig:pwall_neutron_flux}(c): at fixed absorbed plasma heating $P_\mathrm{abs}$ and fixed wall area $A_b$, increasing $Q_\mathrm{plas}$ will eventually increase $ \langle p_\mrm{wall} \rangle$ as the fusion power and therefore alpha heating increase. Assuming that annual feedstock depletion is low ($\mrm{FBR}_a \ll 1$), using \Cref{eq:FBR_heuristic,eq:pwall_initial} we find the feedstock burn rate is
\begin{equation}
    \mathrm{FBR}_a \simeq \frac{\Xi \sigma \langle p_\mrm{wall} \rangle}{\left(f_\alpha + 1/\eta_\mrm{abs} Q_\mrm{plas} \right)} \frac{T_\mrm{year}}{E_\mrm{fus}} = \Xi \, \sigma \, \Phi_0 \, T_\mrm{year}.
    \label{eq:FBR_pwall}
\end{equation}

%there is a maximum possible wall power density $p_\mrm{wall,max}$. This sets an upper bound on the maximum-achievable $\mrm{FBR}_a$,
%\begin{equation}
%    \mathrm{FBR}_a \lesssim \Xi \sigma p_\mrm{wall,max} \frac{T_\mrm{year}}{f_\alpha E_\mrm{fus}}.
%    \label{eq:FBR_max}
%\end{equation}
While there are speculative methods for increasing $\mathrm{FBR}_a$ explored later in this work, \Cref{eq:FBR_pwall} sets an upper limit based on just three variables: the neutron reaction cross section $\sigma$, the maximum-tolerable wall power $p_\mrm{wall,max}$, and the neutron transmutation fraction $\eta_\mrm{pro}$ (which enters through $\Xi$). 

In \Cref{fig:FBR_etapro_pwall}(a) we plot $\mrm{FBR}_a$ versus $\langle p_\mathrm{wall} \rangle$ for different $Q_\mathrm{plas}$ values. While high $\mrm{FBR}_a$ values ($\mrm{FBR}_a \gtrsim 0.1$) are attainable at high $\langle p_\mathrm{wall} \rangle$ and high $Q_\mathrm{plas}$ values, they correspond to $\Phi_0 \gtrsim \qty{e15}{\per\cm\squared}$, which generally far exceeds neutron source, wall power handling, and material neutronics capabilities for the foreseeable future in magnetic confinement fusion. In \Cref{fig:FBR_etapro_pwall}(b) we plot $\mrm{FBR}_a$ versus $Q_\mathrm{plas}$.

These results underscore the importance of wall materials that can tolerate a sufficiently high heat power loading, an area of significant research in the fusion research program \cite{Krasheninnikov2003,Maingi2012,Boyle_2023,Berkery2024,Verhaegh_2024}. They also highlight the important role of $Q_\mathrm{plas}$: in fusion energy systems $Q_\mathrm{plas}$ is a scientific measure of the ratio of fusion power to absorbed heating power. In transmutation systems, $Q_\mathrm{plas}$ describes (i) how efficiently neutrons are produced per unit of absorbed power (and therefore approximately the electricity cost per neutron), and (ii) how much wall power each neutron is responsible for. Therefore, at a fixed neutron rate, increasing $Q_\mathrm{plas}$ is beneficial for feedstock burn rate, as well as reducing the electricity cost per neutron and lowering the wall heat flux. A recent work \cite{parisi2025k} demonstrated how muon-catalyzed fusion offers a path to achieve high NWL without correspondingly high heat flux due to the absence of external heating, although the viability of scaling muon-catalyzed fusion systems to high fusion power remains unclear.

%The fusion power density wall constraint is areal, but neutrons damp volumetrically.

\subsection{Target Thickness Considerations} \label{subsec:target_thickness}

The feedstock inventory, target thickness, and blanket area are related by
\begin{equation}
    N_\mrm{feed,0} = n_\mrm{feed} \, l_b \, A_b,
    \label{eq:Nfeed_geometry}
\end{equation}
where $n_\mrm{feed}$ is the feedstock number density in the blanket. The fusion machine sets a fixed wall flux $\Phi_0$ (determined by machine power and geometry), so the blanket intercepts $\dot{N}_\mrm{n} = \Phi_0 A_b$ neutrons per second.
Using \Cref{eq:FBR_pwall}, the annual feedstock burn rate is
\begin{equation}
    \mrm{FBR}_a = \Xi(\tau) \, \sigma \, \Phi_0 \, T_\mrm{year},
    \qquad \tau \equiv \sigma n_\mrm{feed} l_b = \frac{\sigma N_\mrm{feed,0}}{A_b},
    \label{eq:FBR_geometry}
\end{equation}
where $\Xi(\tau) = (1 - e^{-\tau})/\tau$ is a monotonically decreasing function of $\tau$. Because $\tau = \Sigma l_b$ grows with thickness, $\Xi$ and hence $\mrm{FBR}_a$ both decrease with $l_b$ at fixed $\Phi_0$. The physical reason is that atoms deep in a thick target are shielded by the front layers and see an exponentially attenuated flux, lowering the average per-atom burn rate. Conversely a thin target spread over a large area exposes all atoms to nearly the full wall flux, maximizing FBR$_a$.

At fixed $N_\mrm{feed,0}$, \Cref{eq:Nfeed_geometry} forces a tradeoff: larger area $A_b$ means thinner target (smaller $l_b$, smaller $\tau$, higher $\Xi$, higher $\mrm{FBR}_a$) but also lower single-pass capture efficiency $\eta_\mrm{pro} = 1 - e^{-\tau}$ and lower total product yield.

\begin{figure}[t!]
    \centering
    \begin{subfigure}[t]{0.92\textwidth}
    \centering
    \includegraphics[width=1.0\textwidth]{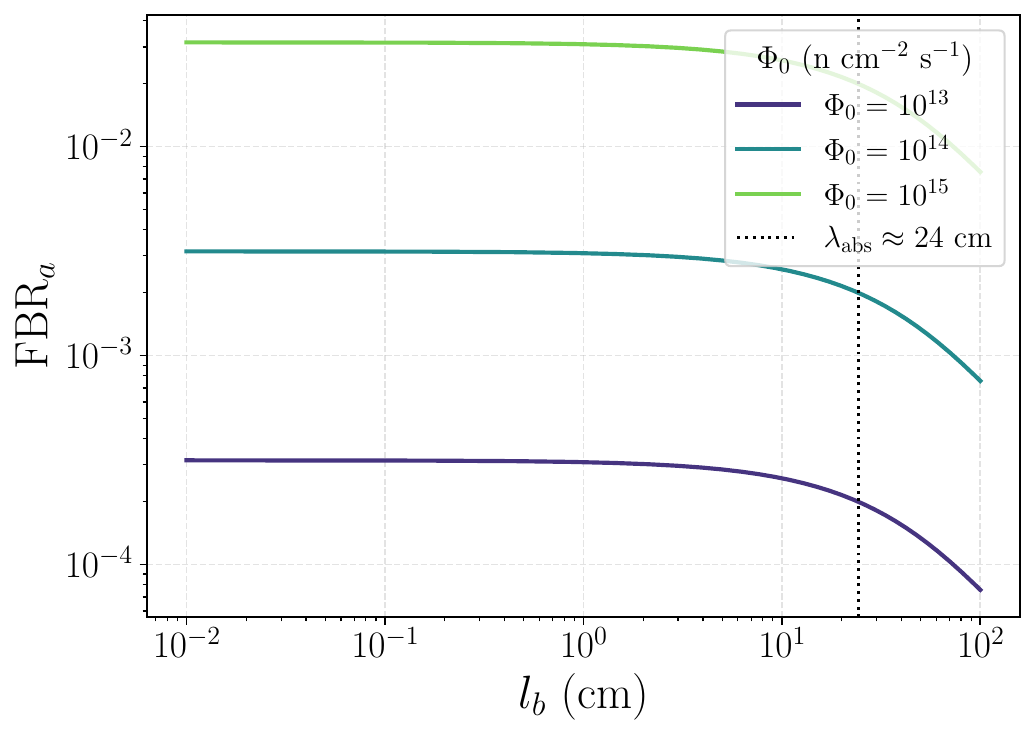}
    \caption{}
    \end{subfigure}
    \begin{subfigure}[t]{0.92\textwidth}
    \centering
    \includegraphics[width=1.0\textwidth]{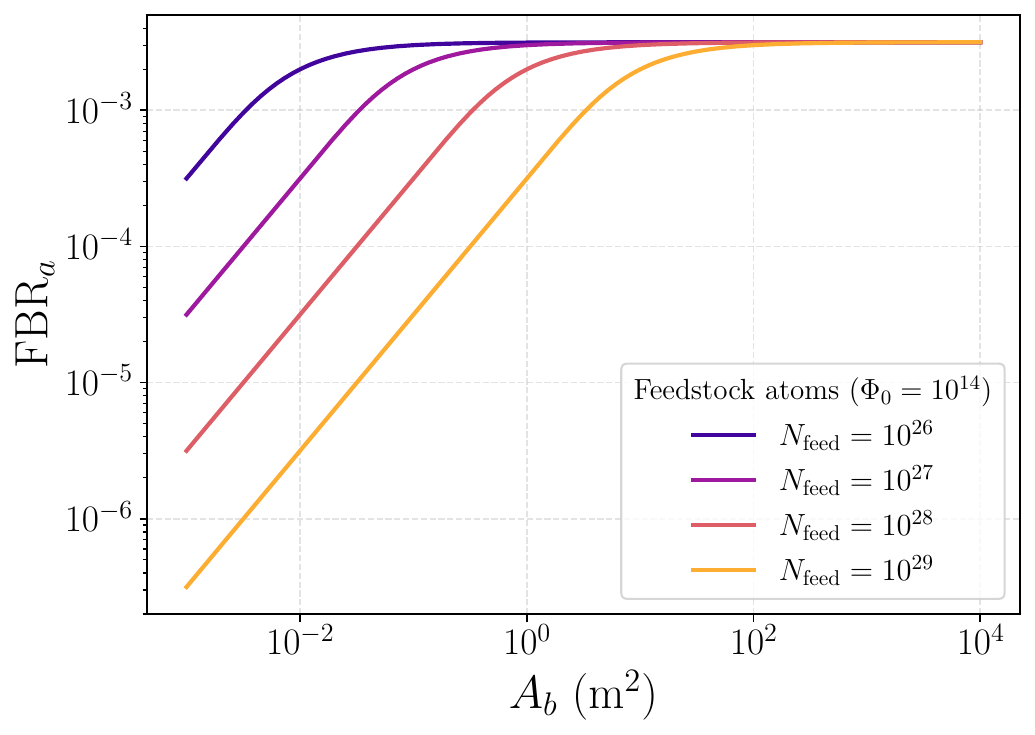}
    \caption{}
    \end{subfigure}
    \caption{$\mrm{FBR}_a$ (\Cref{eq:FBR_geometry}) versus (a) target thickness $l_b$ for
    three wall fluxes $\Phi_0$, and (b) blanket area $A_b$ for four feedstock inventories
    $N_\mrm{feed,0}$ at fixed $\Phi_0 = 10^{14}$~n~cm$^{-2}$~s$^{-1}$.
    We assume $\sigma = 1\,$b and $n_\mrm{feed} \approx 4.1\times 10^{22}$~cm$^{-3}$ (liquid mercury).
    The dotted line in (a) marks the absorption mean free path
    $\lambda_\mrm{abs} = (\sigma n_\mrm{feed})^{-1} \approx 24$~cm.
    In both panels $\mrm{FBR}_a$ decreases with $l_b$ and increases with $A_b$: thinner
    targets spread over larger areas burn feedstock more efficiently per atom.}
    \label{fig:FBR_thickness}
\end{figure}

We illustrate these trends in \Cref{fig:FBR_thickness}. Panel~(a) shows $\mrm{FBR}_a$ versus $l_b$ for three wall flux values spanning compact tokamaks through high-flux mirror devices. $\mrm{FBR}_a$ falls steeply once $l_b$ exceeds $\lambda_\mrm{abs} \approx 24$~cm, the absorption mean free path. Panel~(b) shows $\mrm{FBR}_a$ versus $A_b$ for fixed $\Phi_0$ and four feedstock inventories: larger area (thinner blanket) always gives higher $\mrm{FBR}_a$, with the knee of each curve located near $A_b \approx \sigma N_\mrm{feed,0}$ where $\tau \sim 1$. The design tradeoff is therefore between $\mrm{FBR}_a$ (favoring large $A_b$, thin target) and $\eta_\mrm{pro}$ (favoring large $l_b$, thick target), consistent with \Cref{fig:FBR_general}(b).

%\section{Co-produced Product and Electricity} \label{sec:governing}
\section{Economic Models} \label{sec:governing}

\begin{figure}[b]
    \centering
    \begin{subfigure}[tb]{0.99\textwidth}
    \centering
    \includegraphics[width=1.0\textwidth]{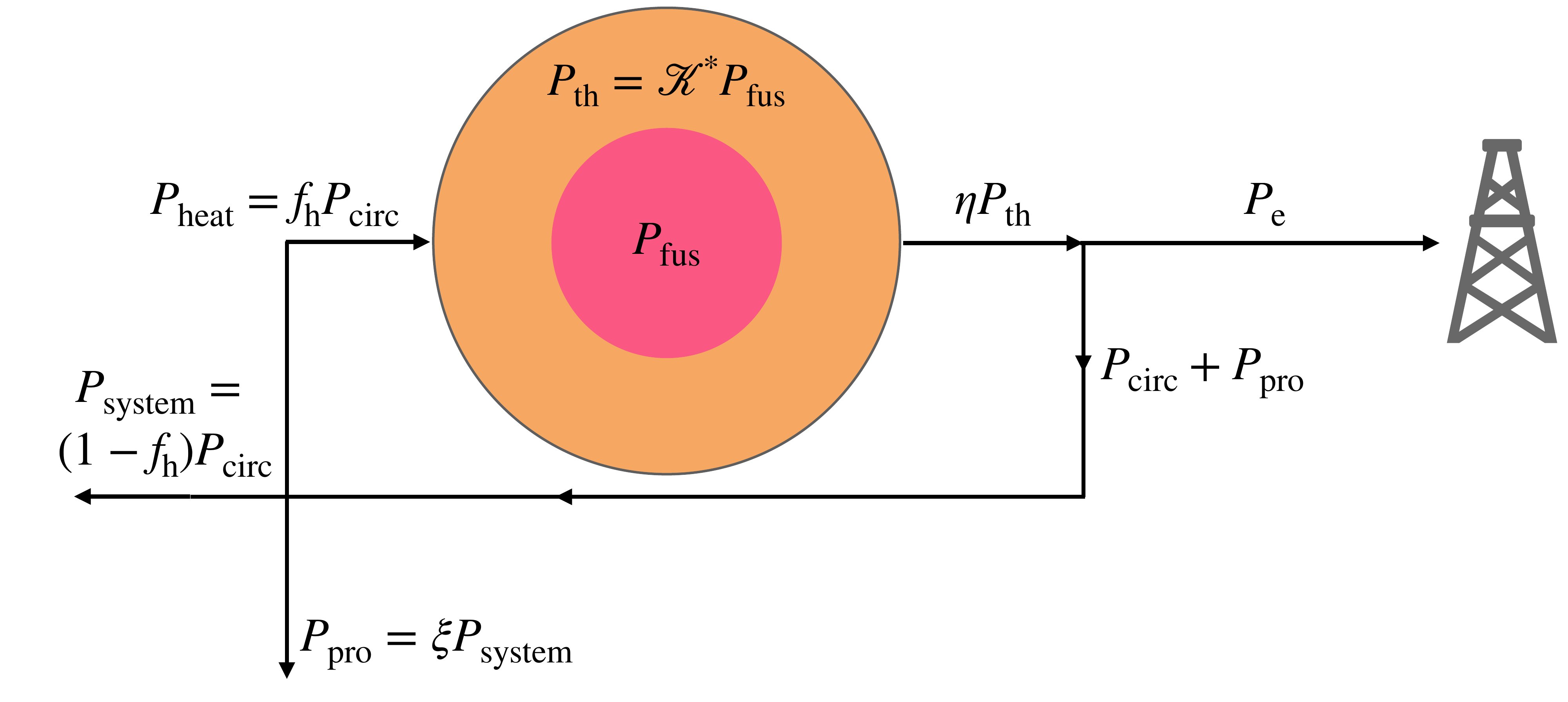}
    \end{subfigure}
    \caption{Power flow diagram for fusion power plant.}
    \label{fig:power_flow_di}
\end{figure}

Fusion transmutation systems can simultaneously generate both valuable isotopes and electrical power. In this section, we formulate the governing relationships that determine the total revenue from co-production. The power flow for a fusion power plant is illustrated in \Cref{fig:power_flow_di}. %accounting for contributions from both the transmuted product and the generated electricity.

The net revenue rate $\dot{R}$ (\$/s) from a fusion power plant that sells both isotopes and electricity is
\begin{equation}
    \dot{R} = \dot{M}_\mrm{pro} \, C_\mrm{pro} + P_\mathrm{e} \, \tilde{C}_\mrm{e}.
    \label{eq:governing_eq_f}
\end{equation}
Here, $P_\mrm{e}$ is the net electric power in watts, $P_\mathrm{pro}$ is the electric power for all transmutation and extraction systems in watts, $\dot{M}_\mrm{pro}$ is the transmutation rate in kg/s, $C_\mrm{pro}$ is the sale price of transmuted product in \unit{\dollar\per\kilo\gram}, and $\tilde{C}_\mrm{e}$ is the sale price of electricity in \unit{\dollar\per\joule},
\begin{equation}
    \tilde{C}_\mrm{e} = C_\mrm{e} / 10^6T_\mrm{hour},
    \label{eq:convenient_C_redef}
\end{equation}
where $C_\mrm{e}$ is the sale price of electricity in \unit{\dollar\per\MWh} and $T_\mrm{hour}=3600$. From \Cref{eq:Mpro_annual}, the mass production rate of transmuted product is
\begin{equation}
    \dot{M}_\mathrm{pro} = \dot{N}_\mathrm{pro} m_\mathrm{pro} = \eta_\mathrm{pro} P_\mathrm{fus} \frac{m_\mathrm{pro}}{E_\mrm{fus}}.
    \label{eq:Mdotpro}
\end{equation}

\begin{figure*}[tb!]
  \centering
  \begin{subfigure}[t]{0.49\textwidth}
  \centering
  \includegraphics[width=1.0\textwidth]{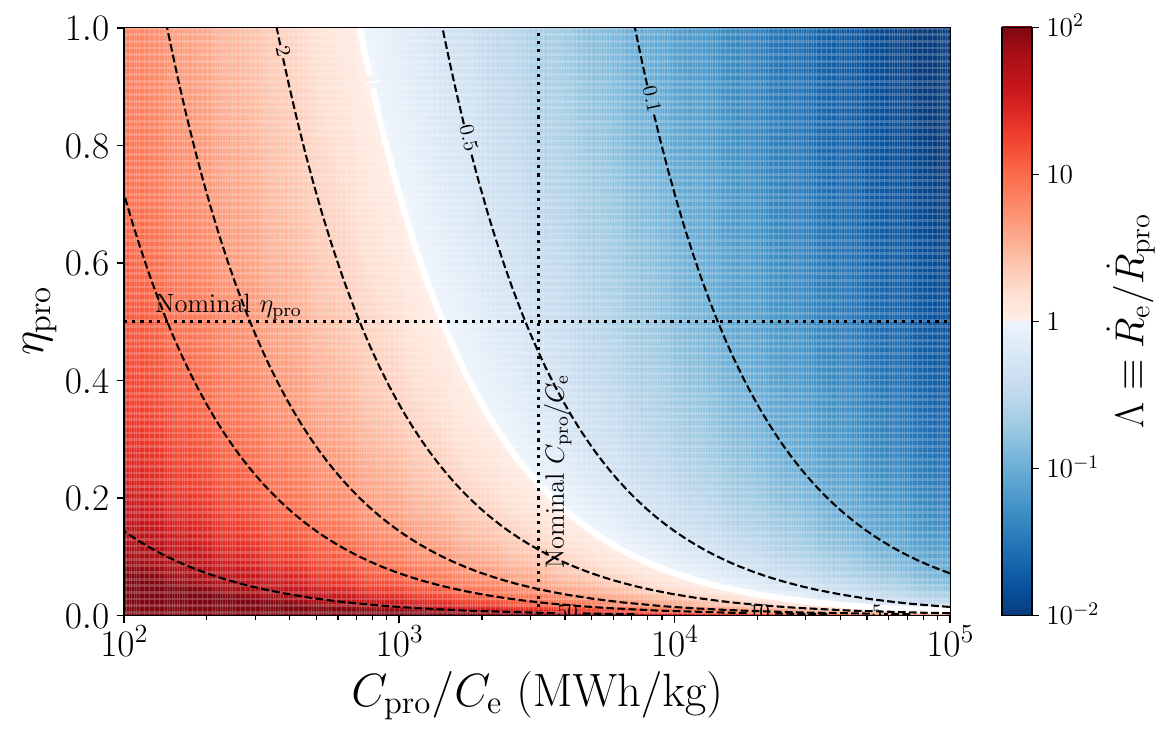}
  \caption{}
  \end{subfigure}
  \centering
  \begin{subfigure}[t]{0.49\textwidth}
  \centering
  \includegraphics[width=1.0\textwidth]{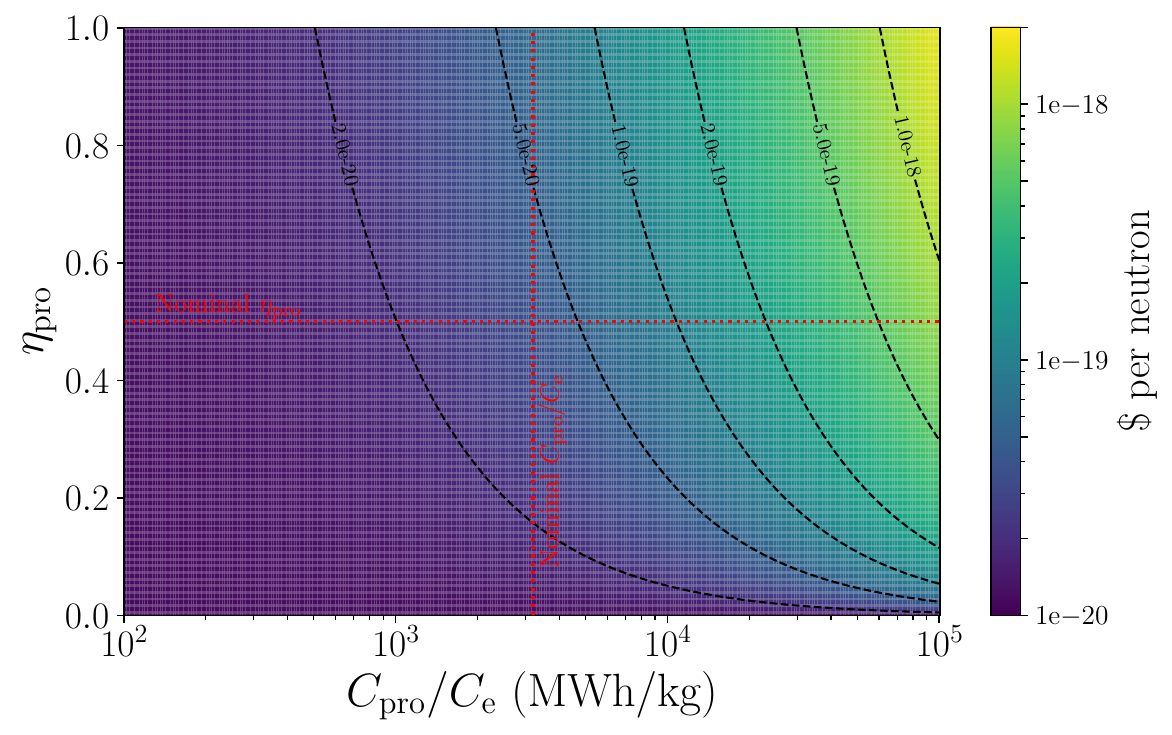}
  \caption{}
  \end{subfigure} 
  \caption{(a) Relative transmutation and electricity revenue, (b) revenue per neutron. We assumed $\mathcal{M}$ = 1.15, $\eta=0.4$, $Q_\mrm{plas} = 40$, $P_\mrm{e} = \qty{340}{MW}$, $P_\mrm{fus} = \qty{1}{GW}$, $P_\mrm{fus} = 0$, and a feedstock of ${}^{198}\mrm{Hg}$. Nominal values in dotted lines indicate $\eta_\mrm{pro} = 0.5$, $C_\mrm{pro} = \mathdollar \qty{160000}{\per\kg}$ and $C_\mrm{e} = \mathdollar 50$/MWh.}
  \label{fig:revenue_value_per_neutron}
\end{figure*}

A simple way to model the net electric power output $P_\mathrm{e}$ of a FPP is
\begin{equation}
    P_\mathrm{e} = \eta P_\mathrm{th} - P_\mathrm{circ} - P_\mrm{pro},
    \label{eq:Pmrm_e_new}
\end{equation}
where $\eta$ is the electricity conversion efficiency and $P_\mathrm{circ}$ is the recirculating power for all non-transmutation systems. We expect that the transmutation-system power will be relatively small $P_\mathrm{pro} \ll P_\mathrm{circ}$. We model the total thermal power $P_\mathrm{th}$ as all of the neutron, alpha, and heating power through the first wall enclosing a fusion plasma,
\begin{equation}
    P_\mathrm{th} = \mathcal{K} P_\mathrm{fus} + P_\mrm{heat} = \mathcal{K}^* P_\mathrm{fus},
    \label{eq:Pth}
\end{equation}
where
\begin{equation}
    \mathcal{K}^* \equiv \frac{P_\mathrm{th}}{P_\mathrm{fus}} = \mathcal{K} + \frac{1}{\eta_\mrm{abs} Q_\mrm{plas}}, \quad \mathcal{K} \equiv \mathcal{M} f_\mathrm{n} + f_\alpha,
\end{equation}
and $\mathcal{M}$ is the power multiplication in the blanket - $\mathcal{M}$ is expected to have values between 1.0 and 1.2 \cite{sawan_physics_2006}. $\mathcal{M}$ describes the net effect of blanket heat sources and sinks, including effects as neutron slowing, exothermic ${}^6\mrm{Li}$ reactions, endothermic $(\mathrm{n},\mathrm{2n})$ reactions, gamma ray absorption, and neutron losses.
%Note that \cite{sawan_physics_2006} found $\mathcal{M} > 1$ even in blanket materials with a relatively high $(\mathrm{n},\mathrm{2n})$ threshold energy such as Pb. 
The power fraction carried by neutrons is
\begin{equation}
    f_\mathrm{n} \equiv \frac{E_\mathrm{n}}{E_\mathrm{fus}} = \frac{4}{5}.
\end{equation}
There is an important distinction between the revenue from transmutation and electricity: transmutation revenue is proportional to the fusion power $P_\mrm{fus}$, whereas electricity revenue is proportional to the net electric power. While there is an additional marginal energy cost of extracting product from blanket material, we assume this marginal energy is small.

Using these results, the revenue rate is
\begin{equation}
\dot{R} = \left(\eta \mathcal{K}^* P_\mrm{fus} - P_\mrm{circ} - P_\mrm{pro}\right) \tilde{C}_\mrm{e} + \eta_\mrm{pro} \frac{m_\mrm{pro}}{E_\mrm{fus}} C_\mrm{pro}\, P_\mrm{fus},
\label{eq:governing_eq_new}
\end{equation}
The absorbed heating power $P_\mrm{abs}$ is related to the wallplug heating power $ P_\mrm{heat}^\mrm{wallplug}$ by
\begin{equation}
    P_\mrm{abs} = \eta_\mrm{abs} \eta_\mrm{heat}  P_\mrm{heat}^\mrm{wallplug},
\end{equation}
where $\eta_\mrm{heat}$ is the heating system power conversion efficiency. We separate the non-transmutation recirculating power into a system and heating term,
\begin{equation}
    P_\mrm{circ} = P_\mrm{system} +  P_\mrm{heat}^\mrm{wallplug},
\end{equation}
where
\begin{equation}
     P_\mrm{heat}^\mrm{wallplug} = f_h \, P_\mrm{circ}, \;\;\; f_h \equiv \frac{ P_\mrm{heat}^\mrm{wallplug}}{P_\mrm{circ}},
\end{equation}
for the heating power fraction $f_h$. We can now relate $P_\mrm{circ}$ to the fusion power and gain,
\begin{equation}
    P_\mrm{circ} = \frac{P_\mrm{fus}}{Q_\mrm{plas}} \frac{1}{\eta_\mrm{abs} \eta_\mrm{heat}} \frac{1}{f_h}.
    \label{eq:Pcircgain}
\end{equation}
It is useful to separate the revenue in \Cref{eq:governing_eq_new} into electricity and transmutation terms,
\begin{equation}
    \dot{R} = \dot{R}_\mrm{e} + \dot{R}_\mrm{pro},
\end{equation}
where
\begin{equation}
\begin{aligned}
\dot{R}_\mrm{e} &= \left(\eta \mathcal{K}^* P_\mrm{fus} - P_\mrm{circ} - P_\mrm{pro}\right) \tilde{C}_\mrm{e},
\end{aligned}
\label{eq:governing_eq_new_ele}
\end{equation}
and
\begin{equation}
\dot{R}_\mrm{pro} = P_\mrm{fus}\, \eta_\mrm{pro} \frac{ m_\mrm{pro}}{E_\mrm{fus}} C_\mrm{pro}.
\label{eq:governing_eq_new_prod}
\end{equation}
The relative electricity to transmutation revenue is
\begin{equation}
    \Lambda \equiv \frac{\dot{R}_\mrm{e}}{\dot{R}_\mrm{pro}} =  \frac{E_\mrm{fus}}{m_\mrm{pro}} \frac{\tilde{C}_\mrm{e}}{C_\mrm{pro}} \frac{\eta \mathcal{K}^* - \left(P_\mrm{circ} + P_\mrm{pro}\right)/P_\mrm{fus}}{\eta_\mrm{pro}}.
\end{equation}
For a pure electric FPP, $\Lambda \to \infty$. For $\Lambda \ll 1$,
\begin{equation}
    \eta_\mrm{pro} \frac{m_\mrm{pro}}{E_\mrm{fus}} \frac{C_\mrm{pro}}{\tilde{C}_\mrm{e}} \gg \eta \mathcal{K}^* - \frac{P_\mrm{circ} + P_\mrm{pro}}{P_\mrm{fus}},
\end{equation}
representing significantly more transmutation revenue than electric revenue. When $\Lambda \ll 1$, it may not be worthwhile selling electricity because the additional capital costs for electric-generating equipment and operating costs likely outweigh the electricity revenue.

In \Cref{fig:revenue_value_per_neutron}(a) we plot $\Lambda$ versus $\eta_\mrm{pro}$ and $C_\mrm{pro} / C_\mrm{e}$ for a fusion plant with a power of \qty{1}{GW}. With the present (2026) price of gold and electricity, selling gold generates twice as much revenue as electricity alone. Co-generation of gold and electricity therefore triples the revenue of a fusion power plant. In \Cref{fig:revenue_value_per_neutron}(b) we plot the value per neutron. With the present (2026) price of gold and electricity and the assumptions described in the caption for \Cref{fig:revenue_value_per_neutron}, each neutron generates \$\num{4e-20} of revenue in a system co-producing electricity and gold. For electricity-only FPPs, the revenue per neutron is \$\num{1.3e-20}.

\subsection{Payback Time} \label{sec:paybacktime}

In this section, we show a simple metric for the viability of transmutation in a wide range of fusion systems: the payback time $T_\mrm{payback}$ required for the total revenue to pay back the cost of feedstock. While this is not as accurate a metric as net present value (presented in the following section), it is simpler, allowing for fast approximate evaluations of the feasibility for a given transmutation pathway.

We include the time value of money by introducing a continuous discount rate $r$. The initial cost of the feedstock inventory is
\begin{equation}
    c_\mrm{feed} = C_\mrm{feed } \, M_\mrm{feed},
\end{equation}
where $C_\mrm{feed}$ is the feedstock price per unit mass and $M_\mrm{feed}$ is the initial feedstock inventory mass. The revenue rate from product sales is
\begin{equation}
    \dot{R}_\mrm{pro} = C_\mrm{pro}\,\dot{M}_\mrm{pro},
\end{equation}
where $C_\mrm{pro}$ is the product price per unit mass and $\dot{M}_\mrm{pro}$ is the production rate. The present value of revenues accumulated over a time interval $[0,T]$ is
\begin{equation}
    \dot{R}_\mrm{pro}^{(\mathrm{PV})}(T) = 
    \int_0^T \dot{R}_\mrm{pro}(t)\,e^{-r t} \mathrm{d}t.
\end{equation}
We define the payback time $T_\mrm{payback}$ as the time at which the present value of revenues equals the initial feedstock cost,
\begin{equation}
    R_\mrm{pro}^{(\mathrm{PV})}(T_\mrm{payback}) = c_\mrm{feed}.
    \label{eq:Tpayback_def_discounted}
\end{equation}
In the thin blanket approximation where the blanket thickness $l_b$ is much shorter than the neutron mean-free path, the production rate is
\begin{equation}
    \dot{M}_\mrm{pro}
    = \dot{N}_\mathrm{n} \, \sigma \, n_\mrm{feed} \, l_b \, m_\mrm{feed}
    = \Phi_0  \, \sigma \, M_\mrm{feed},
\end{equation}
The initial feedstock mass is
\begin{equation}
    M_\mrm{feed} = n_\mrm{feed} \, l_b \, A_b \, m_\mrm{feed}.
\end{equation}
Assuming a constant production rate in time, $\dot{M}_\mrm{pro}$ is independent of $t$, so the present-value revenue is
\begin{equation}
    R_\mrm{pro}^{(\mathrm{PV})}(T)
    = \frac{C_\mrm{pro}\,\dot{M}_\mrm{pro}}{r} \left(1 - e^{-rT}\right).
\end{equation}
Equating this to $c_\mrm{feed} = C_\mrm{feed} M_\mrm{feed}$ and solving for $T = T_\mrm{payback}$ gives
\begin{equation}
    C_\mrm{feed} M_\mrm{feed}
    = \frac{C_\mrm{pro}\,\dot{M}_\mrm{pro}}{r}
      \left(1 - e^{-r T_\mrm{payback}}\right),
\end{equation}
or
\begin{equation}
    T_\mrm{payback}
    = -\frac{1}{r}\ln\!\left[
        1 - r \frac{C_\mrm{feed}M_\mrm{feed}}
                    {C_\mrm{pro}\dot{M}_\mrm{pro}}
      \right].
    \label{eq:Tpayback_discount_general}
\end{equation}
Using $\dot{M}_\mrm{pro} = \Phi_0 \sigma M_\mrm{feed}$, the dependence on the initial feedstock mass again cancels,
\begin{equation}
    T_\mrm{payback}
    = -\frac{1}{r}\ln\!\left[
        1 - r \frac{C_\mrm{feed}}{C_\mrm{pro}}
          \frac{1}{\Phi_0 \, \sigma}
      \right],
    \label{eq:Tpayback_discount_thin}
\end{equation}
which depends only on the price ratio $C_\mrm{feed}/C_\mrm{pro}$, the neutron flux $\Phi_0$, the effective cross section $\sigma$, and the discount rate $r$. The argument of the logarithm must be positive, otherwise the discounted revenue can never repay the initial feedstock cost. In the limit of vanishing discount rate, $r \to 0$, \Cref{eq:Tpayback_discount_thin} reduces to the non-discounted payback time
\begin{equation}
    T_\mrm{payback}
    \xrightarrow{\,r\to 0\,}
    \frac{C_\mrm{feed}}{C_\mrm{pro}} \frac{1}{\Phi_0 \, \sigma}.
    \label{eq:Tpayback_later}
\end{equation}

\begin{figure}[tb]
    \centering
    \begin{subfigure}[tb]{0.98\textwidth}
    \centering
    \includegraphics[width=1.0\textwidth]{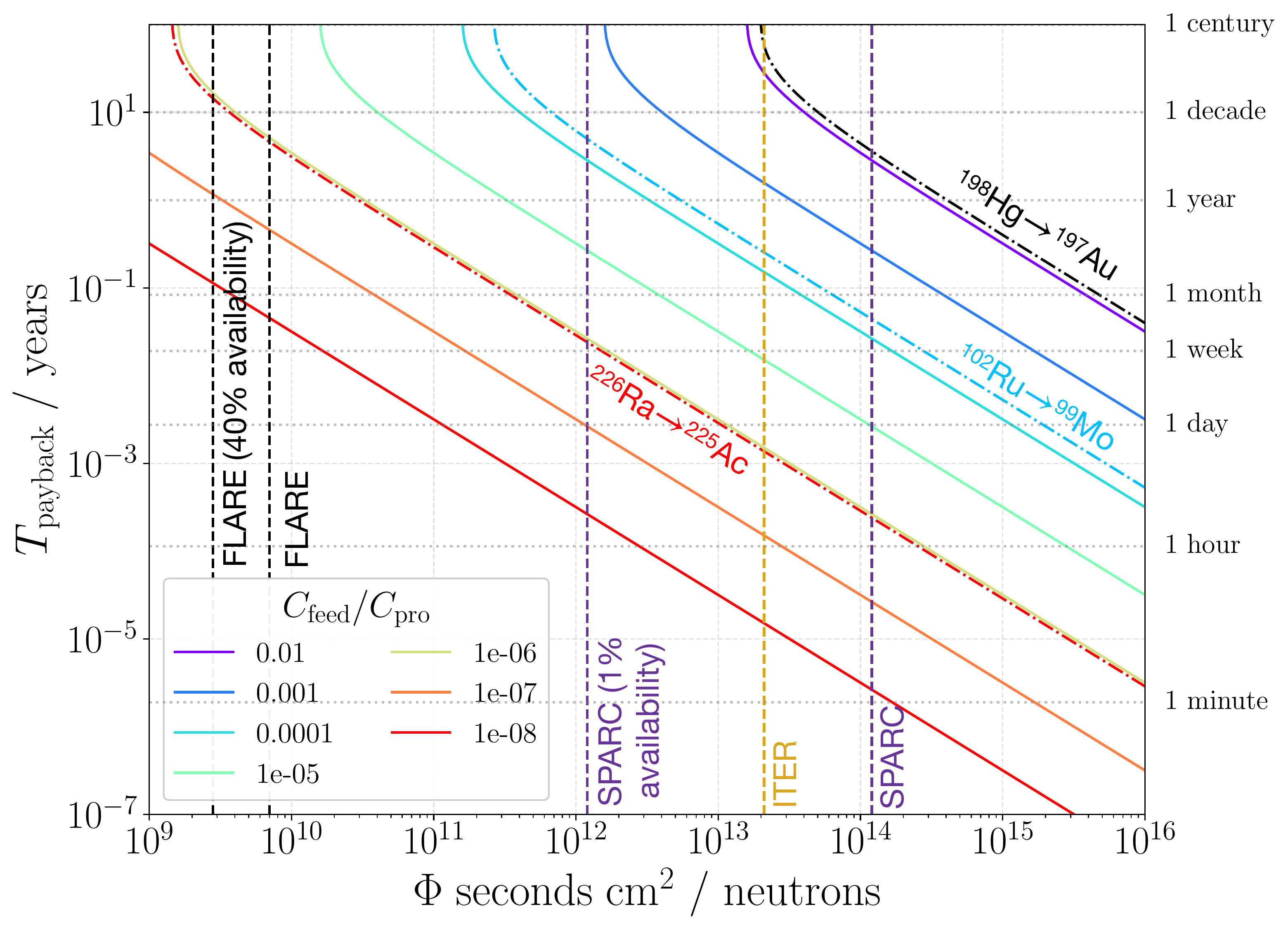}
    \caption{}
    \end{subfigure}
    \centering
    \begin{subfigure}[tb]{0.98\textwidth}
    \centering
    \includegraphics[width=1.0\textwidth]{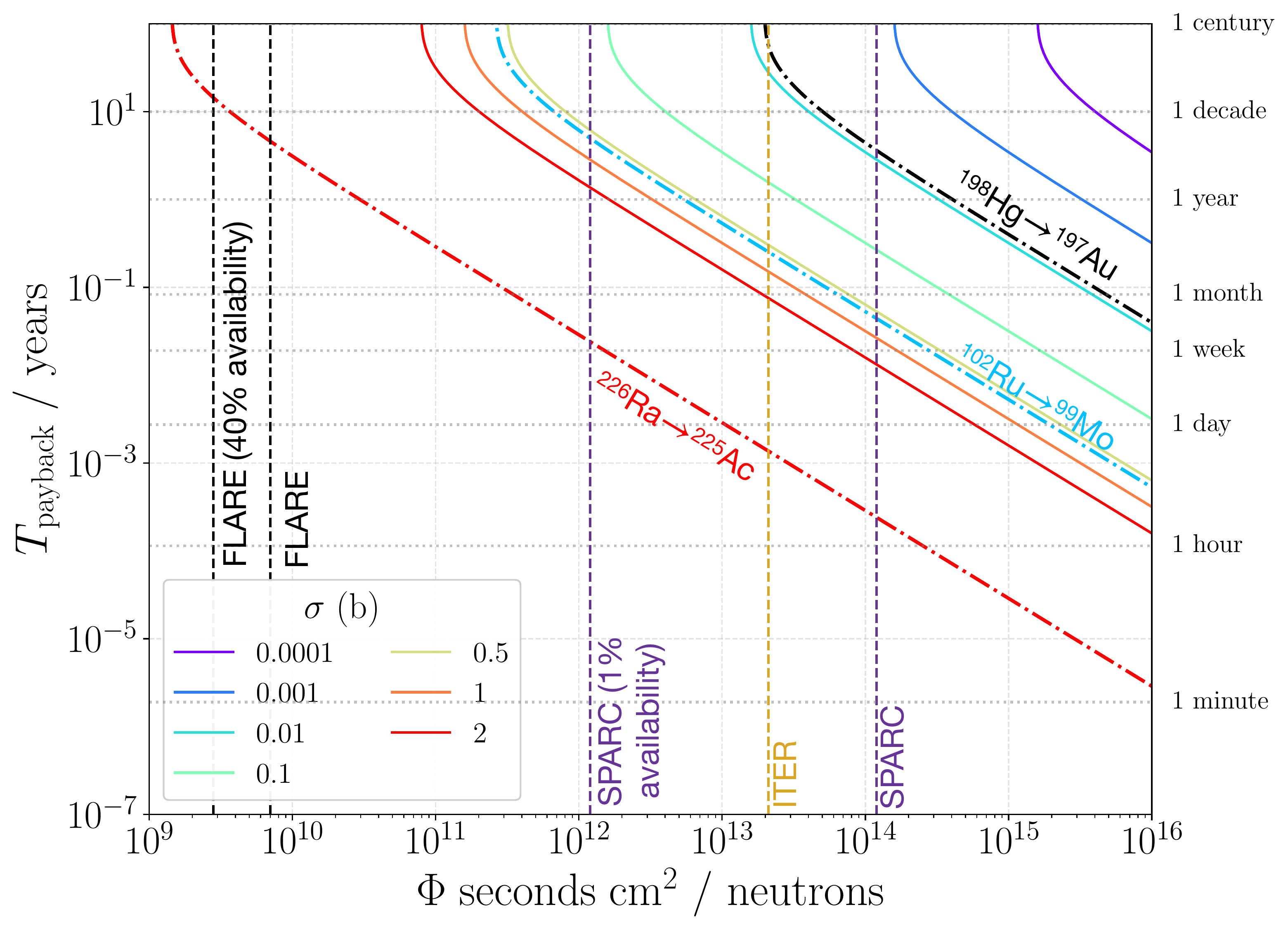}
    \caption{}
    \end{subfigure}
    \caption{Payback time $T_\mrm{payback}$ (\Cref{eq:Tpayback_discount_thin}) versus average first-wall neutron flux $\Phi_0$ for (a) different feedstock-to-product cost ratios with $\sigma = 1$ barn, and (b) different cross sections with $C_\mrm{feed}/C_\mrm{pro} = 0.0001$. Unless indicated otherwise, vertical dashed lines assume 100\% machine availability. Dash-dot lines use realistic values of $C_\mrm{feed}/C_\mrm{pro}$ and $\sigma$. Neutron flux sources: \cite{Ikeda2007,Kinsey2011,Creely2020,shine_flare_facility_2025}.}
    \label{fig:T_payback_simple}
\end{figure}

In \Cref{fig:T_payback_simple}(a) we plot the payback time from \Cref{eq:Tpayback_discount_thin} versus $\Phi_0$ for various $C_\mrm{feed}/C_\mrm{pro}$ values, assuming a constant cross section $\sigma = 1$ barn and a discount rate $r = \qty{0.05}{yr^{-1}}$. The vertical dashed lines show approximate D-T fusion neutron fluxes for different machines. For FLARE-like neutron fluxes, the product must be more than one million times more valuable than the feedstock for the discounted payback time to be less than one year. Higher fluxes such as ITER's relax this requirement by roughly two orders of magnitude: for a payback period shorter than one year, the product must be more than \num{10000} times more valuable than the feedstock. In \Cref{fig:T_payback_simple}(b) we plot the payback time for different cross sections with $C_\mrm{feed}/C_\mrm{pro} = 0.0001$.

\begin{figure}[tb!]
    \centering
    \begin{subfigure}[t]{0.9\textwidth}
    \centering
    \includegraphics[width=1.0\textwidth]{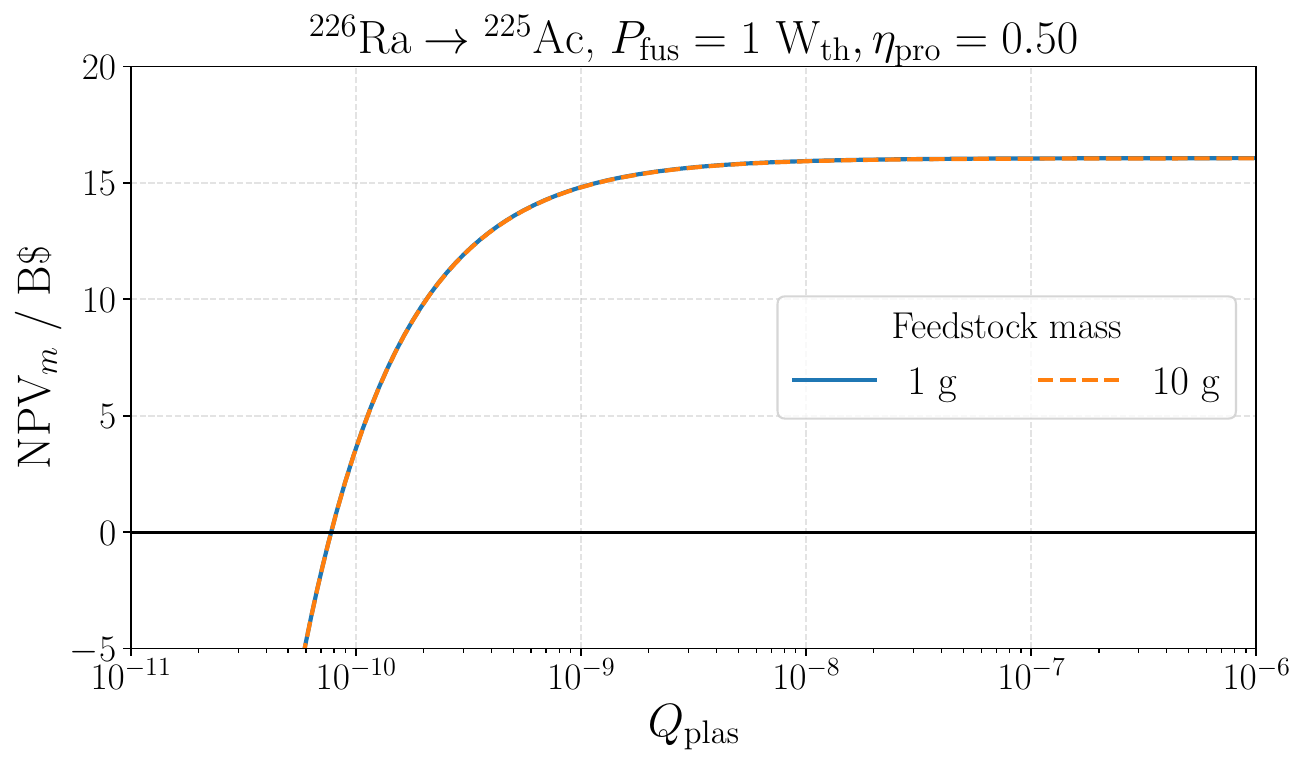}
    \caption{$\mrm{NPV}_m$/\$B.}
    \end{subfigure}
    \centering
    \begin{subfigure}[t]{0.9\textwidth}
    \centering
    \includegraphics[width=1.0\textwidth]{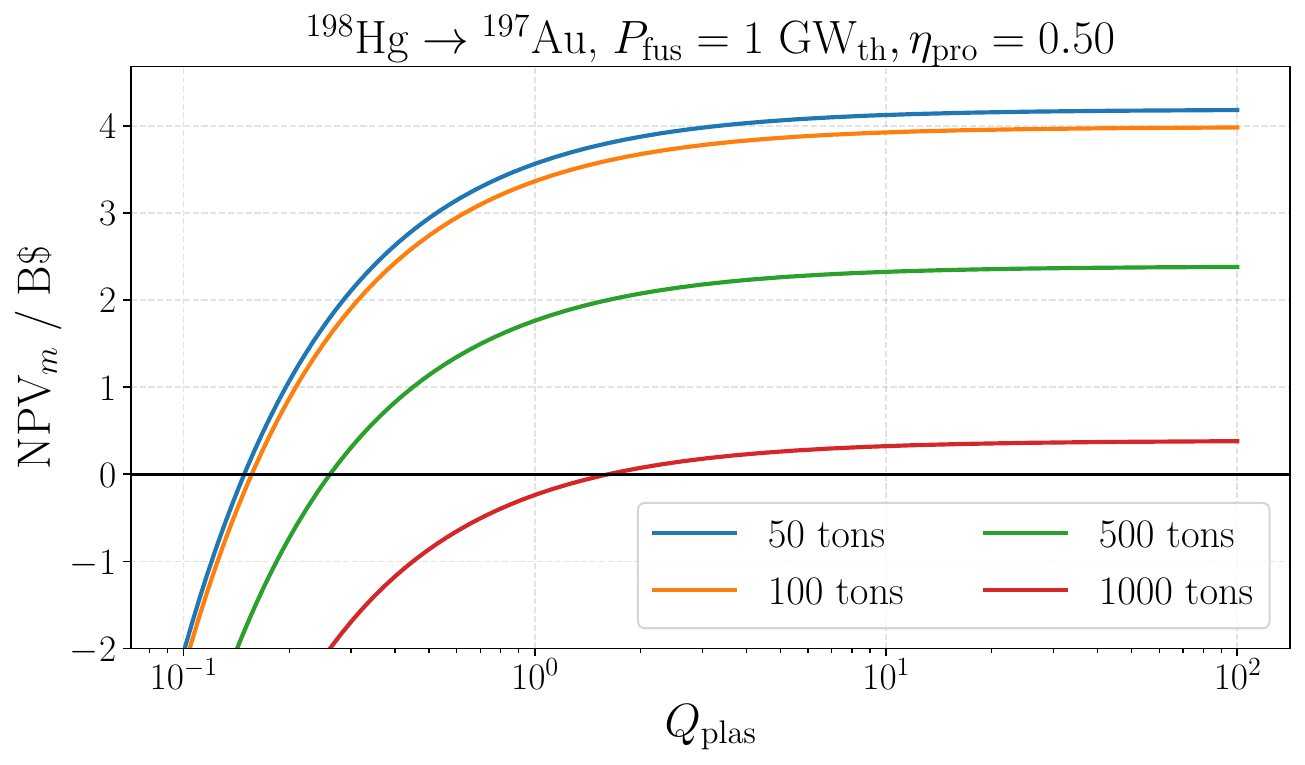}
    \caption{$\mrm{NPV}_m$/\$B.}
    \end{subfigure}
    \centering
    \begin{subfigure}[t]{0.9\textwidth}
    \centering
    \includegraphics[width=1.0\textwidth]{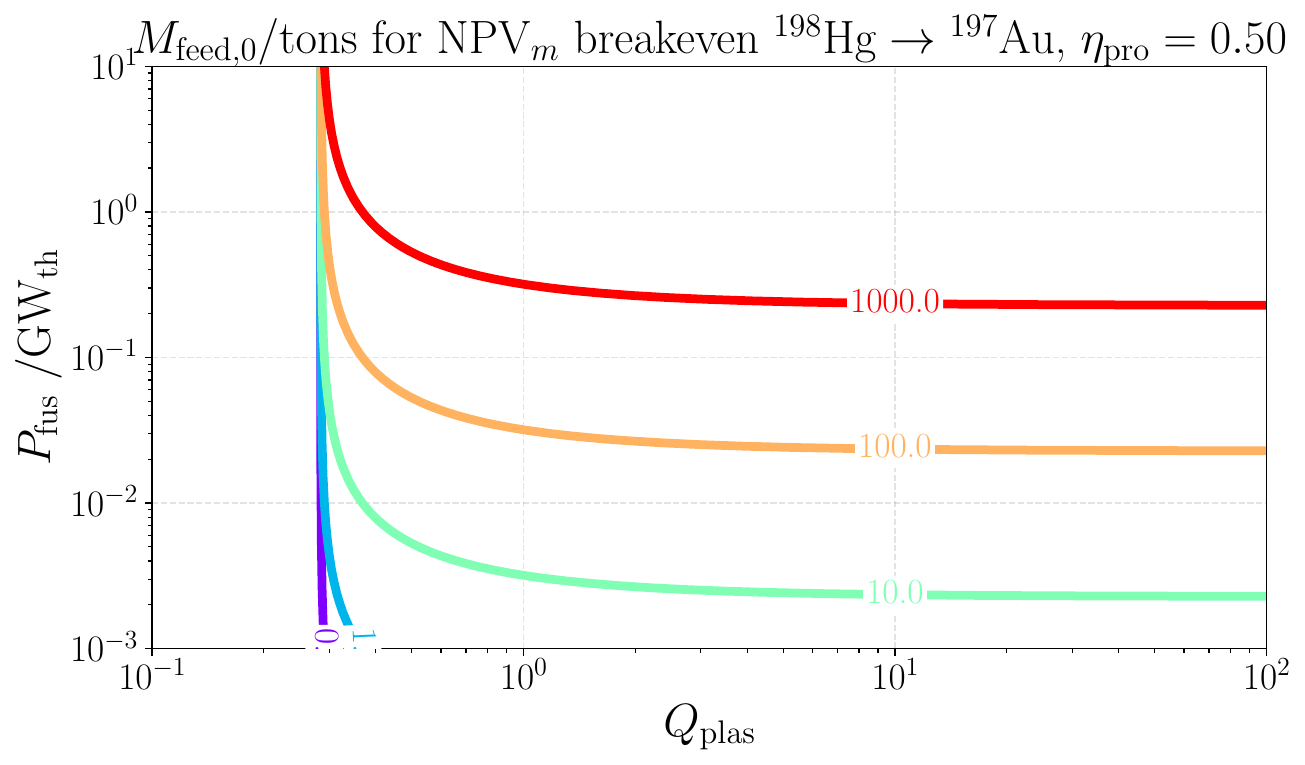}
    \caption{$M_\mrm{feed,0}$/tons.}
    \end{subfigure}
    \caption{$\mathrm{NPV}_m$ versus $P_\mrm{fus}$ for (a) $\ce{^225Ac}$ and (b) $\ce{^197Au}$ production. (c) $M_\mrm{feed,0}$ versus $P_\mrm{fus}$ and $Q_\mrm{plas}$ (contour labels) for mercury to gold transmutation.}
    \label{fig:NPVm_powerQ_scan}
\end{figure}

\iffalse
\begin{figure}[bt!]
    \centering
    \begin{subfigure}[t]{\textwidth}
    \centering
    \includegraphics[width=0.8\textwidth]{NPVmarginal_Pt_Ir_transmutation_breakevenonly}
    \caption{}
    \end{subfigure}
    \centering
    \begin{subfigure}[t]{\textwidth}
    \centering
    \includegraphics[width=0.8\textwidth]{NPVmarginal_Ag_Pd_transmutation_breakevenonly}
    \caption{}
    \end{subfigure}
    \centering
    \begin{subfigure}[t]{\textwidth}
    \centering
    \includegraphics[width=0.8\textwidth]{NPVmarginal_Hg_Au_transmutation_breakevenonly}
    \caption{}
    \end{subfigure}
    \caption{Contours of $\mathrm{NPV}_m$ breakeven versus $\eta_\mrm{pro}$ and $Q_\mrm{plas}$ for three transmutation pathways. We assumed $P_\mrm{fus} = \qty{10}{MW}$, $\eta_\mrm{abs} = 0.95$, $\eta_\mrm{heat} = 0.60$, $\xi = 0.1$, $C_\mrm{Pt} = \$\qty{50000}{\per\kg}$, $C_{\ce{Ir}} = \$\qty{150000}{\per\kg}$, $C_\mrm{Ag} = \$\qty{1230}{\per\kg}$, $C_\mrm{Pd} = \$\qty{40000}{\per\kg}$, $C_\mrm{Hg} = \$\qty{100}{\per\kg}$, $C_\mrm{Au} = \$\qty{140000}{\per\kg}$, $r = 0.08$, $L = 30$ years.}
    \label{fig:NPV_contours}
\end{figure}
\fi

\subsection{Blanket-Only Net Present Value}

Up to now, we have neglected the capital cost of a fusion system, focusing on revenue. In the following subsections, we expand our modeling to include capital and processing costs of transmutation.

%While transmutation produces valuable isotopes, the feedstock can also be expensive, which can make transmutation more economically challenging.

To account for the capital cost of loading a blanket, we first present a simple net present value (NPV) calculation accounting for the capital cost of the blanket. This is not the NPV for the total plant, but rather the marginal NPV corresponding to the blanket loading for an expensive feedstock, denoted as $\mathrm{NPV}_m$. For a given FPP design point, $\mathrm{NPV}_m >0$ indicates that adding the blanket transmutation system adds economic value. Hence, $\mathrm{NPV}_m$ is a useful metric for scoping when transmutation is worth including in a FPP, avoiding the complexity of other capital costs. In the following section we present full, not marginal, NPV calculations for a fusion system.

An example transmutation pathway requiring a high initial capital cost is radium-226 to actinium-225 transmutation
\begin{equation}
\ce{^{226}Ra}(\mrm{n,2n})\,\ce{^{225}Ra}
\;\xrightarrow[\;t_{1/2}=14.8\ \text{d}\;]{\beta^{-}}\;
\ce{^{225}Ac},
\end{equation}
where $\ce{^{226}Ra}$ costs $\approx$\$1/$\mu$g and $\ce{^{225}Ac}$ costs $\approx$\$500000/$\mu$g. Loading a blanket with $M_\mrm{feed,0}$ kilograms has capital cost
\begin{equation}
  C_{\rm cap,feed} = M_{\rm feed} \, C_{\rm feed},
  \label{eq:Ccap}
\end{equation}
so for example, a \qty{10}{g} $\ce{^{226}Ra}$ blanket ties up \$10\,M. The net additional cash generated by transmutation is,
\begin{equation}
\dot{R}_\mrm{pro,marg} = \dot{R}_\mrm{pro} - P_\mrm{pro} \, \tilde{C}_\mrm{e},
\label{eq:revenue_transmutationonly}
\end{equation}
where $\dot{R}_\mrm{pro}$ is the product revenue defined in \Cref{eq:governing_eq_new_prod} and the second term represents the electricity cost of transmutation.
The following cash flow model describes a fusion system with blanket loading capital costs at $t = 0$, plant operations and transmutation sales starting at $t = 1$ and ending at $t = L$,
\begin{equation}
\text{Cash}(t)=
  \begin{cases}
    -C_{\rm cap,feed}, & t=0,\\[4pt]
    \displaystyle \dot{R}_\mrm{pro,marg}, & 1\le t\le L,\\[4pt]
    0, & t>L,
  \end{cases}
\end{equation}
with $L$ the operating lifetime in years. For discount rate $r$, the marginal NPV is
\begin{equation}
\text{NPV}_m(L,r)
  =\sum_{t=0}^{L}\frac{\text{Cash}(t)}{(1+r)^{t}}.
  \label{eq:NPVm}
\end{equation}
We plot $\mrm{NPV}_m$ versus $Q_\mrm{plas}$ in \Cref{fig:NPVm_powerQ_scan}(a) for $\ce{^226Ra} \to \ce{^225Ac}$ with $P_\mrm{fus} = 1$W, showing transmutation of precious metals can achieve $\mrm{NPV}_m > 0$ even at very low plasma gain - here $Q_\mrm{plas} \gtrsim 10^{-10}$. Because $\ce{^225Ac}$ is so valuable per neutron, very low values of $Q_\mrm{plas}$ can be tolerated when considering electricity costs - however, as discussed in \Cref{subsec:powerdensity}, it is necessary to increase $Q_\mrm{plas}$ to a sufficiently high value to keep the wall heat flux loading sufficiently low, and so in reality, $Q_\mrm{plas}$ values much larger than the range plotted in \Cref{fig:NPVm_powerQ_scan}(a) are likely required in order to respect wall material limits.

We plot $\mrm{NPV}_m$ versus $Q_\mrm{plas}$ in \Cref{fig:NPVm_powerQ_scan}(b) for $\ce{^198Hg} \to \ce{^197Au}$ with $P_\mrm{fus} = 1$GW - producing gold from mercury adds value even for $Q_\mrm{plas} < 1$. In \Cref{fig:NPVm_powerQ_scan}(b), $Q_\mrm{plas}$ parameterizes the energy cost of neutrons for transmutation - in the burning plasma regime where $Q_\mrm{plas}\gtrsim 5$, alpha heating provided by fusion reactions rather than external heating dominates and so there is much less increase in $\mrm{NPV}_m$ as $Q_\mrm{plas}$ increases in the burning plasma regime. In \Cref{fig:NPVm_powerQ_scan}(c) we plot breakeven contours for mercury to gold transmutation. Because we assumed the enriched mercury feedstock is much less expensive at $c_\mrm{feed} = \$\qty{1000}{\per\kg}$, a transmutation plant using mercury feedstock can tolerate lower $Q_\mrm{plas}$ and/or higher feedstock inventories (and therefore lower required feedstock burn rate).

\subsection{General NPV}

\iffalse
\begin{figure}[tb!]
    \centering
    \begin{subfigure}[t]{\textwidth}
    \centering
    \includegraphics[width=1.0\textwidth]{NPV0_contours_vs_Cpro_Ccap_Pfus1MW.pdf}
    \end{subfigure}
    \caption{$\mathrm{NPV}$ breakeven contours versus $C_\mrm{cap}$ and $C_\mrm{pro}$ for a transmutation-only system with $P_\mrm{fus} = 1$MW. We assumed $\eta_\mrm{abs} = 0.95$, $\eta_\mrm{heat} = 0.60$, $\eta_\mrm{pro} = 0.5$.}
    \label{fig:NPV_CAPEX_Cpro_scan}
\end{figure}
\fi

\begin{figure}[tb!]
    \centering
    \begin{subfigure}[t]{\textwidth}
    \centering
    \includegraphics[width=1.0\textwidth]{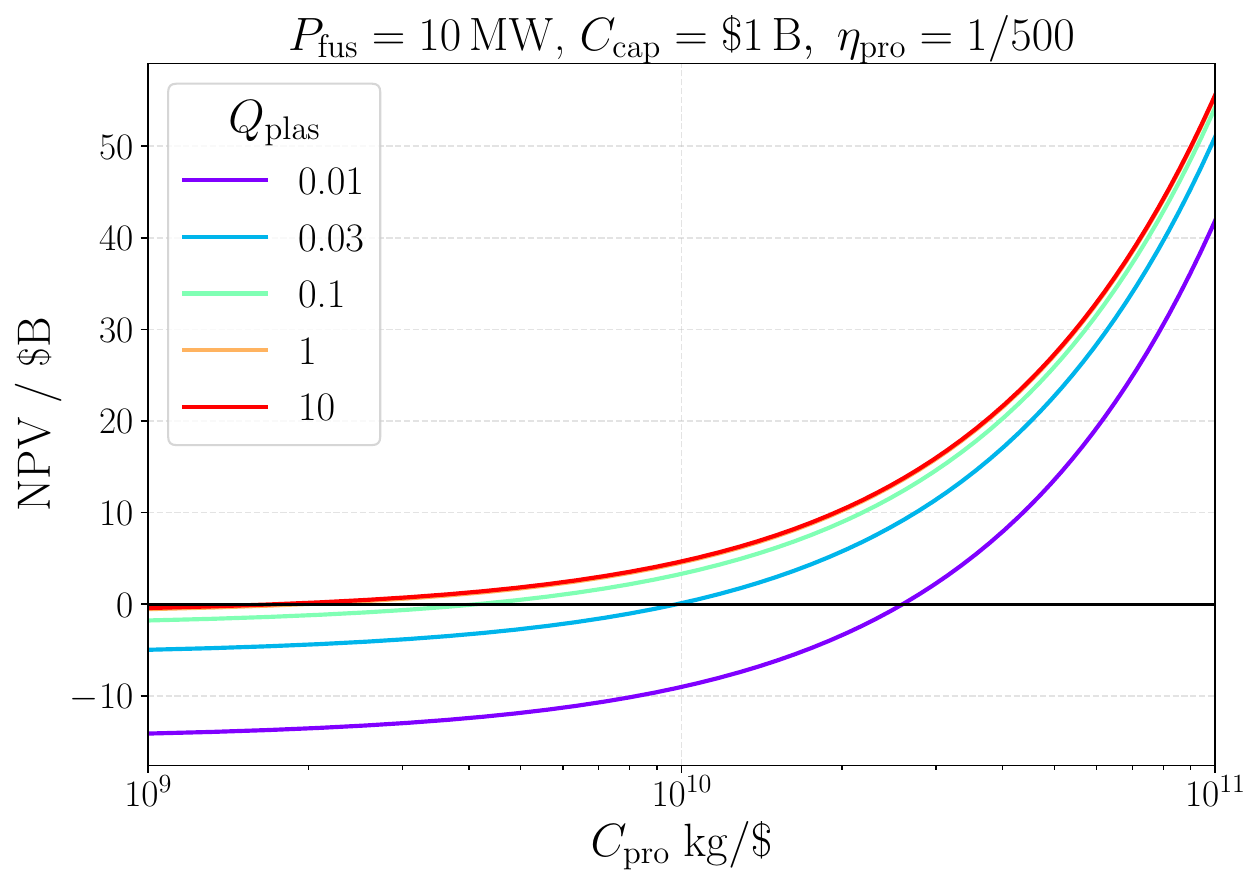}
    \caption{}
    \end{subfigure}
    \begin{subfigure}[t]{\textwidth}
    \centering
    \includegraphics[width=1.0\textwidth]{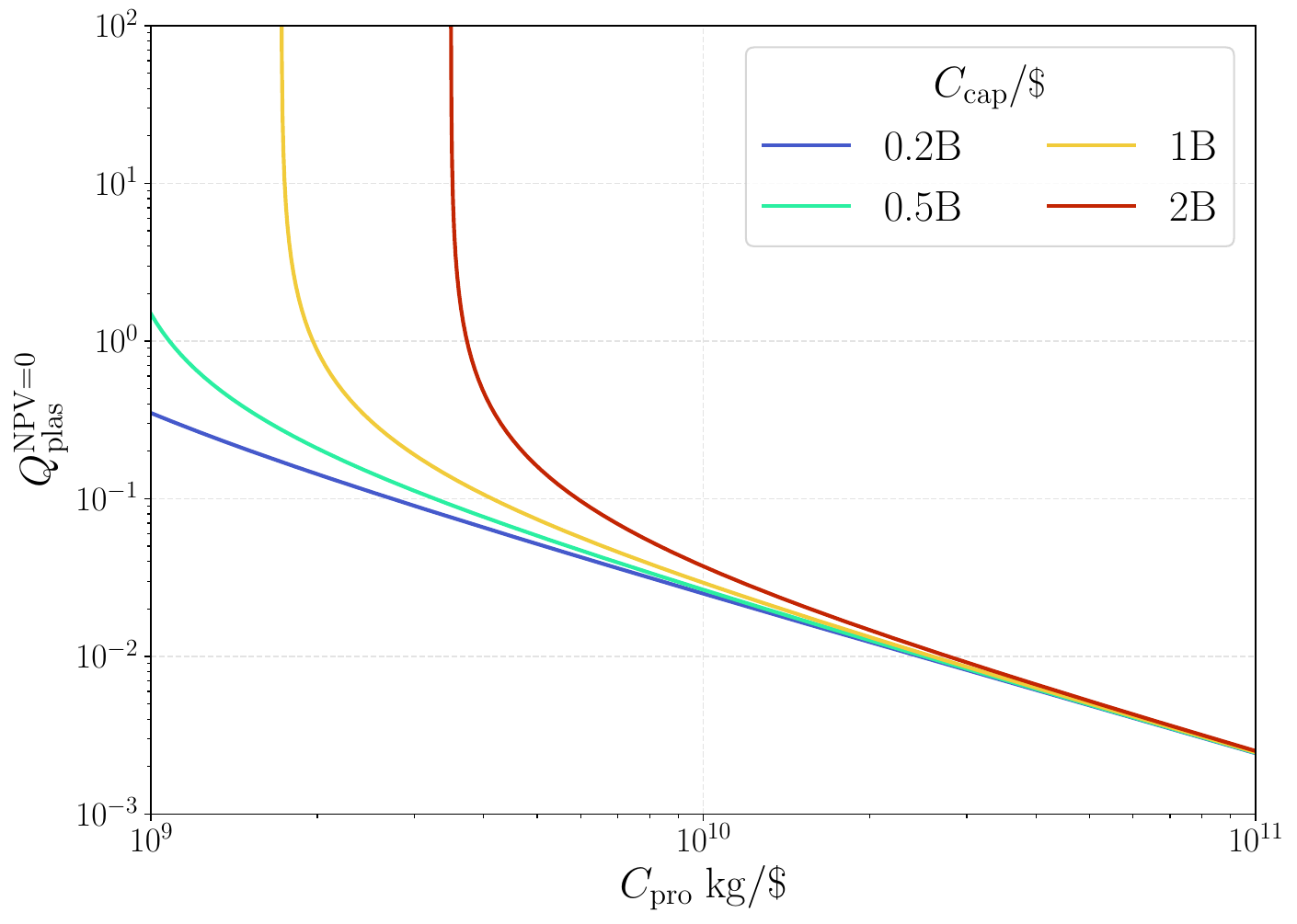}
    \caption{}
    \end{subfigure}
    \caption{(a) $\mathrm{NPV}$ and (b) $Q_\mathrm{plas}^\mathrm{NPV = 0}$ versus $C_\mrm{pro}$ for $P_\mrm{fus} = 10$MW for an isotope with mass 99 amu. We assume $\eta_\mrm{pro} = 1/500$.}
    \label{fig:NPV_Cpro_gain}
\end{figure}

\begin{figure}[tb!]
    \centering
    \begin{subfigure}[t]{\textwidth}
    \centering
    \includegraphics[width=1.0\textwidth]{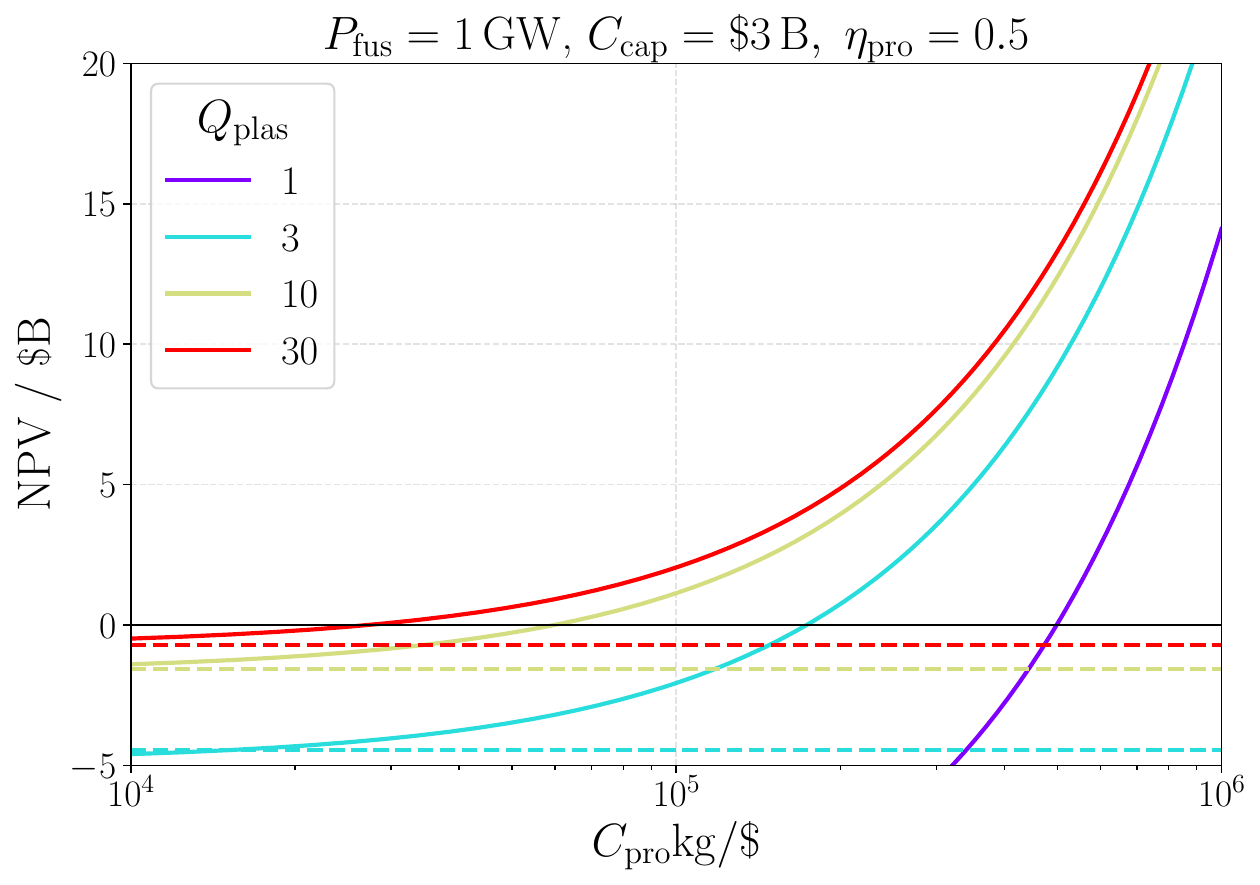}
    \caption{}
    \end{subfigure}
    \begin{subfigure}[t]{\textwidth}
    \centering
    \includegraphics[width=1.0\textwidth]{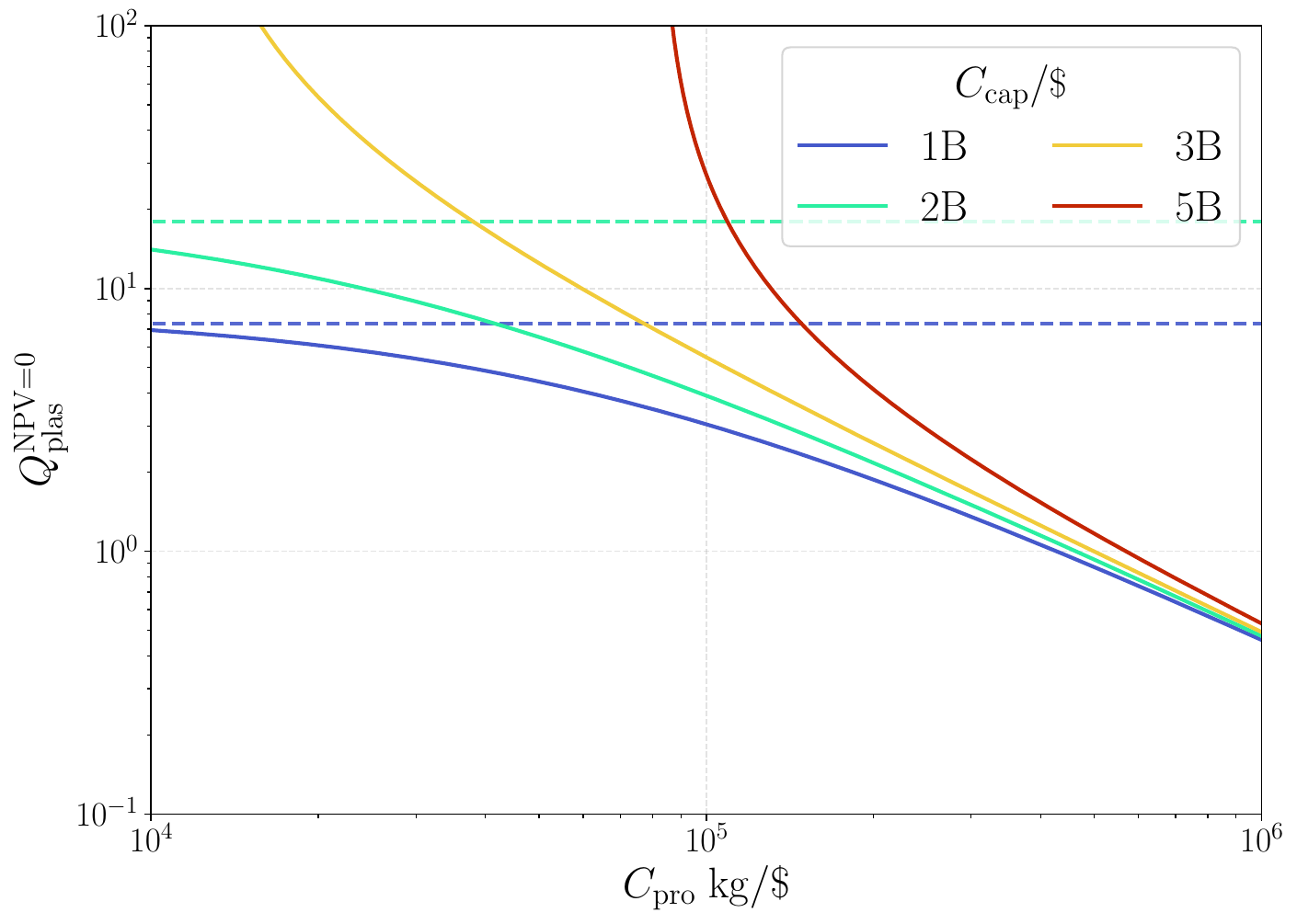}
    \caption{}
    \end{subfigure}
    \caption{(a) $\mathrm{NPV}$ and (b) $Q_\mathrm{plas}^\mathrm{NPV = 0}$ versus $C_\mrm{pro}$ for $P_\mrm{fus} = 1$GW for an isotope with mass 197 amu. We assume $\eta_\mrm{pro} = 0.50$.}
    \label{fig:NPV_Cpro_gain_1GW}
\end{figure}

We now calculate the full, rather than marginal, NPV for a fusion system by including the CAPEX costs. We investigate a transmutation-only 10 MW-class machine and a 1 GW-class machine that co-produces electricity and isotopes. The updated initial capital cost includes both the plant construction cost $C_\mrm{plant}$ (excluding the blanket inventory loading) and the blanket inventory cost
\begin{equation}
  C_{\rm cap}
  = C_\mrm{plant} + M_{\rm feed} \, C_{\rm feed}.  
  \label{eq:Ccap_updated}
\end{equation}
where $C_\mrm{plant}$ is the overall plant cost. There is a subtlety that we neglect in this work, which is that a fusion power plant only producing electricity still has a nonzero blanket loading cost. Later when comparing electricity-only with co-producing systems, we consider the blanket loading cost only of the co-producing system. This will incorrectly make co-producing systems appear more relatively expensive compared to electricity-only systems than they actually are. For simplicity, in this work we also make the simplifying assumption that the electricity sale and purchase price for a fusion system operator are equal; this need not be true in practice.

\begin{figure*}[tb!]
    \centering
    \begin{subfigure}[t]{0.48\textwidth}
    \centering
    \includegraphics[width=0.8\textwidth]{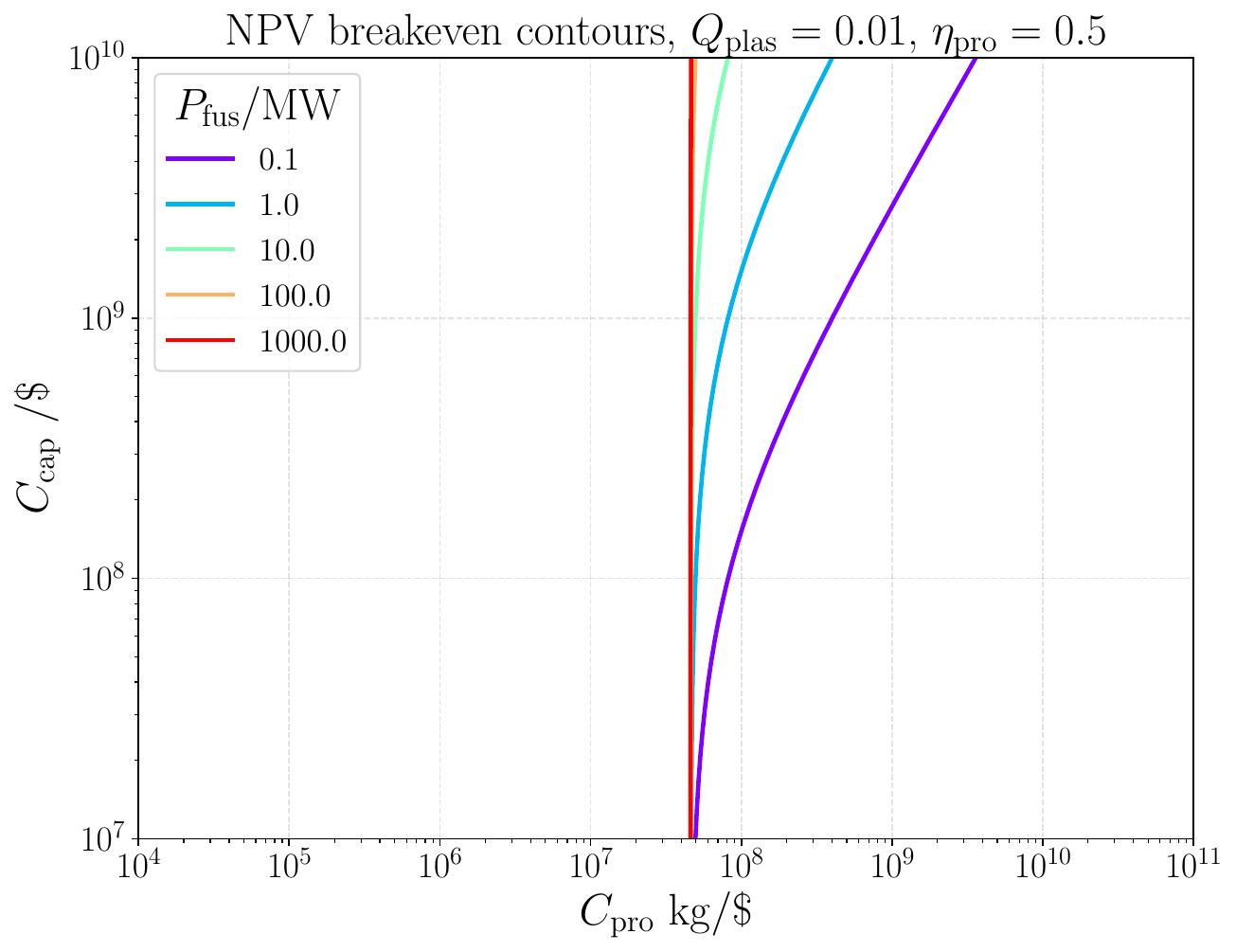}
    \caption{$Q_\mrm{plas} = 0.01$.}
    \end{subfigure}
    \centering
    \begin{subfigure}[t]{0.48\textwidth}
    \centering
    \includegraphics[width=0.8\textwidth]{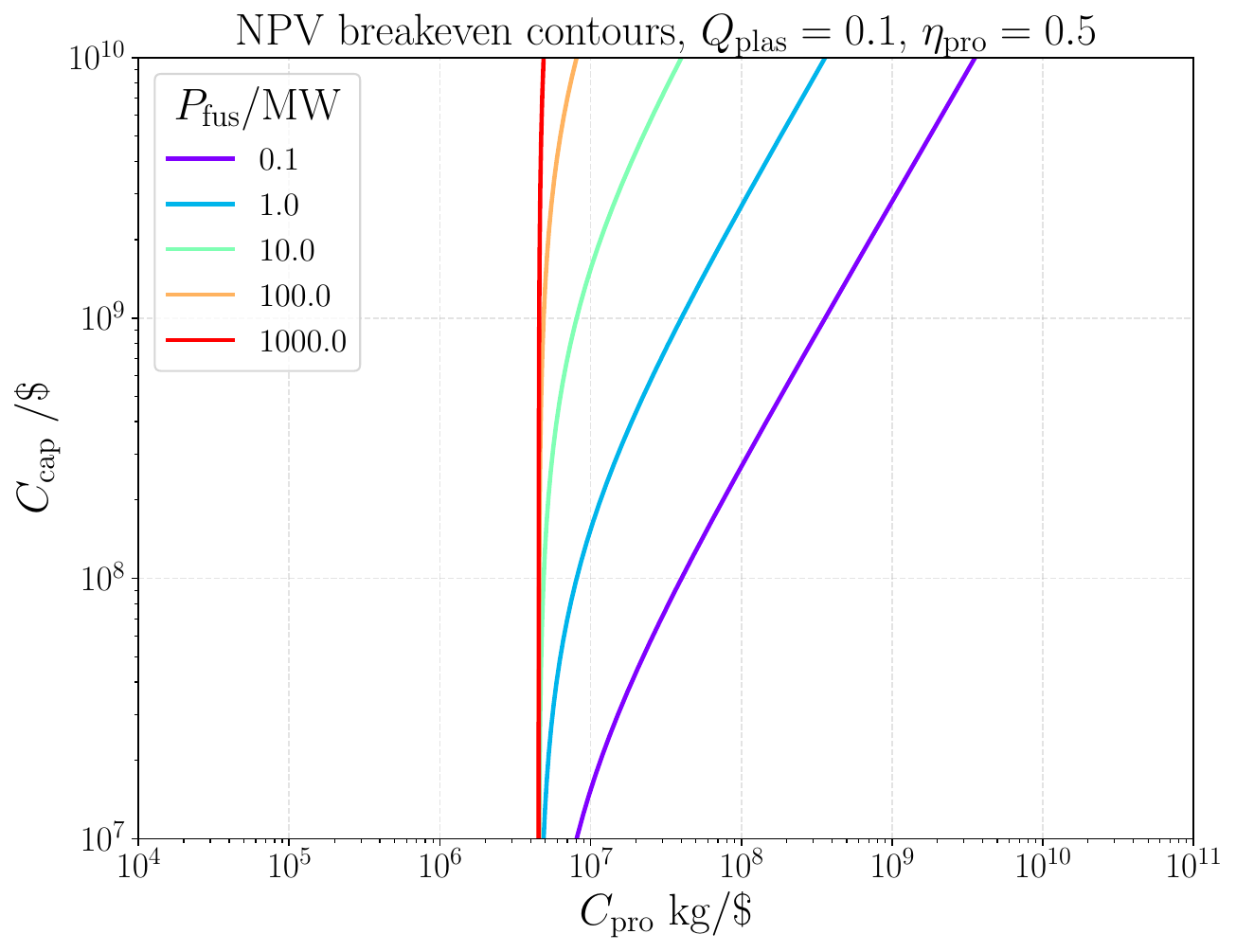}
    \caption{$Q_\mrm{plas} = 0.1$.}
    \end{subfigure}
    \centering
    \begin{subfigure}[t]{0.48\textwidth}
    \centering
    \includegraphics[width=0.8\textwidth]{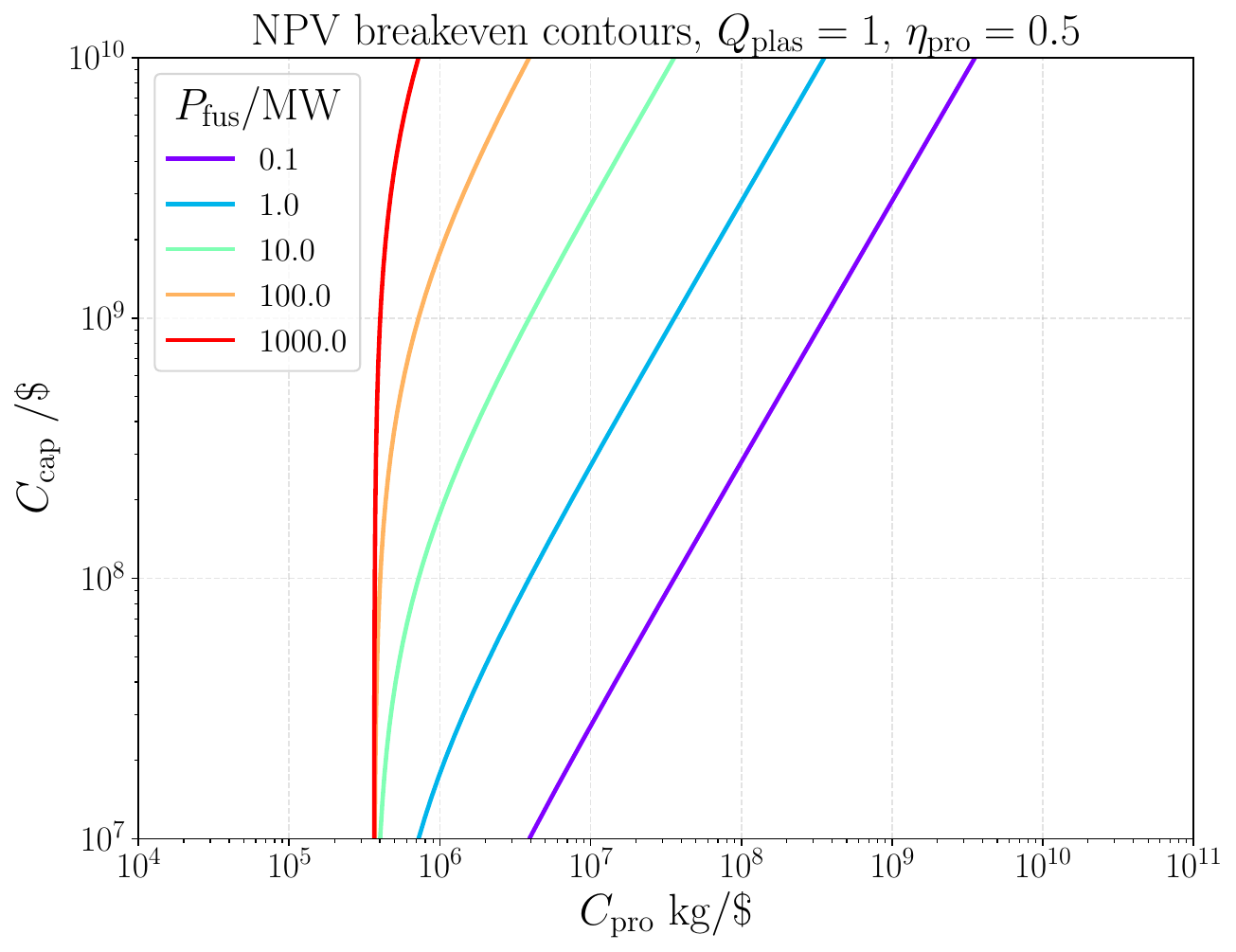}
    \caption{$Q_\mrm{plas} = 1$.}
    \end{subfigure}
    \begin{subfigure}[t]{0.48\textwidth}
    \centering
    \includegraphics[width=0.8\textwidth]{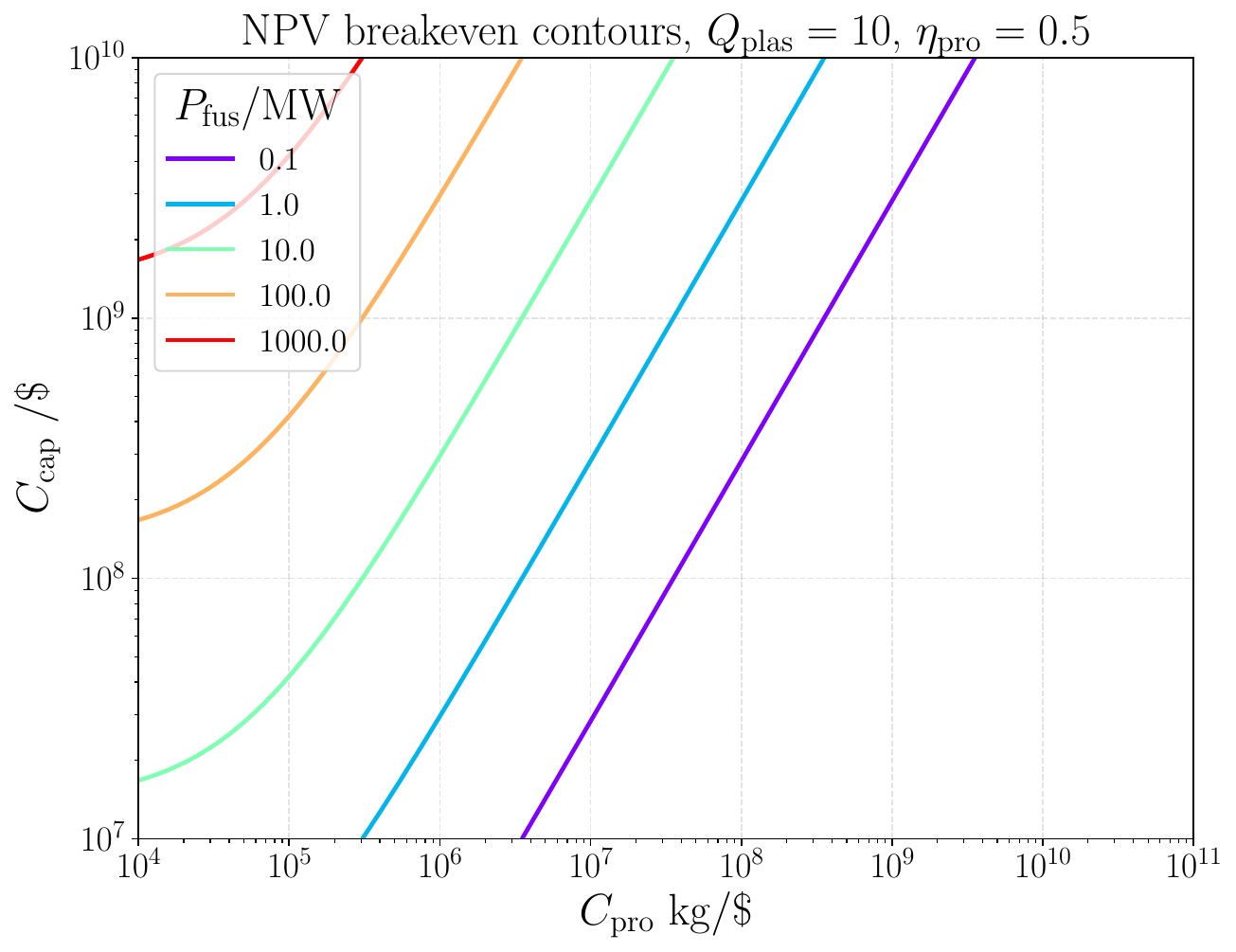}
    \caption{$Q_\mrm{plas} = 10$.}
    \end{subfigure}
    \caption{NPV breakeven contours for co-producing (solid) fusion systems. Each subplot corresponds to a different $Q_\mrm{plas} \in [0.01, 0.1, 1, 10]$.}
    \label{fig:NPV_breakeven_Pfus_scan}
\end{figure*}

\subsubsection{Radioisotope Transmutation-Only 10\,MW System}

We first consider a transmutation-only system for radioisotope production - such systems are likely to have lower absolute power because of the high value of radioisotopes. Our model describes a fusion system that sells transmutation product and pays for the electricity to do this. We assume it generates electricity from blanket heat. The cashflow model also includes the revenue/cost of electricity,
\begin{equation}
\text{Cash}(t)=
  \begin{cases}
    -C_{\rm cap}, & t=0,\\[4pt]
    \displaystyle \dot{R}, & 1\le t\le L,\\[4pt]
    0, & t>L.
  \end{cases}
\end{equation}
In order to show relatively general results, we plot NPV versus product value in \Cref{fig:NPV_Cpro_gain}(a), and the required plasma gain for NPV breakeven in \Cref{fig:NPV_Cpro_gain}(b). With higher product values, concepts with extremely low $Q_\mathrm{plas}$ are viable, although as shown in \Cref{fig:pwall_neutron_flux}, heat flux constraints could prove challenging.

While these calculations do not account for the extra capital costs of transmutation, they do strongly indicate that transmutation of valuable products can be worthwhile at relatively low values of plasma gain, even for plasma gain less than one.

\subsubsection{Electricity + Transmutation 1 GW System}

\iffalse
\begin{figure}[tb!]
    \centering
    \begin{subfigure}[t]{\textwidth}
    \centering
    \includegraphics[width=1.0\textwidth]{NPV0_contours_vs_Cpro_Ccap_Pfus1500MW.pdf}
    \caption{}
    \end{subfigure}
    \centering
    \begin{subfigure}[t]{\textwidth}
    \centering
    \includegraphics[width=1.0\textwidth]{Q_required_for_NPV0_vs_Cpro_by_Ccap.pdf}
    \caption{}
    \end{subfigure}
    \caption{$\mathrm{NPV}$ breakeven contours versus $C_\mrm{cap}$ and $C_\mrm{pro}$ for a transmutation-only system with $P_\mrm{fus} = \qty{1500}{MW}$. We assumed $\eta_\mrm{abs} = 0.95$, $\eta_\mrm{heat} = 0.60$, $\eta_\mrm{pro} = 0.5$, $r = 0.08$, and $L = 30$ years.}
    \label{fig:NPV_CAPEX_Cpro_scan_1500MW}
\end{figure}
\fi

We now consider a 1\,GW$_\mathrm{th}$ plant co-producing electricity and transmutation product. In \Cref{fig:NPV_Cpro_gain_1GW}(a) we plot NPV versus product value for different plasma gain. Compared with the 10 MW plant in the previous section, achieving NPV$>0$ can be done with a much lower product value $C_\mrm{pro}$ provided the gain is sufficiently high. In \Cref{fig:NPV_Cpro_gain_1GW}(b) we plot the plasma gain required for NPV breakeven, $Q_\mrm{plas}^\mrm{NPV = 0}$ for different capital costs. Above a critical capital cost, here roughly \$$2.2$B, there is no economically viable electricity-only plant unless transmutation is used.

%breakeven contours. Compared with the 1 MW transmuter-only system in the previous section, achieving NPV$>0$ can be done with a much lower product value $C_\mrm{pro}$ provided the gain is sufficiently high. In \Cref{fig:NPV_CAPEX_Cpro_scan_1500MW}(b) we plot the plasma gain required for NPV breakeven, $Q_\mrm{plas}^\mrm{NPV = 0}$ for different capital costs. Above a critical capital cost, here roughly \$$2.5$B, there is no economically viable plant unless transmutation is used.

In \Cref{fig:NPV_breakeven_Pfus_scan}, we show the breakeven contours for a scan in the plant capital cost $C_\mrm{cap}$ and product value $C_\mrm{pro}$ for four plasma gain values, $Q_\mrm{plas} \in [0.01, 0.1, 1, 10]$, assuming that $\eta_\mrm{pro} = 0.5$. As the plasma gain increases, the product value required to breakeven decreases primarily because the electricity running costs decrease, and secondarily because at higher gain ($Q_\mrm{plas} \gtrsim 5-10$) the plant generates net electricity to sell. Note that there is no feasible electricity-only fusion system at any capital cost or power for $Q_\mrm{plas} \lesssim 5-10$ because there is no electricity to sell. However, transmutation opens up the viable space for fusion systems at a wide range of plasma gain, even those with gain significantly lower than one.

In this section we have shown how a fusion system's NPV depends on the fusion power, fusion gain, feedstock and product value, and feedstock inventory. We used two types of NPV: (i) the marginal NPV from the blanket system and (ii) the total plant NPV. By focusing on the boundaries where NPV$=0$, we showed how transmutation can significantly enhance a fusion system's value, especially for machines that are marginally economically viable when relying on electricity alone.

\iffalse
\section{Return on Investment}

In this section, we expand the net present value calculation to return on investment.
\fi

In the appendices, we study transmutation scalings in different fusion concepts: tokamaks (Appendix \ref{sec:trans_only}) and magnetic mirrors (Appendix \ref{sec:trans_only_mirror}).

\section{Neutron Wall Loading Non-uniformity} \label{sec:neutron_wall_asymmetry}

\begin{figure}[tb]
    \begin{subfigure}[tb]{0.99\textwidth}    \includegraphics[width=1.0\textwidth]{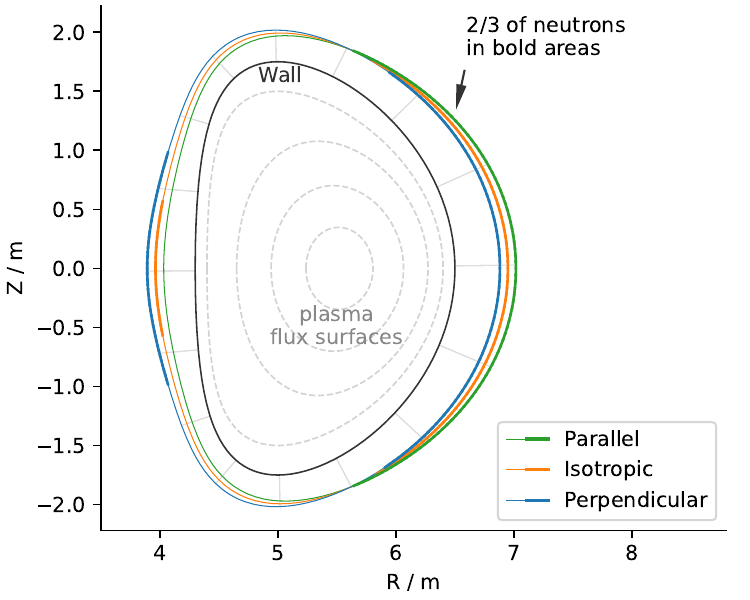}
    \caption{}
    \end{subfigure}
    \begin{subfigure}[tb]{0.99\textwidth}    \includegraphics[width=0.65\textwidth]{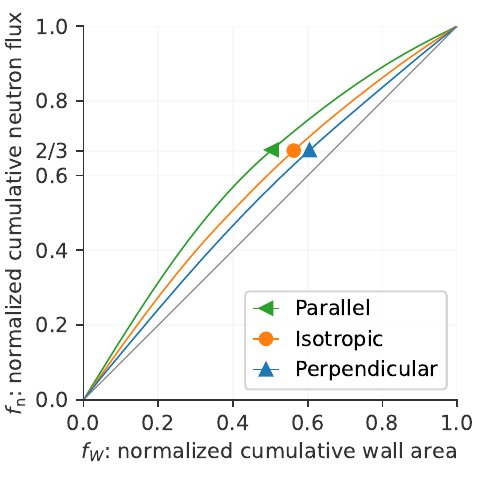}
    \caption{}
    \end{subfigure}
    \caption{(a) Normalized neutron wall loads in an example tokamak geometry, from three fuel spin-polarization mixes. For each curve the neutron wall load is indicated by the distance outward from the wall. Bold arcs along each curve highlight a portion of the wall which captures $2/3$ of the neutrons. The parallel mode moves neutrons toward the outboard side, while the perpendicular mode moves them inboard. (b) For the same geometry, the relationship between cumulative wall area and cumulative neutron flux, assuming that the feedstock is first placed in the highest-flux regions. The parallel mode further concentrates the neutron flux while the perpendicular mode decreases the non-uniformity.}
\label{fig:tokamak_spf_nonuniformity}
\end{figure}

In this section, we investigate methods for enhancing spatial variation of neutron wall loading (NWL) in a toroidal confinement machine. Such non-uniformity might be exploited by placing feedstock only in regions where the neutron flux is sufficiently high, thereby increasing the feedstock burn rate. Generally there are two approaches to changing the NWL loading: (1) using complex geometry, and (2) modifying the distribution of neutrons from the fusion source. In this section, we focus on (2) using spin-polarized fuel\cite{Kulsrud1986,Heidbrink2024,WhanBae_2025,Parisi_2024_spf,Schwartz2025} to alter the angular distribution of neutrons and accentuate \textit{or} dampen the non-uniformity of the neutron wall loading.

We present a simplified tokamak-like geometry as an example. \Cref{fig:tokamak_spf_nonuniformity}(a) shows a tokamak plasma ($A = 5.2$, $\kappa=1.5$) with an dee-shaped last closed flux surface and nearly conformal but slightly elongated dee-shaped wall. Typical density, temperature, and $q$ profiles were assumed and the MHD solution was found using DESC \cite{Panici2023}.
% fusion is calculated the usual way
The curves `wrapped' around the wall show the poloidal variation in the neutron wall load. The distance from the wall is proportional to the flux.
The neutron wall load was calculated using the package \texttt{anarrima} \cite{Schwartz2025}, which assumes that neutrons travel in straight lines to the first wall. This neglects complicated effects of neutron scattering. See \cite{WhanBae_2025} for an example.

The neutrons from typical unpolarized D-T fusion reactions are emitted in an isotropic manner. The load from these, shown by the orange curve, is greater on the outboard side due to the Shafranov shift and also from purely geometric effects.

In this work, we refer to three polarization modes, characterized by the emission bias of neutrons and alphas. 'Perpendicular' refers to the polarization scheme where the deuterium and tritium nuclear spins are aligned, resulting in neutron emission that is preferentially perpendicular to the magnetic field direction. 'Parallel' refers to the polarization scheme where the deuterium nuclear spins is tensor polarized (and tritium remains unpolarized), resulting in neutron emission that is preferentially parallel to the magnetic field direction. 'Isotropic' refers to unpolarized fuel where neutron and alpha emission is spatially isotropic. 

The neutron wall load from fuel in the parallel mode is shown with the green curve. Compared with the isotropic mode, the parallel mode concentrates the neutrons on the outboard side, decreasing the load on the inboard, top and bottom of the machine.

\begin{figure*}[tb]
    \centering
    \begin{subfigure}[tb]{0.8\textwidth}
    \centering
\includegraphics[width=1.0\textwidth]{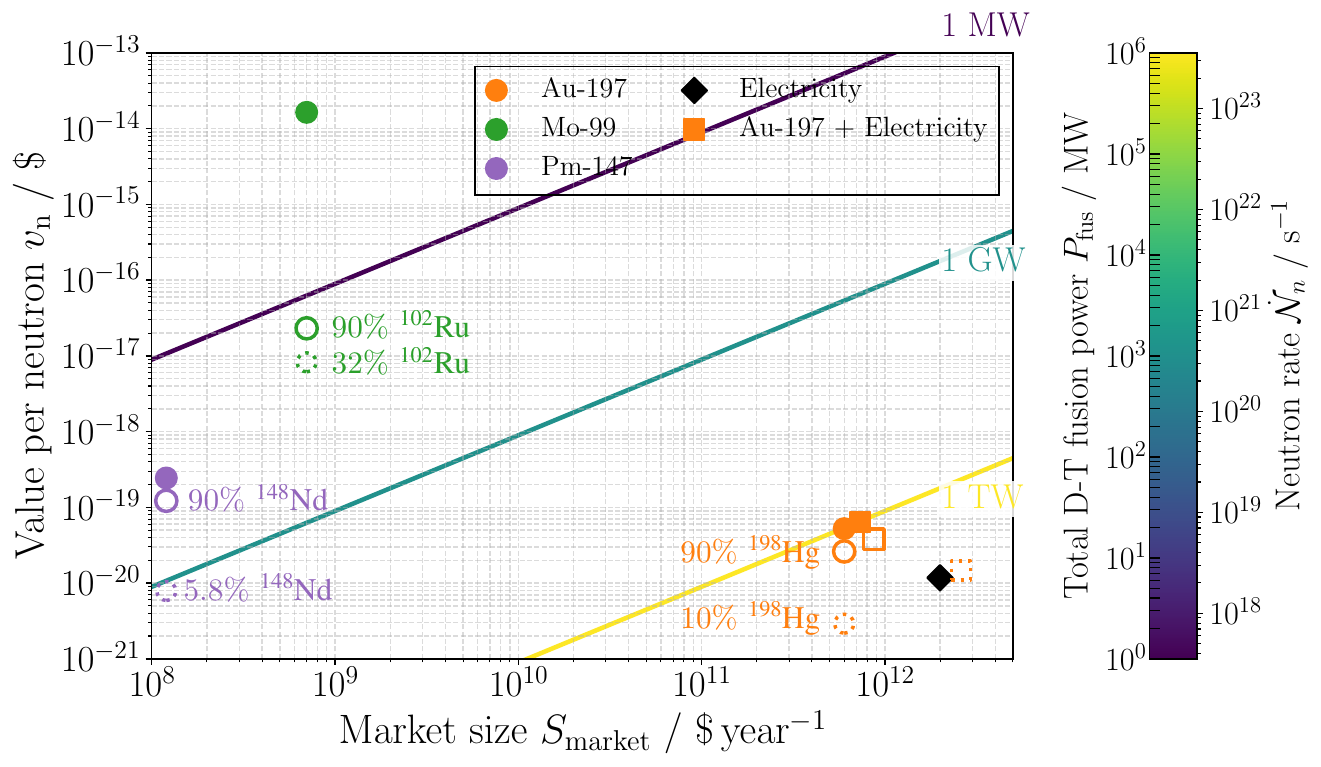}
    \end{subfigure}
    \caption{Value per neutron versus market size (see \Cref{eq:Smarket}). Filled markers correspond to $\eta_\mrm{pro} = 1.0$, dashed and dotted hollow markers correspond to a realistic $\eta_\mrm{pro}$ value for that transmutation pathway according to 90\% and natural abundance feedstock enrichment; for 90\% enrichment we use $\eta_\mrm{pro} = 1.4 \cdot 10^{-3}$ for $^{99}\mrm{Mo}$ and $\eta_\mrm{pro} = 0.5$ for other isotopes.}
\label{fig:val_per_neutron_versus_market_size}
\end{figure*}

The NWL resulting from the perpendicular mode is shown with the blue curve. This mode acts oppositely to the parallel mode, increasing the flux on the center stack. Note that the perpendicular mode is associated with an increase in the overall fusion reaction rate \footnote{A fully polarized perpendicular mode increases the D-T reaction cross section by 50\% \cite{Kulsrud1982} and can increase the overall fusion power by $\sim 80-90\%$ \cite{Smith2018,Heidbrink2024}}, but this has been neglected to emphasize the variation in neutron wall load.

\Cref{fig:tokamak_spf_nonuniformity}(b) shows the variation in the concentration of neutron flux produced by the three modes. Relative to the isotropic mode, the parallel (perpendicular) mode decreases (increases) the fraction of the wall area $f_\mathrm{W}$ on which impinges a given fraction of the neutrons $f_\mathrm{n}$. For example, $2/3$ of the neutrons impinge on $50\%, 56\%$, and $60\%$ of the wall area for the parallel, isotropic, and perpendicular modes, respectively, as indicated by the marker symbols. In \Cref{fig:tokamak_spf_nonuniformity}(a) the bold regions of the curves indicate minimal-area subsets of the wall which capture $2/3$ of the neutron flux. Therefore, using the parallel polarization mode that captures $2/3$ of the neutrons with only $50\%$ of the wall covered, the average feedstock burn rate would be $4/3$ larger than for the unoptimized case.
In this geometry, using the parallel mode and placing the feedstock on the outboard side would allow one to maximize the transmutation rate while minimizing the amount of feedstock.

%This idea is similar to 
A fraction $f_\mrm{n}$ of all neutrons passing through the first wall will be incident on a wall region with feedstock inventory. This requires packing a fraction $f_\mrm{W}$ of first-wall area with feedstock. To leading order the feedstock inventory reduction $\mathcal{R}$ when the neutron mean free path in the blanket is held constant is
\begin{equation}
\mathcal{R} \simeq \frac{1}{f_\mrm{W}}.
\label{eq:feedtock_reduction}
\end{equation}
We have also assumed that transmutation occurs on lines of sight from the neutron birth site to the neutron absorption location.

The new FBR from the nominal feedstock burn rate $\mathrm{FBR}_{a,0}$ is
\begin{equation}
\mathrm{FBR}_a = \mathcal{E} \; \mathrm{FBR}_{a,0},
\end{equation}
where the feedstock burn rate enhancement factor is
\begin{equation}
\mathcal{E} \equiv \mathcal{P} f_\mrm{n} \mathcal{R}.
\end{equation}
Here, $\mathcal{P}$ is an enhancement or suppression of the thermal fusion power due to spin-polarized fuel. Therefore, the goal is to make $\mathcal{E}$ as large as possible. As $f_\mrm{n} \to 1$, the full wall surface area is covered in feedstock inventory, and the only benefit is increased power density described by $\mathcal{P}$, making $\mathcal{\epsilon} \to \mathcal{P}$. In a fusion system, $f_\mrm{W}$ will generally always be lower than 1 due to wall ports and access points taking space. Because these ports will benefit by being in locations with lower neutron flux, these are the regions with the lowest transmutation rate. Significant neutron wall-loading asymmetries can also arise from geometric effects. For example, stellarator designs can feature significant asymmetry in NWL values compared with tokamaks \cite{el1995neutronics,el2008designing,slaybaugh2009three,lion2022deterministic,lion2025stellaris,clark2025breeder}. Such asymmetry offers new opportunities for optimizing isotope production in a blanket.

%We plot $\mathcal{E}$ versus $f_\mathrm{N}$ for different $\mathcal{R}$ and $\mathcal{P}$ values in \Cref{fig:fwall_fN_R}(b). 

\begin{figure}[tb]
    \centering
    \begin{subfigure}[tb]{0.99\textwidth}
    \centering
\includegraphics[width=1.0\textwidth]{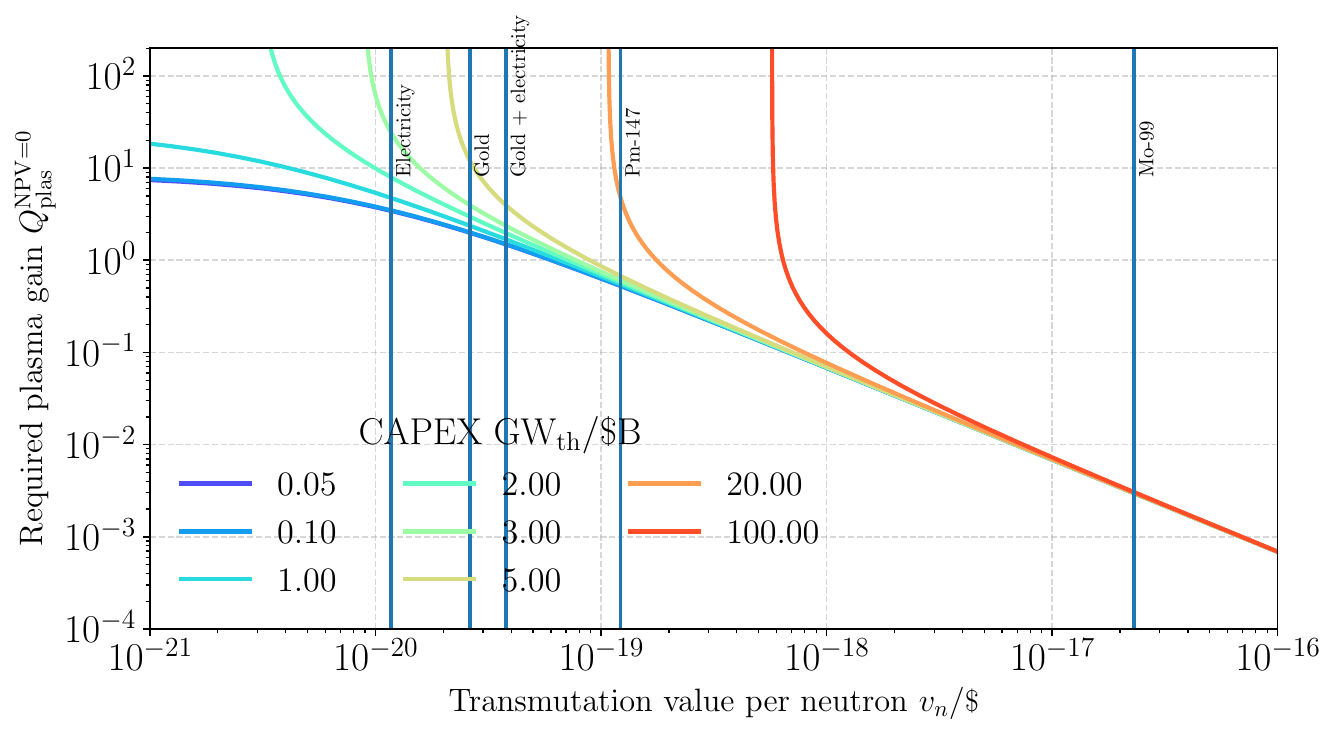}
    \end{subfigure}
    \caption{Contours of NPV breakeven versus plasma gain $Q^\mrm{NPV =0}_\mrm{plas}$ and transmutation value per neutron $v_n$ for different capital costs per $\mrm{GW_{th}}$ of fusion power. For simplicity we assume 90\% enriched feedstock and use realistic values of $\eta_\mrm{pro}$ for each feedstock.}
\label{fig:Qrequired_capex_product}
\end{figure}

\section{Market Size} \label{sec:market_size}

In this section we investigate the relation between product value and market size. If isotopes produced by fusion transmutation supplies the whole market, the annual transmutation market size $S_\mrm{market}$ is approximately,
\begin{equation}
    S_\mrm{market} \approx T_\mrm{year} \, v_\mrm{n} \, \dot{\mathcal{N}}_n,
    \label{eq:Smarket}
\end{equation}
where the value of each D-T neutron is
\begin{equation}
    v_\mrm{n} = C_\mrm{pro} \, m_\mrm{pro} \, \eta_\mrm{pro},
    \label{eq:neutronvalue}
\end{equation}
and a sum over output from all fusion neutron sources producing isotopes in a particular market is
\begin{equation}
    \dot{\mathcal{N}}_n \equiv \sum \dot{N}_\mathrm{n}.
    \label{eq:dotbigN}
\end{equation}
In \Cref{fig:val_per_neutron_versus_market_size} we plot the value per neutron versus annual market size for \ce{^197Au}, \ce{^99Mo}, \ce{^147Pm}, \ce{^194Pt}, and electricity; the filled markers for isotopes correspond to $\eta_\mrm{pro} = 1.0$. The dashed hollow markers correspond to more realistic values of $\eta_\mrm{pro}$ based on the cross section and density for each transmutation pathway.

We can now relate market size to value per neutron and the required plasma gain (see Appendix \ref{app:gain_size_derivation} for a derivation). In this context, the plasma gain $Q_{\mathrm{plas}}$ represents the ratio of fusion power generated to external heating power injected into the plasma, but it enters our market-level model as an effective performance parameter for the fleet of reactors supplying a given transmutation market. Higher $Q_{\mathrm{plas}}$ corresponds to lower recirculating power requirements, since less auxiliary heating and external electricity are needed to sustain the plasma, thereby reducing operating costs. In the fleet-averaged economic model, we assume that each plant has approximately the same plasma gain, such that $Q_{\mathrm{plas}}$ characterizes the mean performance of the deployed systems.

We first plot the contours of NPV breakeven versus plasma gain and value per neutron in \Cref{fig:Qrequired_capex_product}. The higher the neutron value, the higher the capital cost and lower plasma gain at NPV$>0$ system can support. We indicate value per neutron for electricity, gold, $\ce{^147Pm}$, and $\ce{^99Mo}$. For \ce{Au}, a plant with a CAPEX of \$3B\,/\,$\mrm{GW_{th}}$ requires $Q_\mrm{plas} \gtrsim 1.5$. If the CAPEX falls to \$2B\,/\,$\mrm{GW_{th}}$, we require $Q_\mrm{plas} \gtrsim 1$ and for CAPEX of \$1B\,/\,$\mrm{GW_{th}}$, we require $Q_\mrm{plas} \gtrsim 0.8$. \Cref{fig:Qrequired_capex_product} optimistically assumes that all D-T neutrons are captured to produce transmutation, $\eta_\mrm{pro} = 1$.

\begin{figure}[tb]
    \centering
    \begin{subfigure}[bt]{0.99\textwidth}
    \centering
\includegraphics[width=1.0\textwidth]{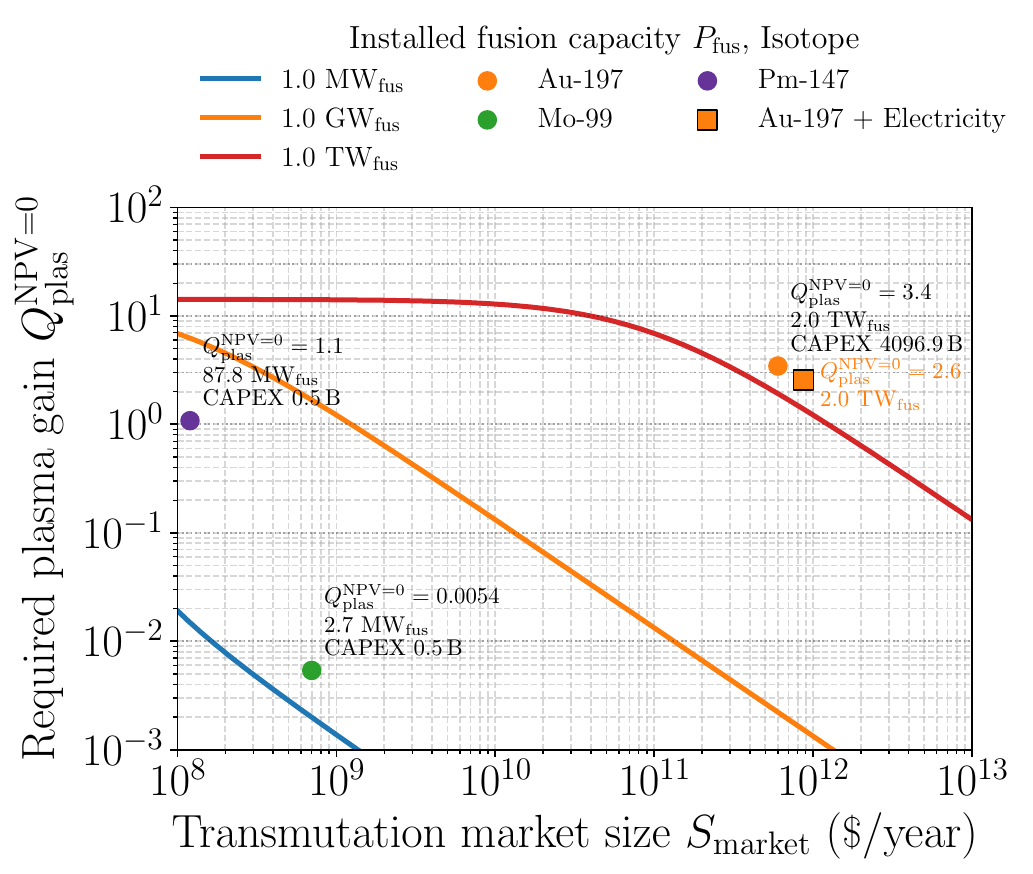}
    \end{subfigure}
    \caption{Contours of NPV breakeven versus plasma gain $Q^\mrm{NPV =0}_\mrm{plas}$ and transmutation market size $S_\mrm{market}$ for systems with different total fusion capacity, assuming a capital cost of per \$2 B per $\mrm{GW_{th}}$ of fusion power for $\ce{^197Au}$. For \ce{^99Mo} and \ce{^147Pm} we assumed a capital cost for a single fusion system of \$$0.5$B. For simplicity we assume 90\% enriched feedstock and use realistic values of $\eta_\mrm{pro}$ for each feedstock.}
\label{fig:Qrequired_marketsize}
\end{figure}

Finally, we combine the results of \Cref{fig:val_per_neutron_versus_market_size,fig:Qrequired_capex_product} to plot the required plasma gain versus the transmutation market size for NPV breakeven, shown in \Cref{fig:Qrequired_marketsize}. We used realistic $\eta_\mrm{pro}$ values corresponding to each transmutation pathway. Under these assumptions (and others listed in the figure caption), if fusion-transmuted gold were to supply the entire current gold market $S_\mrm{market} \approx \$360\mrm{B / year}$, it would support 2.0\,$\mrm{TW_{th}}$ of fusion capacity with each fusion plant requiring $Q = 3.4$ - for co-generating gold and electricity plants, it would support 2.0\,$\mrm{TW_{th}}$ of fusion capacity with each fusion plant requiring $Q = 2.6$. Supplying the entire $\ce{^99Mo}$ market would support 2.7$\mrm{MW_{th}}$ of fusion capacity with $Q = 0.0054$, and supplying the entire $\ce{^147Pm}$ market would support 87.8$\mrm{MW_{th}}$ of fusion capacity with $Q = 1.1$.

In summary, we have shown how transmutation value per neutron and market size affect the required plasma gain for NPV breakeven for a range of isotopes. Under reasonably optimistic conditions and current prices, the entire gold market could support $\sim$1--2\,$\mrm{TW_{th}}$ of fusion power with economic-breakeven fusion gain $Q_\mathrm{plas}\sim$2-3. Therefore even $\sim$10\% of the gold market could support hundreds of gigawatts of fusion power. An important caveat is that we have assumed market size will remain fixed if fusion-neutron-made isotopes were to supply the entire market - in practice it is likely that market size will change due to different production costs than existing methods.

\begin{figure}[tb!]
    \centering
    \begin{subfigure}[t]{0.99\textwidth}
    \centering \includegraphics[width=1.0\textwidth]{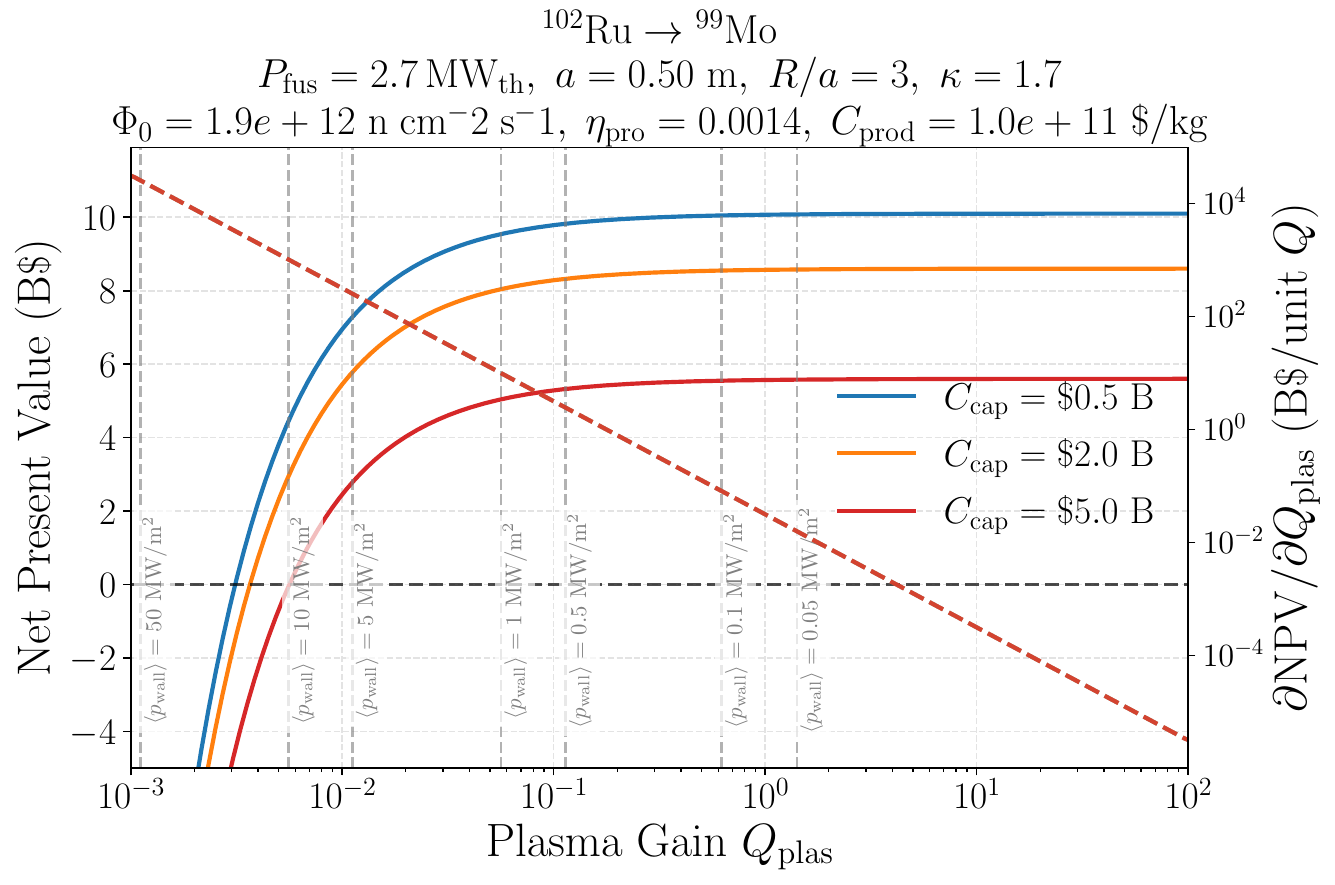}
    \end{subfigure}
    \caption{NPV for \ce{^99Mo} production (with 90\%at enriched $\ce{^102Ru}$ enriched feedstock) versus plasma gain and plant capital cost, and $\partial$NPV/ $\partial Q_\mrm{plas}$ (dashed red line). Lines of constant $\langle p_\mrm{wall} \rangle$ are also shown. We use a toroidal machine with $a = \qty{0.50}{m}$, $R/a = 3$, $\kappa = 1.7$, and first wall neutron flux $\Phi_0 = \qty{1.9e12}{\per\square\centi\meter\per\second}$.}
    \label{fig:NPV_contours_moly_plant}
\end{figure}

\section{Fusion Medical Radioisotope Facility} \label{sec:medicalradioisotopes}

In this section we provide a more detailed overview of a fusion medical isotope facility. Medical radioisotopes in many cases require high specific activity \cite{ketring2002production} --- therefore a transmutation pathway involving fusion-neutrons \cite{li2023feasibility,evitts2025theoretical,parisi2025j} requires a feedstock of a different element to the target isotope so that chemical extraction can be used to rapidly extract the target radioisotope from the feedstock \cite{parisi2025j}. There are several medical radioisotope pathways that satisfy this with reasonably high cross section including \ce{^99Mo} starting from \ce{^102Ru}, which unlike gold production in \Cref{eq:alchemy_pathway} does not use $(\mathrm{n,2n})$ but rather $(\mrm{n},\alpha)$ \cite{gascoine2021towards,evitts2025theoretical,parisi2025j},
\begin{align}
\ce{^{102}_{44}Ru}(\mrm{n},\alpha)\ce{^{99}_{42}Mo}.
\label{eq:moly_pathway}
\end{align}
Because \ce{^99Mo} has such a high value it significantly reduces the plasma physics requirements for an economically viable fusion power system. In \Cref{fig:NPV_contours_moly_plant} we plot NPV versus $Q_\mrm{plas}$ and $C_\mrm{cap}$ for a toroidal fusion configuration with $P_\mrm{fus} =2.7\,\mrm{MW}_\mrm{th}$, a minor radius of \qty{0.5}{m}, aspect ratio $A=3$, $\kappa = 1.7$, and a first-wall neutron flux $\Phi_0 = \qty{1.9e12}{\per\square\centi\meter\per\second}$. The total fusion power was chosen so that with a product price of $C_\mrm{prod} = \$\qty{e11}{\per\gram}$, approximately the entire annual market of \ce{^99Mo} is fulfilled. \Cref{fig:NPV_contours_moly_plant} shows two benefits to increasing $Q_\mrm{plas}$ at fixed fusion power: first, increasing $Q_\mrm{plas}$ reduces electricity cost because less heating power is required; second, increasing $Q_\mrm{plas}$ reduces the average wall heat flux $\langle p_\mrm{wall} \rangle$. \Cref{fig:NPV_contours_moly_plant} also shows $\partial \mrm{NPV}/\partial Q_\mrm{plas}$: the lower the value of $Q_\mrm{plas}$, the larger the financial incentive for increasing it. For example increasing $Q_\mrm{plas}$ from 0.01 to 0.02 has a value of several billion dollars.

\begin{figure}[b]
    \centering
    \begin{subfigure}[tb]{0.99\textwidth}
    \centering
\includegraphics[width=1.0\textwidth]{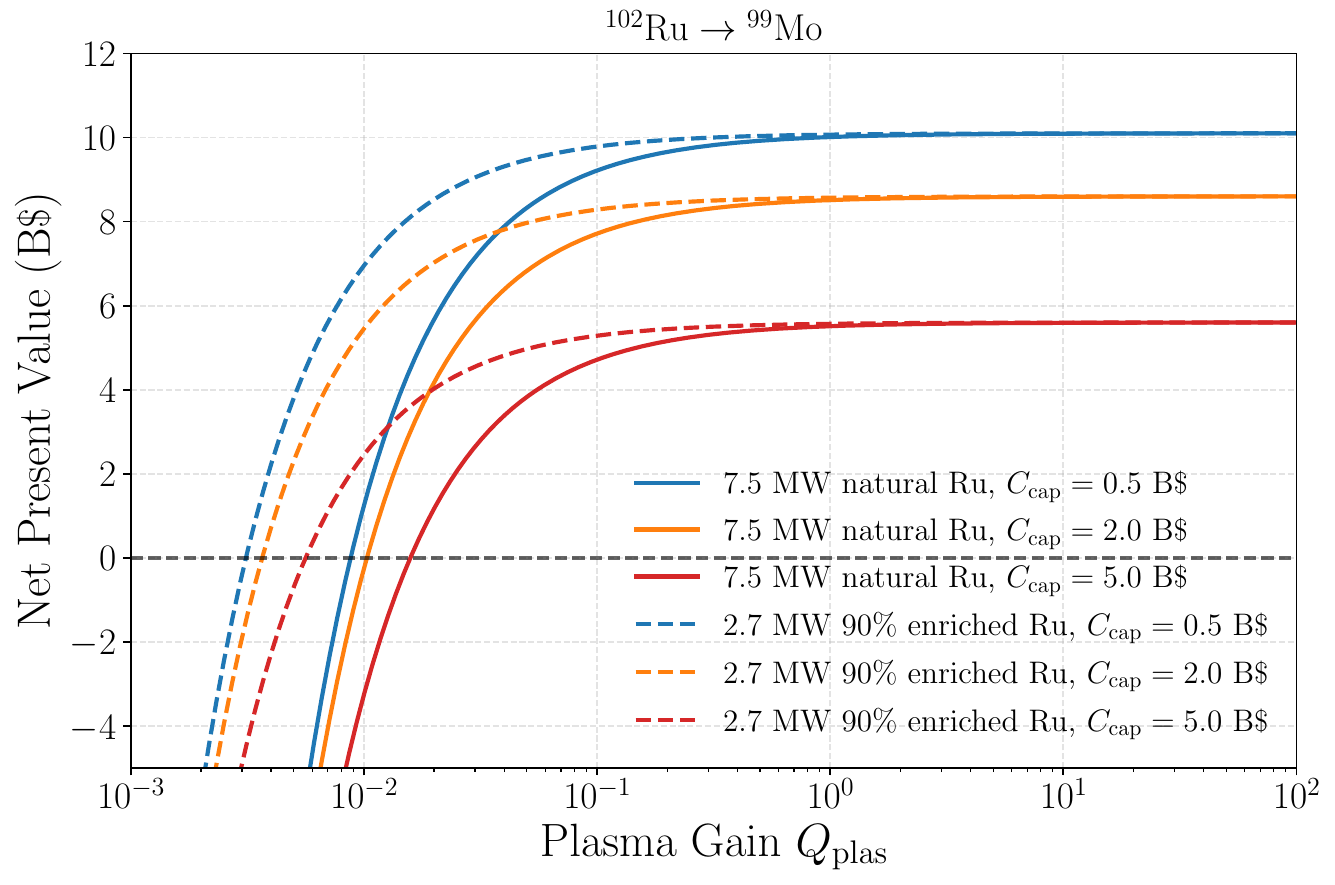}
    \end{subfigure}
    \caption{NPV for \ce{^{99}Mo} production versus plasma gain and plant capital cost for different \ce{^{102}Ru} enrichment levels.}
\label{fig:Ruenrichment}
\end{figure}

Finally, to underscore the importance of enrichment, we calculate the NPV for a $^{99}\mrm{Mo}$ transmuter supplying the entire market but with natural 32\% $^{102}\mrm{Ru}$ enrichment rather than 90\% $^{102}\mrm{Ru}$ enrichment. \Cref{fig:Ruenrichment} shows the results - while enrichment doesn't have a large impact (at fixed $C_\mathrm{cap}$) for higher plasma gain, the beneficial impact of enrichment at lower plasma gain adds several \$B to NPV.

\section{Discussion} \label{sec:discuss}

We have shown that the economic value of fusion systems can be increased substantially by harnessing high-energy fusion neutrons for isotope production within the fusion blanket. Rather than treating transmutation as a passive byproduct, we identify it as a core capability of fusion on par with electricity generation. This expands fusion from a source of power alone into a versatile platform for advanced materials production. From medical radioisotopes in megawatt-class machines to high-value stable elements such as gold at gigawatt scale, neutron-driven transmutation enhances the societal value of fusion technology.

All neutron-producing fusion concepts stand to gain from this approach - even those with lower-energy neutrons such as D-D machines~\cite{swanson2025scoping}. Integrating transmutation enhances the viability of every confinement scheme under development, aligning with broader advances across fusion confinement concepts.

Significant work remains to realize isotope production in fusion systems. This paper focused primarily on heuristic scaling arguments. Future efforts should focus on detailed neutronics and activation modeling, techno-economic assessments of target isotopes and supply chains, and the design and experimental validation of integrated fusion-transmutation blankets. Together, these steps will help establish fusion not only as an energy source but as a platform for a new class of manufacturing technologies.

\section{Acknowledgments}

We are grateful for conversations with A. Ahlholm, A. Sandorfi, K. Schiller, and J. Wexler.

\section{Code and Data Availability}

The data used in this work will be made available upon reasonable request.

\begin{table*}
\caption{Key quantities used in this work.}
\begin{ruledtabular}
\centering
  \begin{tabular}{ cccc  }
   \textbf{Name} & \textbf{Quantity} & \textbf{Units} & \textbf{Equation}  \\
    \hline
    Neutron birth rate & $\dot{N}_\mathrm{n}$ & s$^{-1}$&   \cref{eq:Ndotn_intro} \\
    Fusion power & $P_\mathrm{fus}$ &W& \cref{eq:Ndotn_intro} \\
    D-T fusion energy release & $E_\mathrm{fus}$ & J &  \cref{eq:Efus} \\
    Neutron transmutation fraction & $\eta_\mathrm{pro} $ && \cref{eq:eta_prod} \\
    Target isotope transmutation rate & $\dot{N}_\mathrm{pro}$ & s$^{-1}$&   \cref{eq:eta_prod} \\
    Macroscopic cross section & $\Sigma$ & m$^{-1}$ &   \cref{eq:Sigma_macro} \\
    Blanket thickness & $l_b$ & m &  \cref{eq:blanket_thick} \\
    Annual feedstock burn rate & $\mathrm{FBR}_a$ &  &  \cref{eq:Befficiency} \\
    FBR & $\mathrm{FBR}$ &  &  \cref{eq:Befficiency} \\
    Average first wall heat flux & $\langle p_\mrm{wall} \rangle$ & W m$^{-2}$ &  \cref{eq:pwall_initial} \\
    Average first wall neutron flux & $\Phi_0$ & s$^{-1}$ cm$^{-2}$ &  \cref{eq:pwall_initial} \\
    Plasma heat absorption efficiency & $\eta_\mrm{abs}$ &  &  \cref{eq:etaabs} \\
    Plasma gain & $Q_\mathrm{plas} $ && \cref{eq:Qplas} \\
    Revenue & $R$ & \$ s$^{-1}$ & \cref{eq:governing_eq_f} \\
    Net electric power & $P_\mathrm{e} $ & W & \cref{eq:Pmrm_e_new} \\
    Isotope price & $C_\mathrm{pro} $ & \$ kg$^{-1}$ & \cref{eq:governing_eq_f} \\
    Electricity price & $\tilde{C}_\mathrm{e} $ & \$ J$^{-1}$ & \cref{eq:governing_eq_f} \\
    Transmutation recirculating power & $P_\mathrm{pro} $ & W & \cref{eq:Pmrm_e_new} \\
    Electricity conversion efficiency & $\eta$ & & \cref{eq:Pmrm_e_new} \\
    Non-transmutation recirculating power & $P_\mathrm{circ} $ & W & \cref{eq:Pmrm_e_new} \\
    Fusion power multiplication & $\mathcal{K}^* $ & & \cref{eq:Pth} \\
    Payback time & $T_\mrm{payback}$ & s & \cref{eq:Tpayback_discount_thin} \\
    Capital cost & $C_{\rm cap}$ & \$ & \cref{eq:Ccap} \\
    Marginal net present value & $\text{NPV}_m$ & \$ & \cref{eq:NPVm} \\
    Feedstock inventory reduction & $\mathcal{R}$ & & \cref{eq:feedtock_reduction} \\
    Market size & $S_\mrm{market}$ & \$ & \cref{eq:Smarket} \\
    Value per neutron & $v_\mrm{n}$ & \$ & \cref{eq:neutronvalue} \\
    Recirculating power fraction & $f_r$ & & \cref{eq:fr}\\
    Mirror heat flux spread factor & $\chi_\mathrm{heat}$ & & \cref{eq:chi_alpha} \\
    Hybrid engineering gain & $Q_{\text{eng}}^{\text{hyb}}$ & & \cref{eq:QengTrans} \\
  \end{tabular}
\end{ruledtabular}
\label{tab:tab0}
\end{table*}

\appendix

\section{Transmutation-Only Scalings: Tokamaks} \label{sec:trans_only}

\begin{figure}[tb!]
    \centering
    \begin{subfigure}[t]{\textwidth}
    \centering
\includegraphics[width=1.0\textwidth]{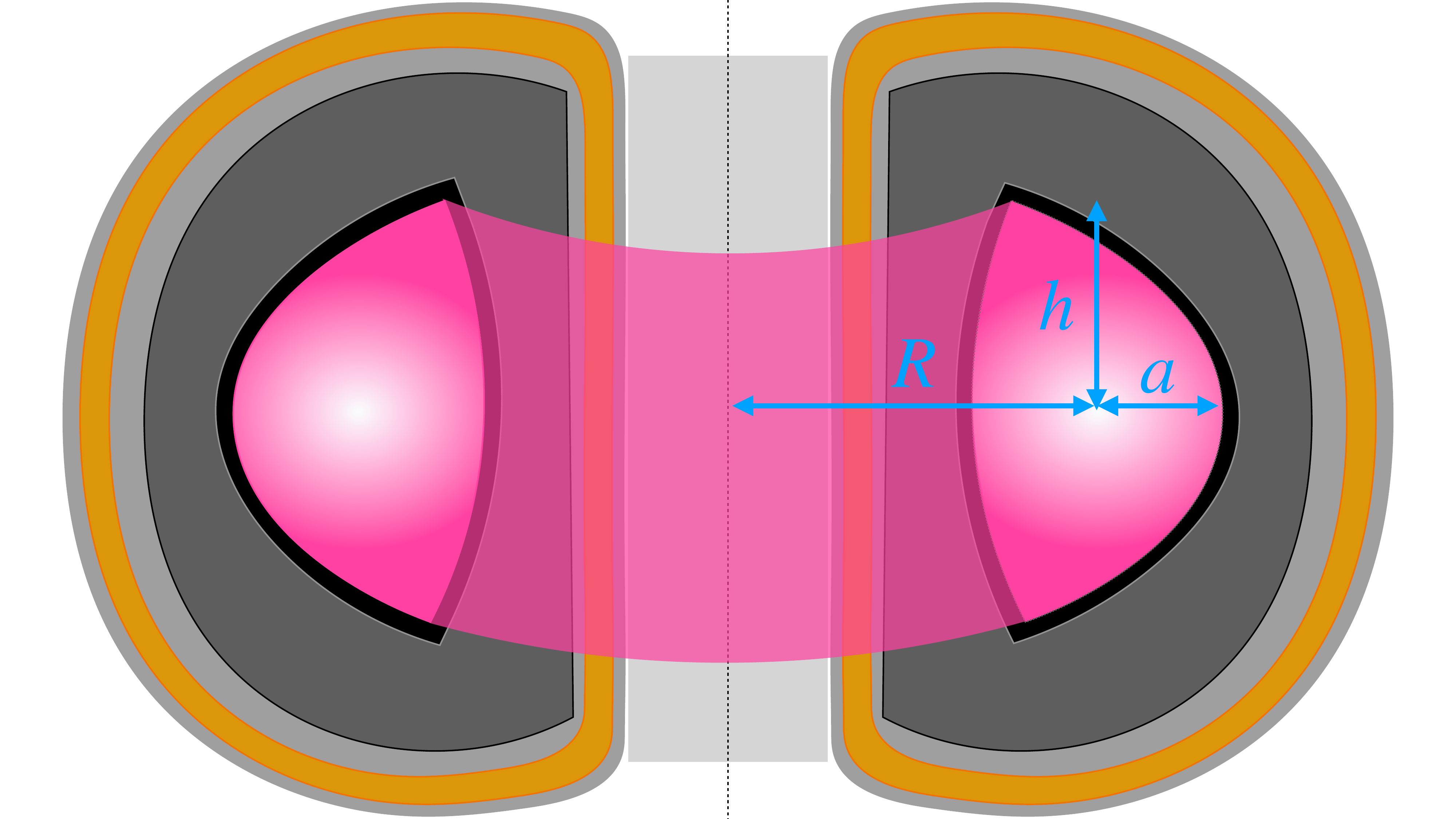}
    \end{subfigure}
    \caption{Major radius is $R$, minor is $a$, plasma height is $h$, and elongation is $\kappa = h/ a$.}
\label{fig:tokamak_schematic}
\end{figure}

In this section, we discuss the scalings for a tokamak co-producing isotopes.

In this analysis we use the following quantities: $R$ is the major radius, $a$ is the minor radius, $\kappa$ is the geometric plasma elongation, $l_b$ is the blanket thickness,
\begin{equation}
    A_b \approx 4\pi^2 R a \kappa,
\end{equation}
is the approximate blanket inner surface area and
\begin{equation}
    V_b \approx A_b l_b \approx 4\pi^2 R a \kappa l_b,
\end{equation}
is the approximate blanket volume. The tokamak geometry parameters are shown in \Cref{fig:tokamak_schematic}.

The blanket thickness is
\begin{equation}
l_b = \frac{V_b}{4\pi^2 R a \kappa}\;\propto\;(R a \kappa)^{-1},
\end{equation}
which gives the transmutation rate in the blanket,
\begin{align}
\dot{N}_{\rm pro}
&= \int_{0}^{l_b} \Phi_0 \Sigma A_b  \exp \left( -\Sigma x \right) dx \\
&= \dot N_n \Bigl[1 - \exp \bigl(-\Sigma\,\tfrac{V}{4\pi^2 R a \kappa}\bigr)\Bigr].
\end{align}
In the limit that the blanket is thin, \(\Sigma\,l_b \ll 1\),
\begin{equation}
  \dot{N}_{\rm pro}
  \approx \dot N_\mrm{n}\,\Sigma\,l_b
  = \dot N_n\,\Sigma\,\frac{V}{4\pi^2 R a \kappa}
  \;\propto\;(R a \kappa)^{-1}.
\end{equation}
In general, the transmutation fraction is
\begin{equation}
\eta_\mrm{pro} = 1 - \exp\bigl[-\Sigma \frac{V_b}{4\pi^2 R a \kappa}\bigr],
\end{equation}
which increases as \(R a \kappa\) decreases. Therefore, to maximize $\dot{N}_{\rm pro}$ at fixed \(V_b\) and \(\dot N_n\), we want to minimize the product $R a \kappa$. Thus a compact, low-aspect-ratio, low-elongation tokamak maximizes the feedstock neutron transmutation fraction $\eta_\mrm{pro}$ given a fixed blanket volume (and fixed feedstock mass).

We now estimate the feedstock burn rate as a function of these geometric parameters. Blanket inventory scales as $N_\mathrm{feed,0} \sim n_\mathrm{feed} V_b \sim n_\mathrm{feed} R a l_b \kappa$ but $\eta_\mrm{pro} = 1 - \exp\bigl[-\Sigma\,l_b\bigr]$. Therefore
\begin{equation}
\mathrm{FBR}_a \simeq \frac{\dot{N}_\mathrm{pro} T_\mathrm{year} }{N_\mathrm{feed,0} } \simeq \frac{\dot N_\mathrm{n} \eta_\mathrm{pro} T_\mathrm{year} }{n_\mathrm{feed} R a l_b \kappa} \sim \mathcal{E}_1 \frac{T_\mathrm{year} \eta_\mathrm{pro} }{l_b n_\mathrm{feed} E_\mathrm{fus} },
\label{eq:FBR_tokamak1}
\end{equation}
where
\begin{equation}
\mathcal{E}_1 \equiv \frac{P_\mathrm{fus} }{R a \kappa} = \frac{P_\mathrm{fus} }{A a^2 \kappa}.
\end{equation}
A volumetric scaling for the fusion power is
\begin{equation}
P_\mathrm{fus} \sim p_\mathrm{fus} V_\mathrm{plas} \sim p_\mathrm{fus} A a^3 \kappa.
\label{eq:Pfus_scaling}
\end{equation}
If such a scaling were true, this implies feedstock burn is more efficient in larger devices because
\begin{equation}
\mathcal{E}_1 \sim a.
\end{equation}
However, the scaling of fusion power with aspect ratio and size is generally more complicated than present in \Cref{eq:Pfus_scaling}. One scaling proposed in \cite{Menard2016} is
\begin{equation}
P_\mathrm{fus} \sim \frac{\left( \kappa \beta_N B_T \right) ^4}{A},
\label{eq:Pfus_Menard}
\end{equation}
which gives
\begin{equation}
\mathcal{E}_1 \sim \frac{\kappa^3 \beta_N^4 B_T^4 }{A^2 a^2},
\end{equation}
where $\beta_N \equiv \beta_T a B_T / I_p$ and $\beta_T \equiv 2 \mu_0 \langle p \rangle / B_{T,0}^2$. Here $B_{T}$ is the toroidal magnetic field and $B_{T,0}$ is $B_T$ evaluated at the magnetic axis. The scaling of $\kappa$, $\beta_N$, and $B_T$ with $A$ and $a$ are not straightforward, but $\kappa$ and $\beta_N$ are observed to increase with $A$ whereas $B_T$ decreases with $A$. Making the conservative assumption that $\kappa^3 \beta_N^4 B_T^4$ has a weak aspect ratio dependence, combining \Cref{eq:FBR_tokamak1,eq:Pfus_Menard} shows that at fixed minor radius
\begin{equation}
\mathrm{FBR}_a \sim \frac{1}{A^2}.
\label{eq:B_menardscaling}
\end{equation}
An integrated systems study \cite{Menard2016} found that above a threshold aspect ratio ($A \simeq 1.8$), $P_\mathrm{fus} \sim 1/A^{\alpha_0}$ (where $\alpha_0 \gtrsim 1$) which therefore demonstrates our assumption of $P_\mathrm{fus} \sim 1/A$ is reasonable. Curiously, the same analysis found $P_\mathrm{fus} / V_b$ increased with $A$. However, this does not contradict our analysis here because the blanket thickness was not kept constant.

Therefore, using the fusion power scaling $P_\mathrm{fus} \sim 1/A$ (\Cref{eq:Pfus_Menard}), we find that combined with the favorable scaling $\mathrm{FBR}_a \sim 1/A$ at fixed fusion power, that lower aspect ratio might significantly increase the feedstock burn rate. There are many challenges associated with lower aspect ratio, not least the high neutron wall loading on the inboard side, which is where many of the neutrons are located. Higher transmutation rates on the inboard side due to higher inboard neutron wall loading produces a higher power density of $\gamma$ ray emission. This may require more inboard shielding, further exacerbating the problem. On the other hand, the scaling $\mathrm{FBR}_a \sim 1/A^2$ is so favorable for low aspect ratio that it might be worthwhile. An even further enhancement of $\mathrm{FBR}_a$ could be obtained with the highly asymmetric neutron wall loading that is typical of spherical tokamaks. We discuss this more in \Cref{sec:neutron_wall_asymmetry}.

The availability of high-temperature superconductors is one of the most significant developments for fusion energy, allowing machines with high magnetic field and low magnet cooling power \cite{Hartwig2012,Sorbom2015,Creely2020}. In this section we show the effect of high field on transmutation.

\begin{figure}[tb]
    \centering
    \begin{subfigure}[tb]{0.99\textwidth}
    \centering
    \includegraphics[width=1.0\textwidth]{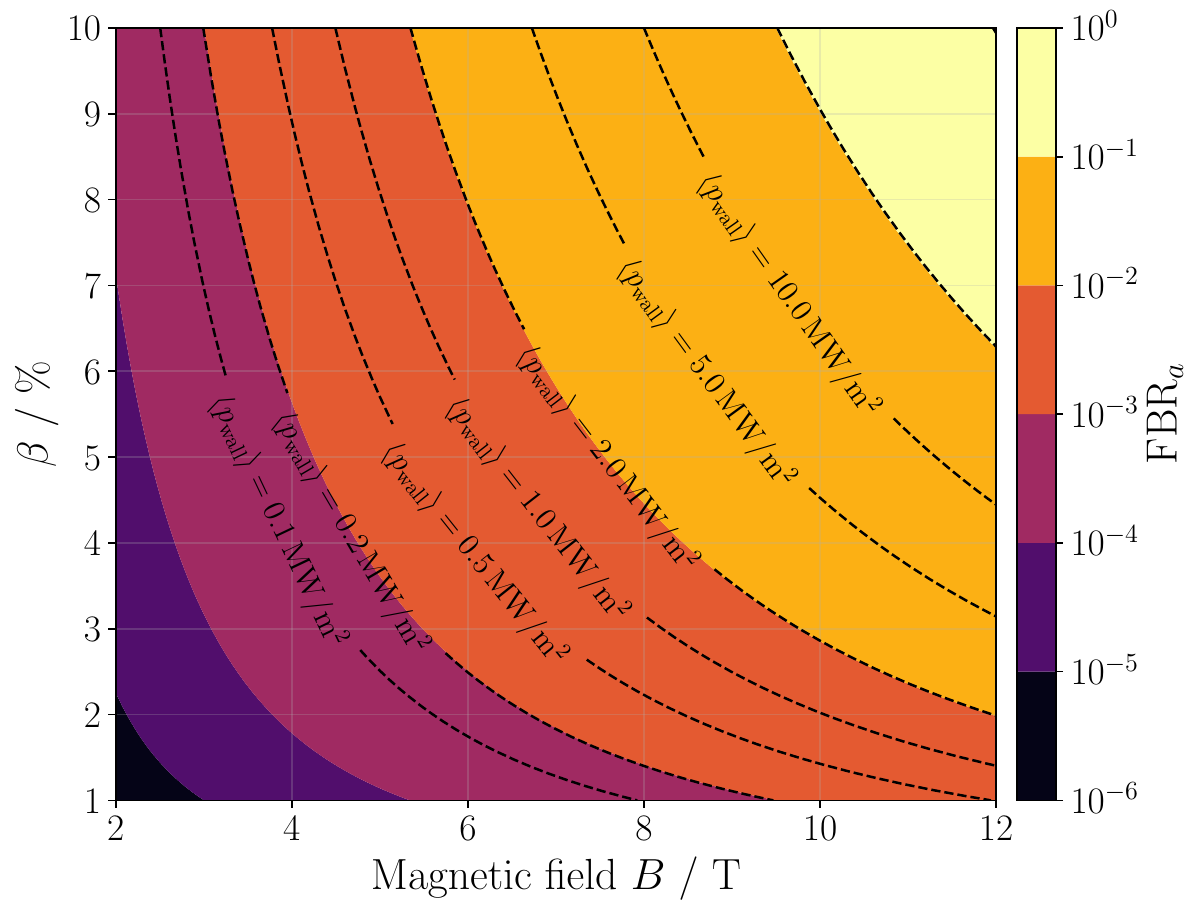}
    \end{subfigure}
    \caption{$\mrm{FBR}_a$ versus $\beta$ and $B$ for a tokamak with minor radius \qty{1.2}{m} and $\kappa = 1.7$ (gives $l_\mrm{geo} = \qty{1.02}{m}$), with contours of constant $\langle p_\mrm{wall} \rangle$ indicated (see \Cref{eq:Qplas_model} for more details). We use $\sigma = 1$b and nominal parameters $Q_\mrm{plas,0} = 0.5$, $\beta_0 = 0.01$, $B_0 = 5.0$T.}
    \label{fig:FBR_B_beta}
\end{figure}

The $\mrm{FBR}_a$ dependence on fusion power density is
\begin{equation}
    \mrm{FBR}_a = \Xi \sigma \langle p_\mrm{fus} \rangle l_\mrm{geo} \frac{T_\mrm{year}}{E_\mrm{fus}},
    \label{eq:FBRintro}
\end{equation}
where the fusion power density is
\begin{equation}
    \langle p_\mrm{fus} \rangle \equiv \frac{P_\mrm{fus}}{V_\mrm{plas}} = \frac{\langle p_\mrm{wall} \rangle}{\left(f_\mrm{\alpha} + 1 / \eta_\mrm{abs} Q_\mathrm{plas} \right) l_\mrm{geo}},
    \label{eq:pfus}
\end{equation}
and
\begin{equation}
    l_\mrm{geo} = \frac{V_\mrm{plas}}{A_b},
\end{equation}
for a plasma with volume $V_\mrm{plas}$. For a tokamak, $l_\mrm{geo} = \kappa r /2$. The fusion power density is
\begin{equation}
    p_\mrm{fus} = n_\mrm{D} n_\mrm{T} \langle \sigma_\mrm{DT} v \rangle E_\mrm{fus},
\end{equation}
$\langle \sigma_\mrm{DT} v \rangle$ is the D-T fusion reactivity averaged over deuterium and tritium distribution functions, and $\sigma_\mrm{DT}$ is the D-T fusion reaction cross section. For simplicity, for a plasma temperature between $T \simeq 10 - 20$ keV, the D-T reactivity is approximately \cite{Wesson2012}
\begin{equation}
    \langle \sigma_\mrm{DT} v \rangle \approx C T_\mrm{keV}^2\,\unit{m^3 s^{-1}},
    \label{eq:fusion_reactivity_approx}
\end{equation}
where $C = 1.1 \cdot 10^{-24}$ $\mrm{m}^3 / \mrm{s} \; \mrm{keV}$ and $T_\mrm{keV}$ is the plasma temperature in keV. We can therefore write
\begin{equation}
    \langle p_\mrm{fus} \rangle \approx K \langle \beta^2 B^4 \rangle E_\mrm{fus},
    \label{eq:pfus_betaB}
\end{equation}
for $K = \num{4.3e17}$ / $\mrm{T}^4\mrm{m}^3\mrm{s}$ (for $n_\mrm{D} = n_\mrm{T}$) where plasma beta is
\begin{equation}
    \beta \equiv \frac{ p}{B^2/2\mu_0},
\end{equation}
for a total plasma pressure $p$ and $\mu_0$ is the vacuum permeability. The FBR is therefore approximately
\begin{equation}
    \mrm{FBR}_a \approx \Xi \sigma K \langle \beta^2 B^4 \rangle l_\mrm{geo} T_\mrm{year},
    \label{eq:FBR_a}
\end{equation}
In \Cref{fig:FBR_B_beta} we plot $\mrm{FBR}_a$ versus $\beta$ and $B$ for a tokamak with a minor radius of \qty{0.8}{m}. If we require $\langle p_\mrm{wall} \rangle \lesssim \qty{10}{MW/m^2}$, it is very challenging to obtain $\mrm{FBR}_a$ greater than $0.06$ without being wall power-limited. In order to calculate contours of constant $\langle p_\mathrm{wall} \rangle$ we use \Cref{eq:FBRintro,eq:pfus} combined with
\begin{equation}
Q_\mathrm{plas} = Q_\mathrm{plas,0} \frac{\beta^2 B^4}{\beta^2_0 B^4_0},
\label{eq:Qplas_model}
\end{equation}
where $Q_\mathrm{plas,0}$, $\beta_0$, and $B_0$ are nominal parameters. In \Cref{fig:FBR_B_beta} we choose $Q_\mathrm{plas,0} = 20$, $\beta_0 = 0.01$, and $B_0 = \qty{5.0}{T}$.

\begin{figure}[tb]
    \centering
    \begin{subfigure}[tb]{0.99\textwidth}
    \centering
    \includegraphics[width=1.0\textwidth]{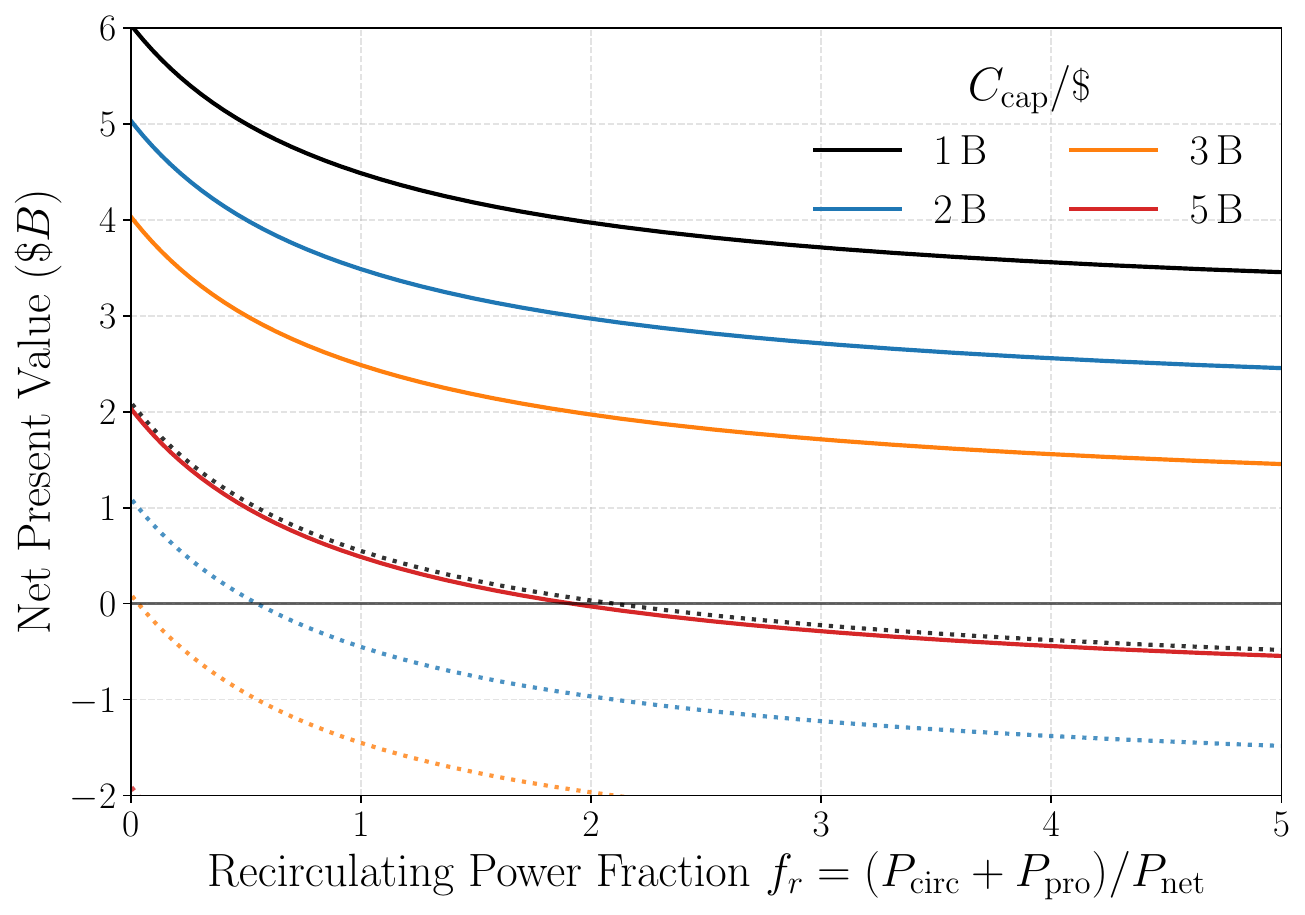}
    \end{subfigure}
    \caption{NPV versus recirculating power fraction for four capital costs. Solid lines: co-production of electricity and gold. Dotted lines: electricity-only. $f_r$ is varied by changing $Q_\mathrm{plas}$. Here, $P_\mathrm{fus} = 1$ GW.}
    \label{fig:recirc_power}
\end{figure}

According to $\mrm{FBR}_a$ in \Cref{eq:FBR_a}, the transmutation rate is
\begin{equation}
    \dot{N}_\mrm{pro} \approx \frac{\mrm{FBR}_a N_\mrm{feed}}{T_\mrm{year}} \approx \Xi \sigma K \langle \beta^2 B^4 \rangle l_\mrm{geo} N_\mrm{feed}.
    \label{eq:Ndot_pro_betaB}
\end{equation}
Therefore, $\mrm{FBR}_a$ is roughly
\begin{equation}
    \mrm{FBR}_a \approx \Xi \sigma K \langle \beta^2 B^4 \rangle l_\mrm{geo} T_\mrm{year}.
    \label{eq:FBR_B_beta_dependence}
\end{equation}

For tokamak FPPs producing electricity, it is crucial to keep the recirculating power fraction $f_r$, the ratio of recirculating power to net electricity production, sufficiently low, where
\begin{equation}
 f_r \equiv \frac{P_\mathrm{circ} + P_\mathrm{pro} }{P_\mathrm{net} }.
 \label{eq:fr}
\end{equation}
However, because transmutation offers an additional valuable product, the recirculating power can become significantly higher. We plot NPV versus $f_r$ in \Cref{fig:recirc_power} for a $P_\mathrm{fus} =$1 GW machine, showing that transmutation allows plants even with very high recirculating power to be economically viable.

\section{Transmutation-Only Scalings: Mirrors} \label{sec:trans_only_mirror}

In this section we consider the effect of one useful feature of magnetic mirrors \cite{Dimov1976AmbipolarMirrors,Baldwin1979,post1987magnetic,frank2025confinement}: the decoupling of the wall heat flux and neutron wall loading. One major difference between a tokamak and a mirror is that the mirror's wall power constraint might be much easier to overcome. This is useful when the achievable transmutation rate is limited by the heating and alpha power to the walls.

\begin{figure}[tb!]
    \centering
    \begin{subfigure}[t]{\textwidth}
    \centering
\includegraphics[width=1.0\textwidth]{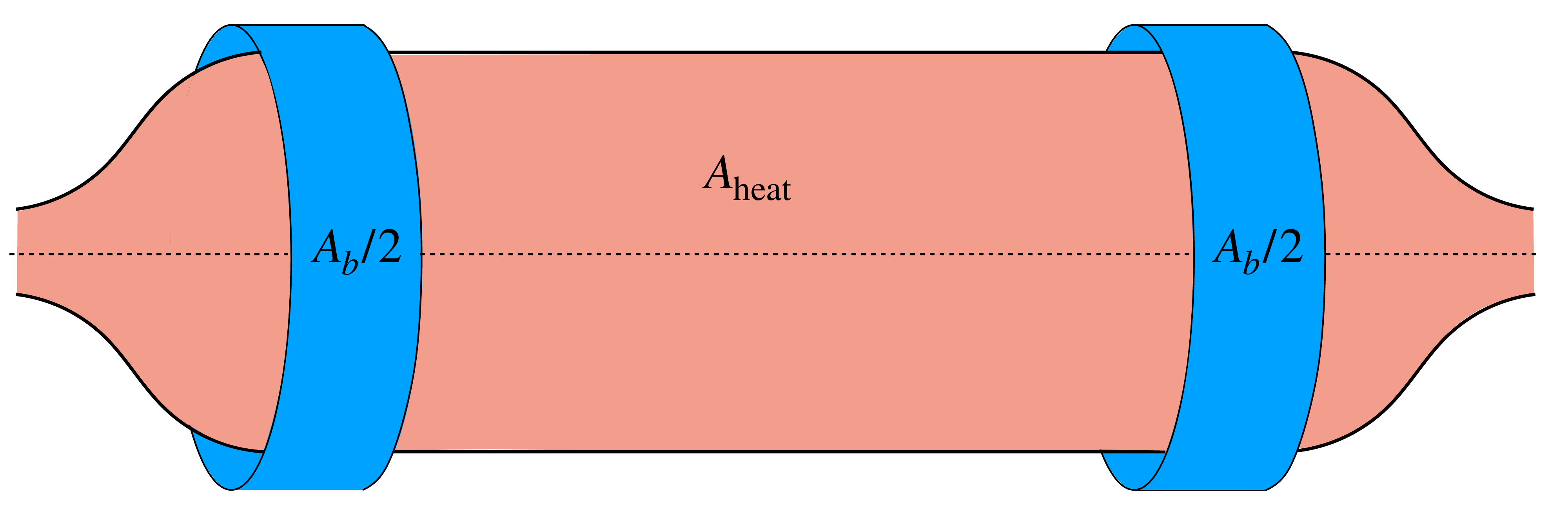}
    \end{subfigure}
    \caption{Mirror schematic: total blanket area is $A_b$ and total wall area for alpha power is $A_\mathrm{heat}$.}
\label{fig:mirror_schematic}
\end{figure}

This can be achieved by decoupling the heat flux wall power from the neutron flux in the blanket region. That is, while for a tokamak the alpha and neutron power go through the same wall area $A_b$, for a mirror the alpha heating power can be distributed over a much larger surface area. We can measure this with a heat flux areal reduction enhancement factor $\chi_\mathrm{heat} $ that relates the neutron first wall loading area $A_b$ to the alpha heating first wall area $A_\mathrm{heat}$,
\begin{equation}
    \chi_\mathrm{heat} \equiv \frac{A_\mathrm{heat}}{A_b}.
    \label{eq:chi_alpha}
\end{equation}
This idea is represented schematically in \Cref{fig:mirror_schematic}. This can be achieved using sloshing ion beam distributions \cite{kesner1980axisymmetric} that focus the power density in the mirror high-field regions, while allowing the alpha power to be distributed along the full mirror length.

Compared to tokamaks, in a mirror the fusion power density can be a factor of $\chi_\mathrm{heat}$ higher relative to the wall power density,
\begin{equation}
    \langle p_\mrm{fus} \rangle = \chi_\mathrm{heat} \frac{\langle p_\mrm{wall} \rangle}{\left(f_\mrm{\alpha} + 1 / Q_\mathrm{plas} \right) l_\mrm{geo}},
    \label{eq:pfus_mirror}
\end{equation}
which enhances the $\mathrm{FBR}_a$ by a factor $\chi_\mathrm{heat}$ for a given average wall power,
\begin{equation}
\begin{aligned}
\mathrm{FBR}_a^\mathrm{mirror} &  \simeq \chi_\mathrm{heat} \frac{\Xi \sigma \langle p_\mrm{wall} \rangle}{\left(f_\alpha + 1/\eta_\mrm{abs} Q_\mrm{plas} \right)} \frac{T_\mrm{year}}{E_\mrm{fus}} \\ & = \chi_\mathrm{heat} \Xi \sigma \Phi_0 T_\mrm{year}.
\end{aligned}
\label{eq:FBR_pwall_mirror}
\end{equation}
Note that $\Phi_0$ describes the neutron flux averaged over the entire mirror area $A_\mathrm{heat}$. However, because only a small region of the mirror has any appreciable neutron flux (approximately $A_b$), the plasma region enclosed by the blanket, the effective neutron flux on feedstock $\Phi_\mathrm{eff}$ is
\begin{equation}
\Phi_\mathrm{eff} = \chi_\mathrm{heat} \Phi_0,
\end{equation}
giving an enhanced feedstock burn rate $\mathrm{FBR}_a^\mathrm{mirror}$.

\begin{figure}[tb]
    \centering
    \begin{subfigure}[tb]{0.99\textwidth}
    \centering
    \includegraphics[width=1.0\textwidth]{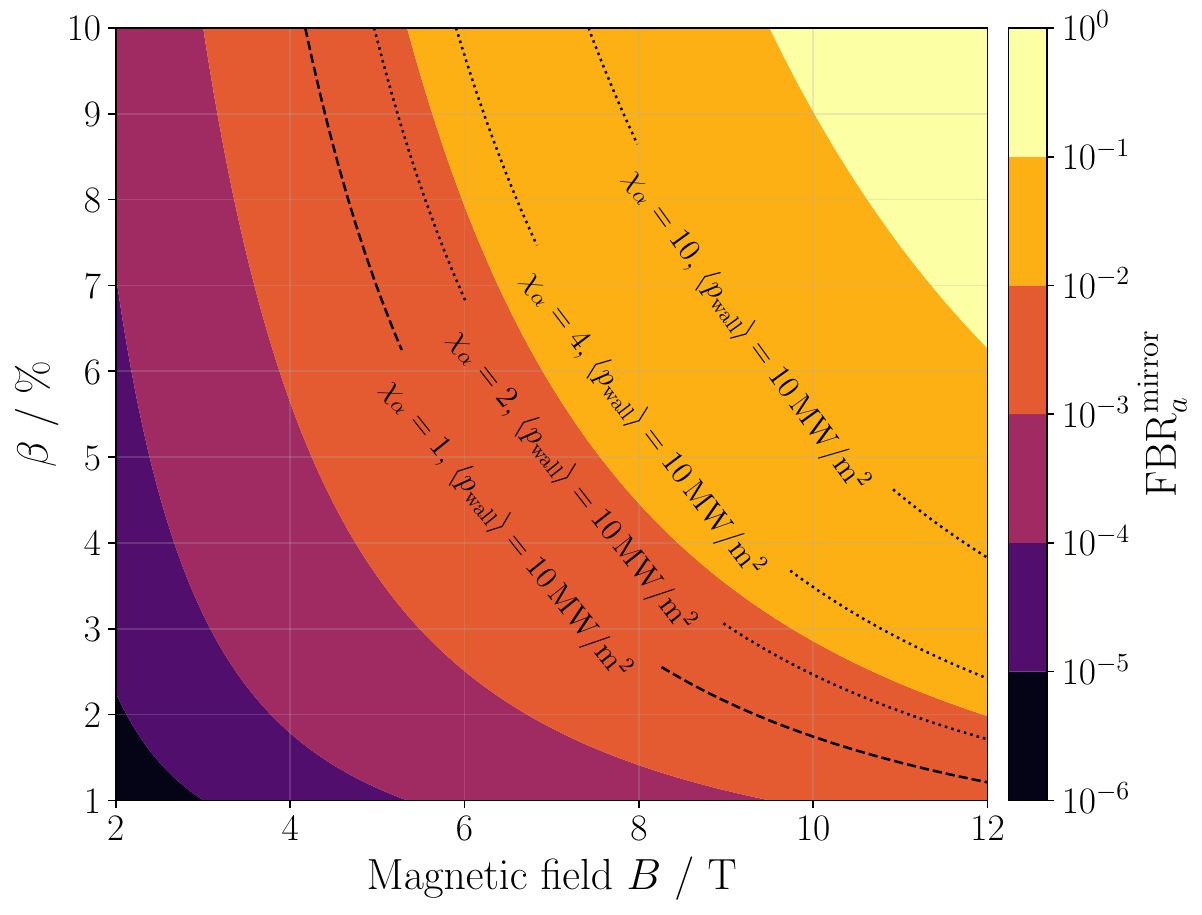}
    \end{subfigure}
    \caption{$\mrm{FBR}_a^\mrm{mirror}$ (\Cref{eq:FBR_pwall_mirror}) versus $\beta$ and $B$ for mirror with nominal $Q_\mrm{plas,0} = 0.5$, $\beta_0 = 0.01$, $B_0 = 5.0$T, and $l_\mrm{geo} =1.02$, with contours of different $\chi_\alpha$ (\Cref{eq:chi_alpha}) at $\langle p_\mrm{wall}\rangle = \qty{10}{MW/m^2}$ indicated.}
    \label{fig:FBR_B_beta_mirror}
\end{figure}

In \Cref{fig:FBR_B_beta_mirror} we plot $\mrm{FBR}_a$ versus $\beta$ and $B$ with contours of $\langle p_\mrm{wall} \rangle = \qty{10}{MW/m^2}$ and different $\chi_\alpha$ values, demonstrating how high $\chi_\alpha$ could allow for very high FBR while ensuring compatibility with alpha heating on the walls. Note that the contours in \Cref{fig:FBR_B_beta_mirror} are highly approximate for a mirror because most of the fusion power is expected to come from beam-target fusion, not thermonuclear fusion. Therefore our approximation for the fusion reactivity in \Cref{eq:fusion_reactivity_approx} will lead to significant errors.

Nonetheless, decoupling the wall power and feedstock burn rate using $\chi_\mathrm{heat}$ could be an effective way to achieve higher transmutation rates in a blanket without hitting heat flux limits. This concept is also applicable to electricity generation because most of the fusion energy recovery occurs in the fusion blanket.

\section{Hybrid Engineering Breakeven} \label{app:hybrid_breakeven}

The goal of this section is to demonstrate how transmutation reduces plasma performance requirements, here measured by a single parameter, the plasma gain $Q_\mrm{plas}$, compared with electricity-only fusion systems.

By co-producing isotopes and electricity, the `revenue breakeven' condition becomes less stringent. For an electricity-only fusion plant the usual `engineering breakeven' condition is net electricity generation \cite{Wurzel2022} $Q_\mrm{eng} > 0$ where
\begin{equation}
  Q_\mrm{eng} \equiv \frac{P_e}{P_{\text{circ}}}.
  \end{equation}
We consider $Q_\mrm{eng}$ here as a milestone against which to compare with the modified `hybrid breakeven' quantity $Q_\mrm{eng}^\mrm{hyb}$, which we introduce shortly, corresponding to total net revenue in a fusion system.

The hybrid breakeven quantity $Q_\mrm{eng}^\mrm{hyb}$ \cite{parisi2025k} also considers the parasitic power $P_{\mrm{pro}}$ needed by transmutation systems in addition to the additional revenue from isotope sales. Converting product revenue $\dot{R}_\mrm{pro}$ (\Cref{eq:governing_eq_new_prod}) to an electric-equivalent power gives,
\begin{equation}
    \widetilde P_{\mrm{pro}}
  \equiv \frac{\dot{R}_{\mrm{pro}}}{\tilde C_{e}}
  = \eta_{\mrm{pro}}
    \frac{P_{\mrm{fus}}}{E_{\text{fus}}}
    \frac{m_{\mrm{pro}}\,C_{\mrm{pro}}}{\tilde C_{e}}.
\end{equation}
A net `power' $P_{\text{net}}$ that includes the power-equivalent $\widetilde P_{\mrm{pro}}$ is
\begin{equation}
  P_{\text{net}}
  = P_e
    + \widetilde P_{\mrm{pro}}.
\end{equation}
To obtain `hybrid engineering' gain, divide $P_\mathrm{net}$ by the total electricity power $P_\mrm{circ}+P_\mrm{pro}$,
\begin{equation}
  Q_{\text{eng}}^{\text{hyb}}
  \;\equiv\;
  \frac{P_\mrm{net}}
       {P_{\text{circ}}+P_{\mrm{pro}}},
  \label{eq:QengTrans}
\end{equation}
which reduces to the standard engineering gain when $P_\mrm{pro} \to 0$ and $\tilde{P}_\mrm{pro} \to 0$.

\begin{figure}[btt!]
    \centering
    \begin{subfigure}[t]{\textwidth}
    \centering
    \includegraphics[width=1.0\textwidth]{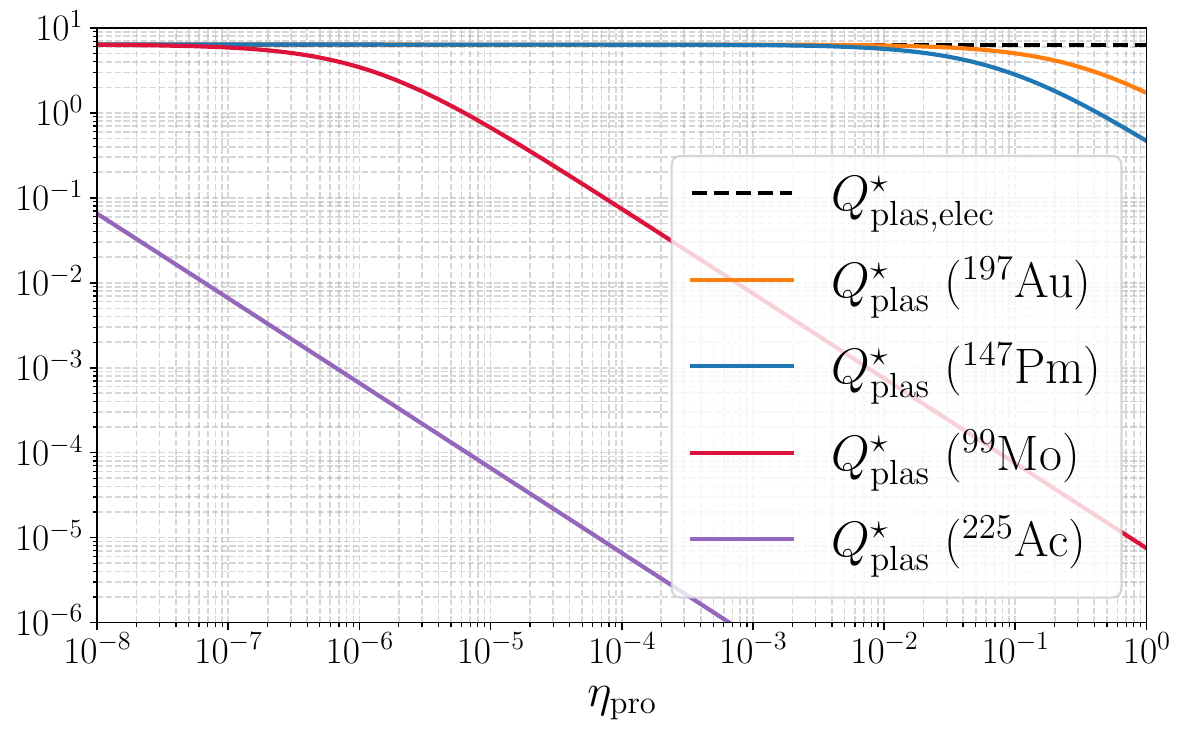}
    \caption{}
    \end{subfigure}
    \begin{subfigure}[t]{\textwidth}
    \centering
    \includegraphics[width=1.0\textwidth]{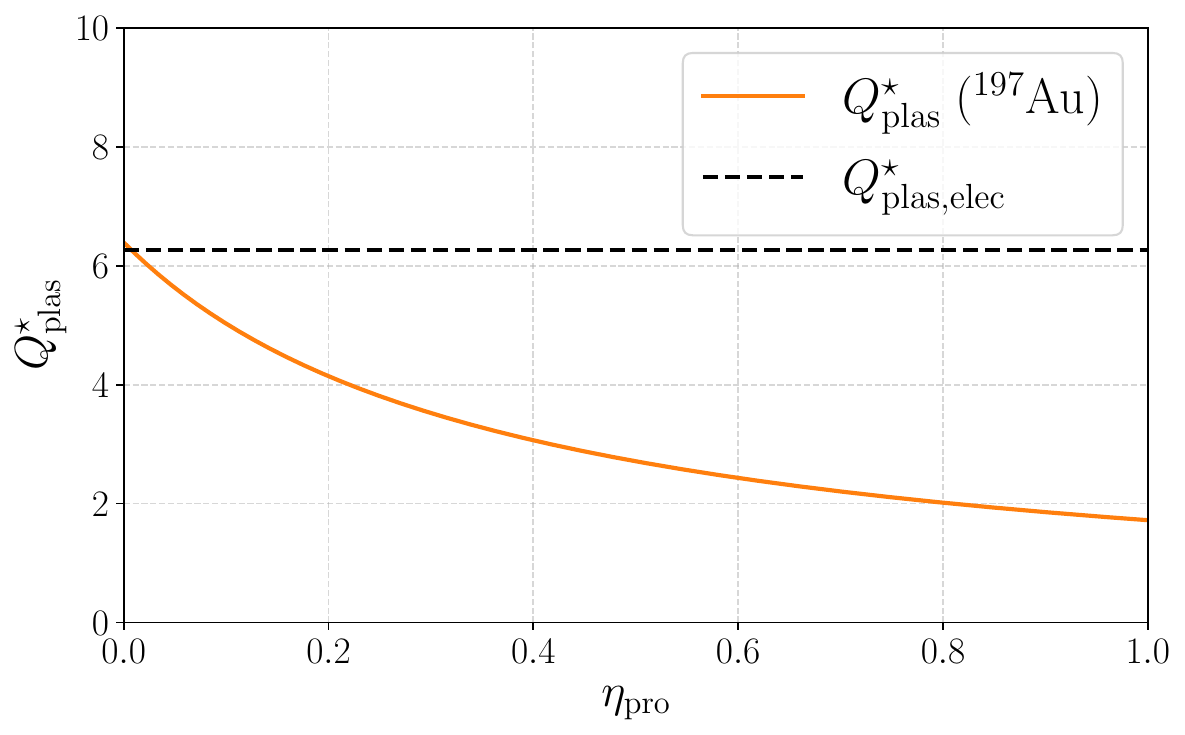}
    \caption{}
    \end{subfigure}
    \caption{Required plasma gain $Q^*_\mrm{plas}$ versus $\eta_\mrm{pro}$ for hybrid engineering breakeven ($Q_{\text{eng}}^{\text{hyb}} = 0$) (\Cref{eq:Qenghyb_simpl,eq:Qstar_breakeven,eq:Qstar_elec_breakeven} for various isotope products with $\xi=0.05$. $Q^*_\mrm{plas,elec}$ is electricity-only. Lower plot is zoomed and linear scaled for gold only.}
\label{fig:Qstar_engineering_breakeven}
\end{figure}

We now calculate the modified plasma gain requirement by relating each parasitic term through a heating term.
Similarly, assuming that the transmutation recirculating power is related to the system power by
\begin{equation}
    P_\mrm{pro} = \xi P_\mrm{system} =  (1 -f_h) \xi P_\mrm{circ},
\end{equation}
we find
\begin{equation}
    P_\mrm{pro} = \frac{P_\mrm{fus}}{Q_\mrm{plas}} \frac{(1-f_h) \xi}{\eta_\mrm{abs} \eta_\mrm{heat} f_h}.
    \label{eq:Pprogain}
\end{equation}
Generally, we expect $\xi \ll 1$. After some algebra, we find
\begin{equation}
    Q_{\text{eng}}^{\text{hyb}} = Q_\mrm{plas} G - 1.
    \label{eq:Qenghyb_simpl}
\end{equation}
where
\begin{equation}
    G \equiv \frac{\eta_\mrm{abs} \eta_\mrm{heat} f_h}{ 1 + (1- f_h) \xi} \left( \eta_\mrm{pro} \frac{m_\mrm{pro}}{E_\mrm{fus}} \frac{C_\mrm{pro}}{\widetilde C_e} + \eta \mathcal{K}^* \right).
\end{equation}
The breakeven plasma gain $Q^*_\mrm{plas}$ occurs when $Q_{\text{eng}}^{\text{hyb}} = 0$,
\begin{equation}
    Q^*_\mrm{plas} = 1 / G.
    \label{eq:Qstar_breakeven}
\end{equation}
For a pure electric output FPP $\xi = C_\mrm{pro} = 0$, the breakeven plasma gain is
\begin{equation}
    Q^*_\mrm{plas,elec} =  \frac{1}{\eta_\mrm{abs} \eta_\mrm{heat} f_h \eta \mathcal{K}^*}.
    \label{eq:Qstar_elec_breakeven}
\end{equation}
In \Cref{fig:Qstar_engineering_breakeven}, we plot $Q^*_\mrm{plas}$ for three product isotopes as well as $Q^*_\mrm{plas,elec}$. Note that because $\xi > 0$, at very low neutron capture efficiency, $Q^*_\mrm{plas} > Q^*_\mrm{plas,elec}$. With $\xi = 0.1$, transmutation fractions of $\eta_\mrm{pro} \gtrsim 0.003$, $\eta_\mrm{pro} \gtrsim 0.00048$, and $\eta_\mrm{pro} \gtrsim 6\cdot10^{-7}$ are required for the extra transmutation revenue to exceed the additional electricity cost of transmutation for \ce{^{197}Au}, \ce{^{147}Pm}, and \ce{^{99}Mo}, respectively. At $\eta_\mrm{pro} \simeq 0.5$, the required gain for hybrid breakeven with a plant co-producing \ce{^{197}Au} is two times lower. For a product as valuable as \ce{^{99}Mo}, the plasma gain for breakeven is extremely small (provided there is a sufficiently large market to sell all the product) because the transmutation product is so valuable in comparison to the electricity.

\section{Default Parameters} \label{app:default_params}

In this appendix - in particular, in Table I - we show standard parameter values used in this work such as isotope prices, market sizes, plant efficiency parameters, and cash flow model parameters.

\iffalse
\begin{table*}[htbp]
    \centering
    \caption{Isotope prices and market sizes used in this work. $C_\mathrm{pro}$ is the isotope price in million US dollars per kilogram and $S_\mrm{market}$ is the market size in billion US dollars.}
    \label{tab:vars_prices}
    \sisetup{table-format = 6.2, table-alignment-mode = format}
    \begin{tabular}{@{}r@{}lSScr@{}lS[table-format=2.1]c@{}}
        \toprule
        \multicolumn{2}{c}{Isotope} & {$C_\mathrm{pro}$ (\$M/kg)} & {$S_\mrm{market}$ (\$B/yr)} & $\eta_\mrm{pro}$ & \multicolumn{2}{c}{Feedstock} & {Natural Abundance (\%)} & Achievable Feedstock Enrichment (\%)  \\
        \midrule
        $^{99}$ & Mo & 100000 & 0.7 & 1/711 & $^{102}$ & Ru & 32 & 90  \\
        $^{147}$ & Pm & 1.0 & 0.11 & 0.5 & $^{148}$ & Nd & 5.8 & 90  \\
        $^{177}$ & Lu &  10000  &    &  1/100   & $^{176}$ & Yb & 13 & 90 \\
        $^{193}$ & Ir &   0.15  &    &   0.5  & $^{194}$ & Pt & 33 & 90  \\
        $^{197}$& Au & 0.14 & 360 & 0.5 & $^{198}$&Hg & 10 & 90 \\
        $^{225}$& Ac & 500000000 & 0.1 & 0.5 & $^{226}$&Ra & XX & 90 \\
        \bottomrule
    \end{tabular}
\end{table*}
\fi

\begin{table*}[htbp]
    \centering
    \caption{Isotope prices and market sizes used in this work. $C_\mathrm{pro}$ is the isotope price in million US dollars per kilogram and $S_\mrm{market}$ is the market size in billion US dollars.}
    \label{tab:vars_prices}
    \sisetup{table-format = 6.2, table-alignment-mode = format}
    \begin{tabular}{@{}r@{}lSScr@{}lS[table-format=2.1]c@{}}
        \toprule
        \multicolumn{2}{c}{Isotope} & {$C_\mathrm{pro}$ (\$M/kg)} & {$S_\mrm{market}$ (\$B/yr)} & $\eta_\mrm{pro}$ & \multicolumn{2}{c}{Feedstock} & {Natural Abundance (\%)} & Achievable Feedstock Enrichment (\%)  \\
        \midrule
        $^{99}$ & Mo & 100000 & 0.7 & $1.4 \cdot 10^{-3}$ & $^{102}$ & Ru & 32 & 90  \\
        $^{147}$ & Pm & 1.0 & 0.11 & 0.5 & $^{148}$ & Nd & 5.8 & 90  \\
        $^{177}$ & Lu &  10000  &    &  1/100   & $^{176}$ & Yb & 13 & 90 \\
        $^{193}$ & Ir &   0.15  &    &   0.5  & $^{194}$ & Pt & 33 & 90  \\
        $^{197}$& Au & 0.16 & 600 & 0.5 & $^{198}$&Hg & 10 & 90 \\
        $^{225}$& Ac & 500000000 & 0.1 & 0.5 & $^{226}$&Ra &  & 90 \\
        \bottomrule
    \end{tabular}
\end{table*}

\begin{table}[htbp]
    \centering
    \caption{Standard parameters used in this work.}
    \label{tab:stand_vals}
    \sisetup{table-format=1.2, table-number-alignment=center}
    \begin{tabular}{cc}
        \toprule
        \textbf{Quantity} & \textbf{Standard Value} \\
        \midrule
        $\eta$ & 0.40 \\
        $\eta_\mathrm{heat}$ & 0.60 \\
        $\eta_\mathrm{abs}$ & 0.95 \\
        $\eta_\mathrm{pro}$ & 0.50 \\
        $\sigma$ & 1 barn \\
        $f_\mathrm{h} $ & 0.6 \\
        $\kappa$ & 1.1 \\
        $\xi$ & 0.1 \\
        $L$ & 30 years \\
        $r$ & 0.05 \\
        $\xi$ & 0.10 \\
        $C_\mathrm{e} $ & 50 \$ / MWh \\
        \bottomrule
    \end{tabular}
\end{table}

\section{Extra Heating Power} \label{app:extra_power}

%The distribution of wall-plug power between heating and electrical export governs the overall performance of a co-producing fusion system. 

If magnetic confinement device is operating below neutron-wall-loading limits, drawing additional power to plasma heating can, in principle, boost the fusion rate and product generation, but at the cost of reduced electrical output. We now evaluate whether an incremental reallocation \(\delta P_{\text{heat}}^{\text{in}}>0\) increases total revenue.

Every watt diverted cannot be sold as electricity, so the electricity revenue is
\begin{equation}
\delta R_{e} \;=\; \tilde{C}_{e} \left(-\delta P_{\text{heat}}^{\text{in}} + \eta \mathcal{K}^* \delta P_\mrm{fus} \right).
\label{eq:dRe}
\end{equation}
The extra fusion power produced is
\begin{equation}
\delta P_{\text{fus}} \;=\;Q_\mrm{plas}\eta_{\text{abs}}\,\eta_{\text{heat}}\,
  \delta P_{\text{heat}}^{\text{in}},
\end{equation}
so the incremental transmutation rate is
\begin{equation}
\delta \dot N_{\mrm{pro}}
  \;=\;
  \eta_{\mrm{pro}}
  \frac{\delta P_{\text{fus}}}{E_{\text{fus}}}
  \;=\;
  \eta_{\mrm{pro}}\,
  \eta_{\text{abs}}\,\eta_{\text{heat}}\,Q_\mrm{plas}\,
  \frac{\delta P_{\text{heat}}^{\text{in}}}{E_{\text{fus}}}.
\end{equation}
Multiplying by the product mass and price gives the incremental product
revenue
\begin{align}
\delta R_{\mrm{pro}}
  &= 
  \delta\dot N_{\mrm{pro}}\,m_{\mrm{pro}}\,C_{\mrm{pro}} \\
  & =
  \eta_{\mrm{pro}}\,
\eta_{\text{abs}}\,\eta_{\text{heat}}\,Q_\mrm{plas}\,
  \frac{m_{\mrm{pro}}}{E_{\text{fus}}}\,
  C_{\mrm{pro}}\,
  \delta P_{\text{heat}}^{\text{in}}.
\label{eq:dRpro}
\end{align}

\begin{figure}[bt!]
    \centering
    \begin{subfigure}[t]{\textwidth}
    \centering
    \includegraphics[width=1.0\textwidth]{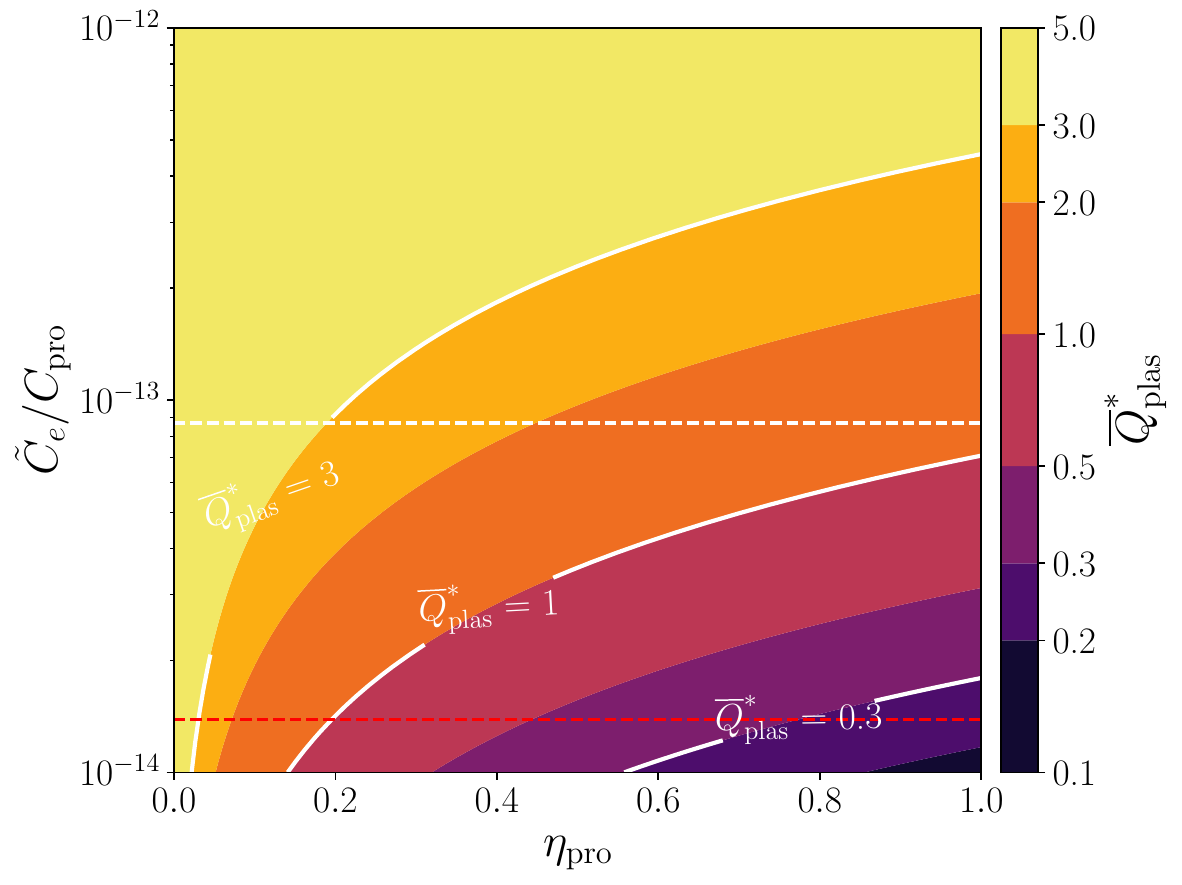}
    \end{subfigure}
    \caption{Threshold $\overline{Q}^*_\mrm{plas}$ (\Cref{eq:Qstar_full}) versus  $\tilde{C}_\mrm{e}/C_\mrm{pro}$, $\eta_\mrm{pro}$. Dashed white horizontal line indicates nominal value for ${}^{197}\mrm{Au}$ and dashed red line indicates ${}^{147}\mrm{Pm}$. We assume $\eta = 0.40$, $\eta_\mrm{abs} = 0.80$, and $\eta_\mrm{heat} = 0.60$.}
    \label{fig:Qstar_vals}
\end{figure}

Therefore the total incremental revenue is
\begin{equation}
\delta R
= \left\{
\eta_{\text{abs}}\,\eta_{\text{heat}}\,Q_\mrm{plas}
\left[
\eta\,\mathcal{K}^*\,\tilde C_e
+ \eta_{\mrm{pro}}\,
  \frac{m_{\mrm{pro}}}{E_{\text{fus}}}\,
  C_{\mrm{pro}}
\right]
- \tilde C_e
\right\}
\delta P_{\text{heat}}^{\text{in}}.
\label{eq:dR_total}
\end{equation}
\iffalse
\begin{equation}
\delta R
  \;=\;
  \biggl[
     \eta_{\mrm{pro}}\,
     \eta_{\text{abs}}\,\eta_{\text{heat}}\,Q\,
     \frac{m_{\mrm{pro}}}{E_{\text{fus}}}\,
     C_{\mrm{pro}}
     \;-\;\tilde{C}_{e}
  \biggr]
  \delta P_{\text{heat}}^{\text{in}}.
\end{equation}
\fi
Because \(\delta P_{\text{heat}}^{\text{in}}>0\), the diversion of electricity to additional heating increases total revenue
(\(\delta R>0\)) when
Because \(\delta P_{\text{heat}}^{\text{in}}>0\), diversion is beneficial (\(\delta R>0\)) when
\begin{equation}
Q_\mrm{plas} > \overline{Q}^{*}_\mrm{plas},
\label{eq:Qstar_full}
\end{equation}
where
\begin{equation}
\overline{Q}^{*}_{\rm plas}
=
\frac{1}{
\eta_{\text{abs}}\,\eta_{\text{heat}}\,
\Bigl[\eta\,\mathcal{K}^*
+\eta_{\mrm{pro}}\,
   \tfrac{m_{\mrm{pro}}}{E_{\text{fus}}}\,
   \tfrac{C_{\mrm{pro}}}{\tilde C_e}
\Bigr]}.
\label{eq:Qstar_norm}
\end{equation}
Thus, \eqref{eq:Qstar_full} is the condition for redirecting power to additional heating: any $Q_\mrm{plas}>\overline{Q}^*_\mrm{plas}$ makes redirecting additional wallplug power into heating financially advantageous; for $Q_\mrm{plas}<\overline{Q}^*_\mrm{plas}$, it is better to sell the electricity.

In \Cref{fig:Qstar_vals} we plot $\overline{Q}^*_\mrm{plas}$ versus $\tilde{C}_{e}/C_{\mrm{pro}}$ and $\eta_\mrm{pro}$. The dashed horizontal lines indicate the values of $\tilde{C}_{e}/C_{\mrm{pro}}$ for\ce{^{197}Au} (white) and \ce{^{147}Pm} (red) at present (2025) prices. If the product is \ce{^{197}Au} and $\eta_\mrm{pro} =0.5$, we require $Q_\mrm{plasma} \gtrsim 2$ for it to be worthwhile redirecting electricity for extra heating to boost transmutation. If the product is \ce{^{147}Pm} and $\eta_\mrm{pro} =0.5$, we only require $Q_\mrm{plasma} \gtrsim 0.3$ for it to be worthwhile redirecting electricity for extra heating to boost transmutation.

\section{Blankets with radioisotopes} \label{app:radioisotopes}

\Cref{eq:Npro_annual} describes production of isotopes that are either stable or are radioisotopes extracted from the blanket faster than their activity $\lambda = \ln 2 / \tau_\mrm{1/2}$, where $\tau_\mrm{1/2}$ is half-life. The net change is radioisotope inventory is
\begin{equation}
\dot{N}_\mrm{pro} = N_\mathrm{feed,0} \int \phi \sigma d E - N_\mathrm{pro} \left( \lambda - f_\mrm{ext} \right),
\end{equation}
where $\phi$ is neutron flux, $f_\mrm{ext}$ is radioisotope extraction rate, and $E$ is neutron energy.

When extraction is faster than decay, $f_\mrm{ext} \gg \lambda$, steady-state inventory ($\dot{N}_\mrm{pro} = 0$, ) is determined by a balance of transmutation and extraction,
\begin{equation}
    M_\mathrm{pro,f_\mrm{ext} \gg \lambda} \simeq \eta_\mathrm{pro} \frac{P_\mathrm{fus} }{f_\mathrm{ext}} \frac{m_\mathrm{pro}}{E_\mrm{fus}}.
    \label{eq:Mpro_annual_half_fastextract}
\end{equation}
When decay is faster than extraction $\lambda \gg f_\mrm{ext}$, steady-state inventory is determined by a balance of transmutation and radioactive decay,
\begin{equation}
    M_\mathrm{pro,\lambda \gg f_\mrm{ext}} \simeq \eta_\mathrm{pro} \frac{P_\mathrm{fus} }{\lambda} \frac{m_\mathrm{pro}}{E_\mrm{fus}}.
    \label{eq:Mpro_annual_half_fastdecay}
\end{equation}
Throughout this work we assumed $\lambda \ll f_\mrm{ext}$. However, it is important to remember that it can be practically difficult to implement a fast isotope extraction system with $\lambda \ll f_\mrm{ext}$ for radioisotopes with short half-life. Transmutation pathways that change the proton number of the product compared with the feedstock are likely to have higher $f_\mrm{ext}$ because chemical separation is generally much faster than isotope separation.

\section{Plasma Gain and Market Size} \label{app:gain_size_derivation}

In this section we derive the relation between plasma gain and market size.

We first update the total revenue $\dot{R}$ from \Cref{eq:governing_eq_new} to include market size,
\begin{equation}
\dot{R}_\mathrm{market}  = \left(\eta \mathcal{K}^* P_\mrm{fus} - P_\mrm{circ} - P_\mrm{pro}\right) \tilde{C}_\mrm{e} + \frac{S_\mrm{market}}{\dot{\mathcal{N}}_n} \frac{P_\mrm{fus}}{E_\mrm{fus} T_\mrm{year}},
\label{eq:governing_eq_new_update}
\end{equation}
We perform an effective cash flow model for the entire market,
\begin{equation}
\text{Cash}(t)=
  \begin{cases}
    -C_{\rm cap}, & t=0,\\[4pt]
    \dot{R}_\mathrm{market}, & 1\le t\le L,\\[4pt]
    0, & t>L.
  \end{cases}
\end{equation}

In order to obtain the required plasma gain, we substitute the total recirculating power using \Cref{eq:Pcircgain,eq:Pprogain},
\begin{equation}
P_\mrm{circ} + P_\mrm{pro} = \frac{P_\mrm{fus}}{Q_\mrm{plas}} \frac{1 + (1 - f_h)\xi}{\eta_\mrm{abs}\,\eta_\mrm{heat}\, f_h}.
\end{equation}
We use $I_{\mathrm{capex}}$ for the capital intensity in \$/GW$_{\mathrm{th}}$ defined as the total plant capital cost per unit of installed fusion power,
\begin{equation}
I_{\mathrm{capex}} = \frac{C_{\mathrm{cap}}}{P_{\mathrm{fus}} / 10^{9}} \quad [\$/\mathrm{GW_{th}}],
\end{equation}
such that $I_{\mathrm{capex}} = \$2\mathrm{B/GW_{th}}$ implies a total cost of \$2B for a $1~\mathrm{GW_{th}}$ fusion system. Next, we write the discount term,
\begin{equation}
S_\mathrm{disc}  =\sum_{t=0}^{L}\frac{1}{(1+r)^{t}}.
  \label{eq:Sdisc}
\end{equation}
which gives a fleet-averaged NPV-breakeven plasma gain,
\begin{align}
Q_{\mathrm{plas}}^{(\mathrm{NPV}=0)} = \frac{1 + (1 - f_h)\xi}{\eta_\mrm{abs}\, \eta_\mrm{heat}\, f_h \left[\eta \mathcal{K}^* + \frac{S_{\mathrm{market}}}{\tilde{C}_{e}\, \dot{\mathcal{N}}_{n}\, E_{\mathrm{fus}}\, T_{\mathrm{year}}} - \frac{I_{\mathrm{capex}}}{10^9\, \tilde{C}_e\, S_{\mathrm{disc}}} \right]}.
\label{eq:Qplas_market}
\end{align}

\bibliography{Master_EverythingPlasmaBib} %

\begin{thebibliography}{64}%
\makeatletter
\providecommand \@ifxundefined [1]{%
 \@ifx{#1\undefined}
}%
\providecommand \@ifnum [1]{%
 \ifnum #1\expandafter \@firstoftwo
 \else \expandafter \@secondoftwo
 \fi
}%
\providecommand \@ifx [1]{%
 \ifx #1\expandafter \@firstoftwo
 \else \expandafter \@secondoftwo
 \fi
}%
\providecommand \natexlab [1]{#1}%
\providecommand \enquote  [1]{``#1''}%
\providecommand \bibnamefont  [1]{#1}%
\providecommand \bibfnamefont [1]{#1}%
\providecommand \citenamefont [1]{#1}%
\providecommand \href@noop [0]{\@secondoftwo}%
\providecommand \href [0]{\begingroup \@sanitize@url \@href}%
\providecommand \@href[1]{\@@startlink{#1}\@@href}%
\providecommand \@@href[1]{\endgroup#1\@@endlink}%
\providecommand \@sanitize@url [0]{\catcode `\\12\catcode `\$12\catcode
  `\&12\catcode `\#12\catcode `\^12\catcode `\_12\catcode `\%12\relax}%
\providecommand \@@startlink[1]{}%
\providecommand \@@endlink[0]{}%
\providecommand \url  [0]{\begingroup\@sanitize@url \@url }%
\providecommand \@url [1]{\endgroup\@href {#1}{\urlprefix }}%
\providecommand \urlprefix  [0]{URL }%
\providecommand \Eprint [0]{\href }%
\providecommand \doibase [0]{https://doi.org/}%
\providecommand \selectlanguage [0]{\@gobble}%
\providecommand \bibinfo  [0]{\@secondoftwo}%
\providecommand \bibfield  [0]{\@secondoftwo}%
\providecommand \translation [1]{[#1]}%
\providecommand \BibitemOpen [0]{}%
\providecommand \bibitemStop [0]{}%
\providecommand \bibitemNoStop [0]{.\EOS\space}%
\providecommand \EOS [0]{\spacefactor3000\relax}%
\providecommand \BibitemShut  [1]{\csname bibitem#1\endcsname}%
\let\auto@bib@innerbib\@empty
%</preamble>
\bibitem [{\citenamefont {Sheffield}\ \emph {et~al.}(2001)\citenamefont
  {Sheffield}, \citenamefont {Brown}, \citenamefont {Garrett}, \citenamefont
  {Hilley}, \citenamefont {McCloud}, \citenamefont {Ogden}, \citenamefont
  {Shields},\ and\ \citenamefont {Waganer}}]{sheffield2001study}%
  \BibitemOpen
  \bibfield  {author} {\bibinfo {author} {\bibfnamefont {J.}~\bibnamefont
  {Sheffield}}, \bibinfo {author} {\bibfnamefont {W.}~\bibnamefont {Brown}},
  \bibinfo {author} {\bibfnamefont {G.}~\bibnamefont {Garrett}}, \bibinfo
  {author} {\bibfnamefont {J.}~\bibnamefont {Hilley}}, \bibinfo {author}
  {\bibfnamefont {D.}~\bibnamefont {McCloud}}, \bibinfo {author} {\bibfnamefont
  {J.}~\bibnamefont {Ogden}}, \bibinfo {author} {\bibfnamefont
  {T.}~\bibnamefont {Shields}},\ and\ \bibinfo {author} {\bibfnamefont
  {L.}~\bibnamefont {Waganer}},\ }\bibfield  {title} {\bibinfo {title} {A study
  of options for the deployment of large fusion power plants},\ }\href@noop {}
  {\bibfield  {journal} {\bibinfo  {journal} {Fusion science and technology}\
  }\textbf {\bibinfo {volume} {40}},\ \bibinfo {pages} {1} (\bibinfo {year}
  {2001})}\BibitemShut {NoStop}%
\bibitem [{\citenamefont {{U.S. Energy Information
  Administration}}(2025)}]{EIA2025ElectricityCapacity}%
  \BibitemOpen
  \bibfield  {author} {\bibinfo {author} {\bibnamefont {{U.S. Energy
  Information Administration}}},\ }\href@noop {} {\bibinfo {title}
  {International electricity capacity — world electricity capacity}},\
  \bibinfo {howpublished}
  {\url{https://www.eia.gov/international/data/world/electricity/electricity-capacity}}
  (\bibinfo {year} {2025}),\ \bibinfo {note} {accessed 2025-12-03}\BibitemShut
  {NoStop}%
\bibitem [{\citenamefont {Parisi}\ and\ \citenamefont
  {Schiller}(2026)}]{parisi2026valuecostfusionneutrons}%
  \BibitemOpen
  \bibfield  {author} {\bibinfo {author} {\bibfnamefont {J.~F.}\ \bibnamefont
  {Parisi}}\ and\ \bibinfo {author} {\bibfnamefont {K.}~\bibnamefont
  {Schiller}},\ }\href {https://arxiv.org/abs/2603.00835} {\bibinfo {title}
  {The value and cost of fusion neutrons}} (\bibinfo {year} {2026}),\ \Eprint
  {https://arxiv.org/abs/2603.00835} {arXiv:2603.00835 [physics.plasm-ph]}
  \BibitemShut {NoStop}%
\bibitem [{\citenamefont {Engholm}\ \emph {et~al.}(1986)\citenamefont
  {Engholm}, \citenamefont {Cheng},\ and\ \citenamefont
  {Schultz}}]{engholm1986radioisotope}%
  \BibitemOpen
  \bibfield  {author} {\bibinfo {author} {\bibfnamefont {B.~A.}\ \bibnamefont
  {Engholm}}, \bibinfo {author} {\bibfnamefont {E.~T.}\ \bibnamefont {Cheng}},\
  and\ \bibinfo {author} {\bibfnamefont {K.~R.}\ \bibnamefont {Schultz}},\
  }\bibfield  {title} {\bibinfo {title} {Radioisotope production in fusion
  reactors},\ }\href@noop {} {\bibfield  {journal} {\bibinfo  {journal} {Fusion
  technology}\ }\textbf {\bibinfo {volume} {10}},\ \bibinfo {pages} {1290}
  (\bibinfo {year} {1986})}\BibitemShut {NoStop}%
\bibitem [{\citenamefont {Bourque}\ \emph {et~al.}(1988)\citenamefont
  {Bourque}, \citenamefont {Schultz},\ and\ \citenamefont
  {Staff}}]{Bourque1988FAME}%
  \BibitemOpen
  \bibfield  {author} {\bibinfo {author} {\bibfnamefont {R.}~\bibnamefont
  {Bourque}}, \bibinfo {author} {\bibfnamefont {K.}~\bibnamefont {Schultz}},\
  and\ \bibinfo {author} {\bibfnamefont {P.}~\bibnamefont {Staff}},\
  }\href@noop {} {\emph {\bibinfo {title} {Fusion Applications and Market
  Evaluation (FAME) Study}}},\ \bibinfo {type} {Technical Report}\ \bibinfo
  {number} {GA\mbox{-}A18658 / UCRL\mbox{-}21073 / UC\mbox{-}420 /
  UC\mbox{-}424 / UC\mbox{-}712}\ (\bibinfo  {institution} {GA Technologies,
  Inc.\ (General Atomics)},\ \bibinfo {address} {San Diego, CA},\ \bibinfo
  {year} {1988})\ \bibinfo {note} {prepared under Subcontract 8236305 for
  Lawrence Livermore National Laboratory; DTIC accession AD\mbox{-}A243
  768}\BibitemShut {NoStop}%
\bibitem [{\citenamefont {Leung}\ \emph {et~al.}(2018)\citenamefont {Leung},
  \citenamefont {Leung},\ and\ \citenamefont {Melville}}]{Leung2018_CompactNG}%
  \BibitemOpen
  \bibfield  {author} {\bibinfo {author} {\bibfnamefont {K.~N.}\ \bibnamefont
  {Leung}}, \bibinfo {author} {\bibfnamefont {J.~K.}\ \bibnamefont {Leung}},\
  and\ \bibinfo {author} {\bibfnamefont {G.}~\bibnamefont {Melville}},\
  }\bibfield  {title} {\bibinfo {title} {Feasibility study on medical isotope
  production using a compact neutron generator},\ }\href
  {https://doi.org/10.1016/j.apradiso.2018.02.026} {\bibfield  {journal}
  {\bibinfo  {journal} {Applied Radiation and Isotopes}\ }\textbf {\bibinfo
  {volume} {137}},\ \bibinfo {pages} {23} (\bibinfo {year} {2018})}\BibitemShut
  {NoStop}%
\bibitem [{\citenamefont {Handley}\ \emph {et~al.}(2021)\citenamefont
  {Handley}, \citenamefont {Slesinski},\ and\ \citenamefont
  {Hsu}}]{Handley2021EarlyFusionMarkets}%
  \BibitemOpen
  \bibfield  {author} {\bibinfo {author} {\bibfnamefont {M.~C.}\ \bibnamefont
  {Handley}}, \bibinfo {author} {\bibfnamefont {D.}~\bibnamefont {Slesinski}},\
  and\ \bibinfo {author} {\bibfnamefont {S.~C.}\ \bibnamefont {Hsu}},\
  }\bibfield  {title} {\bibinfo {title} {Potential early markets for fusion
  energy},\ }\bibfield  {journal} {\bibinfo  {journal} {Journal of Fusion
  Energy}\ }\textbf {\bibinfo {volume} {40}},\ \href
  {https://doi.org/10.1007/s10894-021-00306-4} {10.1007/s10894-021-00306-4}
  (\bibinfo {year} {2021}),\ \bibinfo {note} {published online 10 July
  2021}\BibitemShut {NoStop}%
\bibitem [{\citenamefont {Honney}(2023)}]{Honney2023FusionNeutrons}%
  \BibitemOpen
  \bibfield  {author} {\bibinfo {author} {\bibfnamefont {T.}~\bibnamefont
  {Honney}},\ }\bibfield  {title} {\bibinfo {title} {New value from fusion
  neutrons},\ }\href
  {https://www.neimagazine.com/analysis/new-value-from-fusion-neutrons-10924344/}
  {\bibfield  {journal} {\bibinfo  {journal} {Nuclear Engineering
  International}\ } (\bibinfo {year} {2023})},\ \bibinfo {note} {interview with
  Greg Piefer (SHINE Technologies)}\BibitemShut {NoStop}%
\bibitem [{shi({\natexlab{a}})}]{shine_neutron_imaging_2025}%
  \BibitemOpen
  \href {https://www.shinefusion.com/phase-1} {\bibinfo {title} {{Neutron
  Imaging for Industrial Components | Phase 1}}} ({\natexlab{a}}),\ \bibinfo
  {note} {accessed 24 May 2025}\BibitemShut {NoStop}%
\bibitem [{\citenamefont {Evitts}\ \emph {et~al.}(2025)\citenamefont {Evitts},
  \citenamefont {Miller}, \citenamefont {{Da Pieve}}, \citenamefont {Turner},\
  and\ \citenamefont {Borini}}]{evitts2025theoretical}%
  \BibitemOpen
  \bibfield  {author} {\bibinfo {author} {\bibfnamefont {L.~J.}\ \bibnamefont
  {Evitts}}, \bibinfo {author} {\bibfnamefont {P.~W.}\ \bibnamefont {Miller}},
  \bibinfo {author} {\bibfnamefont {C.}~\bibnamefont {{Da Pieve}}}, \bibinfo
  {author} {\bibfnamefont {A.}~\bibnamefont {Turner}},\ and\ \bibinfo {author}
  {\bibfnamefont {S.}~\bibnamefont {Borini}},\ }\bibfield  {title} {\bibinfo
  {title} {Theoretical novel medical isotope production with deuterium-tritium
  fusion technology},\ }\href
  {https://doi.org/https://doi.org/10.1016/j.apradiso.2025.112163} {\bibfield
  {journal} {\bibinfo  {journal} {Applied Radiation and Isotopes}\ }\textbf
  {\bibinfo {volume} {226}},\ \bibinfo {pages} {112163} (\bibinfo {year}
  {2025})}\BibitemShut {NoStop}%
\bibitem [{\citenamefont {Parisi}\ \emph {et~al.}(2025)\citenamefont {Parisi},
  \citenamefont {Rutkowski}, \citenamefont {Harter}, \citenamefont {Schwartz},\
  and\ \citenamefont {Chen}}]{parisi2025j}%
  \BibitemOpen
  \bibfield  {author} {\bibinfo {author} {\bibfnamefont {J.~F.}\ \bibnamefont
  {Parisi}}, \bibinfo {author} {\bibfnamefont {A.}~\bibnamefont {Rutkowski}},
  \bibinfo {author} {\bibfnamefont {J.}~\bibnamefont {Harter}}, \bibinfo
  {author} {\bibfnamefont {J.~A.}\ \bibnamefont {Schwartz}},\ and\ \bibinfo
  {author} {\bibfnamefont {S.}~\bibnamefont {Chen}},\ }\href
  {https://arxiv.org/abs/2511.02814} {\bibinfo {title} {Production of
  high-specific-activity radioisotopes using high-energy fusion neutrons}}
  (\bibinfo {year} {2025}),\ \Eprint {https://arxiv.org/abs/2511.02814}
  {arXiv:2511.02814 [nucl-ex]} \BibitemShut {NoStop}%
\bibitem [{\citenamefont {Rutkowski}\ \emph {et~al.}(2025)\citenamefont
  {Rutkowski}, \citenamefont {Harter},\ and\ \citenamefont
  {Parisi}}]{rutkowski2025scalable}%
  \BibitemOpen
  \bibfield  {author} {\bibinfo {author} {\bibfnamefont {A.}~\bibnamefont
  {Rutkowski}}, \bibinfo {author} {\bibfnamefont {J.}~\bibnamefont {Harter}},\
  and\ \bibinfo {author} {\bibfnamefont {J.}~\bibnamefont {Parisi}},\
  }\bibfield  {title} {\bibinfo {title} {Scalable chrysopoeia via $(n, 2n) $
  reactions driven by deuterium-tritium fusion neutrons},\ }\href@noop {}
  {\bibfield  {journal} {\bibinfo  {journal} {arXiv preprint arXiv:2507.13461}\
  } (\bibinfo {year} {2025})}\BibitemShut {NoStop}%
\bibitem [{\citenamefont {Brown}\ \emph {et~al.}(2018)\citenamefont {Brown},
  \citenamefont {Chadwick}, \citenamefont {Capote}, \citenamefont {Kahler},
  \citenamefont {Trkov}, \citenamefont {Herman}, \citenamefont {Sonzogni},
  \citenamefont {Danon}, \citenamefont {Carlson}, \citenamefont {Dunn},
  \citenamefont {Smith}, \citenamefont {Hale}, \citenamefont {Arbanas},
  \citenamefont {Arcilla}, \citenamefont {Bates}, \citenamefont {Beck},
  \citenamefont {Becker}, \citenamefont {Brown}, \citenamefont {Casperson},
  \citenamefont {Conlin}, \citenamefont {Cullen}, \citenamefont {Descalle},
  \citenamefont {Firestone}, \citenamefont {Gaines}, \citenamefont {Guber},
  \citenamefont {Hawari}, \citenamefont {Holmes}, \citenamefont {Johnson},
  \citenamefont {Kawano}, \citenamefont {Kiedrowski}, \citenamefont {Koning},
  \citenamefont {Kopecky}, \citenamefont {Leal}, \citenamefont {Lestone},
  \citenamefont {Lubitz}, \citenamefont {Dami{\'a}n}, \citenamefont {Mattoon},
  \citenamefont {McCutchan}, \citenamefont {Mughabghab}, \citenamefont
  {Navratil}, \citenamefont {Neudecker}, \citenamefont {Nobre}, \citenamefont
  {Noguere}, \citenamefont {Paris}, \citenamefont {Pigni}, \citenamefont
  {Plompen}, \citenamefont {Pritychenko}, \citenamefont {Pronyaev},
  \citenamefont {Roubtsov}, \citenamefont {Rochman}, \citenamefont {Romano},
  \citenamefont {Schillebeeckx}, \citenamefont {Simakov}, \citenamefont {Sin},
  \citenamefont {Sirakov}, \citenamefont {Sleaford}, \citenamefont {Sobes},
  \citenamefont {Soukhovitskii}, \citenamefont {Stetcu}, \citenamefont {Talou},
  \citenamefont {Thompson}, \citenamefont {van~der Marck}, \citenamefont
  {Welser-Sherrill}, \citenamefont {Wiarda}, \citenamefont {White},
  \citenamefont {Wormald}, \citenamefont {Wright}, \citenamefont {Zerkle},
  \citenamefont {\v{Z}erovnik},\ and\ \citenamefont {Zhu}}]{Brown20181}%
  \BibitemOpen
  \bibfield  {author} {\bibinfo {author} {\bibfnamefont {D.}~\bibnamefont
  {Brown}}, \bibinfo {author} {\bibfnamefont {M.}~\bibnamefont {Chadwick}},
  \bibinfo {author} {\bibfnamefont {R.}~\bibnamefont {Capote}}, \bibinfo
  {author} {\bibfnamefont {A.}~\bibnamefont {Kahler}}, \bibinfo {author}
  {\bibfnamefont {A.}~\bibnamefont {Trkov}}, \bibinfo {author} {\bibfnamefont
  {M.}~\bibnamefont {Herman}}, \bibinfo {author} {\bibfnamefont
  {A.}~\bibnamefont {Sonzogni}}, \bibinfo {author} {\bibfnamefont
  {Y.}~\bibnamefont {Danon}}, \bibinfo {author} {\bibfnamefont
  {A.}~\bibnamefont {Carlson}}, \bibinfo {author} {\bibfnamefont
  {M.}~\bibnamefont {Dunn}}, \bibinfo {author} {\bibfnamefont {D.}~\bibnamefont
  {Smith}}, \bibinfo {author} {\bibfnamefont {G.}~\bibnamefont {Hale}},
  \bibinfo {author} {\bibfnamefont {G.}~\bibnamefont {Arbanas}}, \bibinfo
  {author} {\bibfnamefont {R.}~\bibnamefont {Arcilla}}, \bibinfo {author}
  {\bibfnamefont {C.}~\bibnamefont {Bates}}, \bibinfo {author} {\bibfnamefont
  {B.}~\bibnamefont {Beck}}, \bibinfo {author} {\bibfnamefont {B.}~\bibnamefont
  {Becker}}, \bibinfo {author} {\bibfnamefont {F.}~\bibnamefont {Brown}},
  \bibinfo {author} {\bibfnamefont {R.}~\bibnamefont {Casperson}}, \bibinfo
  {author} {\bibfnamefont {J.}~\bibnamefont {Conlin}}, \bibinfo {author}
  {\bibfnamefont {D.}~\bibnamefont {Cullen}}, \bibinfo {author} {\bibfnamefont
  {M.-A.}\ \bibnamefont {Descalle}}, \bibinfo {author} {\bibfnamefont
  {R.}~\bibnamefont {Firestone}}, \bibinfo {author} {\bibfnamefont
  {T.}~\bibnamefont {Gaines}}, \bibinfo {author} {\bibfnamefont
  {K.}~\bibnamefont {Guber}}, \bibinfo {author} {\bibfnamefont
  {A.}~\bibnamefont {Hawari}}, \bibinfo {author} {\bibfnamefont
  {J.}~\bibnamefont {Holmes}}, \bibinfo {author} {\bibfnamefont
  {T.}~\bibnamefont {Johnson}}, \bibinfo {author} {\bibfnamefont
  {T.}~\bibnamefont {Kawano}}, \bibinfo {author} {\bibfnamefont
  {B.}~\bibnamefont {Kiedrowski}}, \bibinfo {author} {\bibfnamefont
  {A.}~\bibnamefont {Koning}}, \bibinfo {author} {\bibfnamefont
  {S.}~\bibnamefont {Kopecky}}, \bibinfo {author} {\bibfnamefont
  {L.}~\bibnamefont {Leal}}, \bibinfo {author} {\bibfnamefont {J.}~\bibnamefont
  {Lestone}}, \bibinfo {author} {\bibfnamefont {C.}~\bibnamefont {Lubitz}},
  \bibinfo {author} {\bibfnamefont {J.~M.}\ \bibnamefont {Dami{\'a}n}},
  \bibinfo {author} {\bibfnamefont {C.}~\bibnamefont {Mattoon}}, \bibinfo
  {author} {\bibfnamefont {E.}~\bibnamefont {McCutchan}}, \bibinfo {author}
  {\bibfnamefont {S.}~\bibnamefont {Mughabghab}}, \bibinfo {author}
  {\bibfnamefont {P.}~\bibnamefont {Navratil}}, \bibinfo {author}
  {\bibfnamefont {D.}~\bibnamefont {Neudecker}}, \bibinfo {author}
  {\bibfnamefont {G.}~\bibnamefont {Nobre}}, \bibinfo {author} {\bibfnamefont
  {G.}~\bibnamefont {Noguere}}, \bibinfo {author} {\bibfnamefont
  {M.}~\bibnamefont {Paris}}, \bibinfo {author} {\bibfnamefont
  {M.}~\bibnamefont {Pigni}}, \bibinfo {author} {\bibfnamefont
  {A.}~\bibnamefont {Plompen}}, \bibinfo {author} {\bibfnamefont
  {B.}~\bibnamefont {Pritychenko}}, \bibinfo {author} {\bibfnamefont
  {V.}~\bibnamefont {Pronyaev}}, \bibinfo {author} {\bibfnamefont
  {D.}~\bibnamefont {Roubtsov}}, \bibinfo {author} {\bibfnamefont
  {D.}~\bibnamefont {Rochman}}, \bibinfo {author} {\bibfnamefont
  {P.}~\bibnamefont {Romano}}, \bibinfo {author} {\bibfnamefont
  {P.}~\bibnamefont {Schillebeeckx}}, \bibinfo {author} {\bibfnamefont
  {S.}~\bibnamefont {Simakov}}, \bibinfo {author} {\bibfnamefont
  {M.}~\bibnamefont {Sin}}, \bibinfo {author} {\bibfnamefont {I.}~\bibnamefont
  {Sirakov}}, \bibinfo {author} {\bibfnamefont {B.}~\bibnamefont {Sleaford}},
  \bibinfo {author} {\bibfnamefont {V.}~\bibnamefont {Sobes}}, \bibinfo
  {author} {\bibfnamefont {E.}~\bibnamefont {Soukhovitskii}}, \bibinfo {author}
  {\bibfnamefont {I.}~\bibnamefont {Stetcu}}, \bibinfo {author} {\bibfnamefont
  {P.}~\bibnamefont {Talou}}, \bibinfo {author} {\bibfnamefont
  {I.}~\bibnamefont {Thompson}}, \bibinfo {author} {\bibfnamefont
  {S.}~\bibnamefont {van~der Marck}}, \bibinfo {author} {\bibfnamefont
  {L.}~\bibnamefont {Welser-Sherrill}}, \bibinfo {author} {\bibfnamefont
  {D.}~\bibnamefont {Wiarda}}, \bibinfo {author} {\bibfnamefont
  {M.}~\bibnamefont {White}}, \bibinfo {author} {\bibfnamefont
  {J.}~\bibnamefont {Wormald}}, \bibinfo {author} {\bibfnamefont
  {R.}~\bibnamefont {Wright}}, \bibinfo {author} {\bibfnamefont
  {M.}~\bibnamefont {Zerkle}}, \bibinfo {author} {\bibfnamefont
  {G.}~\bibnamefont {\v{Z}erovnik}},\ and\ \bibinfo {author} {\bibfnamefont
  {Y.}~\bibnamefont {Zhu}},\ }\bibfield  {title} {\bibinfo {title}
  {{ENDF/B-VIII.0}: The {8$^{th}$} major release of the nuclear reaction data
  library with {CIELO}-project cross sections, new standards and thermal
  scattering data},\ }\href
  {https://doi.org/https://doi.org/10.1016/j.nds.2018.02.001} {\bibfield
  {journal} {\bibinfo  {journal} {Nuclear Data Sheets}\ }\textbf {\bibinfo
  {volume} {148}},\ \bibinfo {pages} {1 } (\bibinfo {year} {2018})},\ \bibinfo
  {note} {special Issue on Nuclear Reaction Data}\BibitemShut {NoStop}%
\bibitem [{\citenamefont {Pietropaolo}\ \emph {et~al.}(2021)\citenamefont
  {Pietropaolo}, \citenamefont {Contessa}, \citenamefont {Farini},
  \citenamefont {Fonnesu}, \citenamefont {Marinari}, \citenamefont {Moro},
  \citenamefont {Rizzo}, \citenamefont {Scaglione}, \citenamefont {Terranova},
  \citenamefont {Utili} \emph {et~al.}}]{pietropaolo2021sorgentina}%
  \BibitemOpen
  \bibfield  {author} {\bibinfo {author} {\bibfnamefont {A.}~\bibnamefont
  {Pietropaolo}}, \bibinfo {author} {\bibfnamefont {G.~M.}\ \bibnamefont
  {Contessa}}, \bibinfo {author} {\bibfnamefont {M.}~\bibnamefont {Farini}},
  \bibinfo {author} {\bibfnamefont {N.}~\bibnamefont {Fonnesu}}, \bibinfo
  {author} {\bibfnamefont {R.}~\bibnamefont {Marinari}}, \bibinfo {author}
  {\bibfnamefont {F.}~\bibnamefont {Moro}}, \bibinfo {author} {\bibfnamefont
  {A.}~\bibnamefont {Rizzo}}, \bibinfo {author} {\bibfnamefont
  {S.}~\bibnamefont {Scaglione}}, \bibinfo {author} {\bibfnamefont
  {N.}~\bibnamefont {Terranova}}, \bibinfo {author} {\bibfnamefont
  {M.}~\bibnamefont {Utili}}, \emph {et~al.},\ }\bibfield  {title} {\bibinfo
  {title} {Sorgentina-rf project: Fusion neutrons for 99mo medical
  radioisotope},\ }\href@noop {} {\bibfield  {journal} {\bibinfo  {journal}
  {Eur. Phys. J. Plus}\ }\textbf {\bibinfo {volume} {136}},\ \bibinfo {pages}
  {1140} (\bibinfo {year} {2021})}\BibitemShut {NoStop}%
\bibitem [{\citenamefont {Li}\ and\ \citenamefont
  {Zheng}(2023)}]{li2023feasibility}%
  \BibitemOpen
  \bibfield  {author} {\bibinfo {author} {\bibfnamefont {J.}~\bibnamefont
  {Li}}\ and\ \bibinfo {author} {\bibfnamefont {S.}~\bibnamefont {Zheng}},\
  }\bibfield  {title} {\bibinfo {title} {Feasibility study to byproduce medical
  radioisotopes in a fusion reactor},\ }\href@noop {} {\bibfield  {journal}
  {\bibinfo  {journal} {Molecules}\ }\textbf {\bibinfo {volume} {28}},\
  \bibinfo {pages} {2040} (\bibinfo {year} {2023})}\BibitemShut {NoStop}%
\bibitem [{\citenamefont {Parisi}\ and\ \citenamefont
  {Rutkowski}(2025)}]{parisi2025k}%
  \BibitemOpen
  \bibfield  {author} {\bibinfo {author} {\bibfnamefont {J.~F.}\ \bibnamefont
  {Parisi}}\ and\ \bibinfo {author} {\bibfnamefont {A.}~\bibnamefont
  {Rutkowski}},\ }\href {https://arxiv.org/abs/2511.02814} {\bibinfo {title}
  {Isotope production in muon-catalyzed-fusion systems}} (\bibinfo {year}
  {2025}),\ \Eprint {https://arxiv.org/abs/2511.02814} {arXiv:2511.02814
  [nucl-ex]} \BibitemShut {NoStop}%
\bibitem [{\citenamefont {Fischer}(1957)}]{fischer1957cross}%
  \BibitemOpen
  \bibfield  {author} {\bibinfo {author} {\bibfnamefont {G.~J.}\ \bibnamefont
  {Fischer}},\ }\bibfield  {title} {\bibinfo {title} {Cross section for the (n,
  2 n) reaction in be 9},\ }\href@noop {} {\bibfield  {journal} {\bibinfo
  {journal} {Physical Review}\ }\textbf {\bibinfo {volume} {108}},\ \bibinfo
  {pages} {99} (\bibinfo {year} {1957})}\BibitemShut {NoStop}%
\bibitem [{\citenamefont {Ihli}\ \emph {et~al.}(2008)\citenamefont {Ihli},
  \citenamefont {Basu}, \citenamefont {Giancarli}, \citenamefont {Konishi},
  \citenamefont {Malang}, \citenamefont {Najmabadi}, \citenamefont {Nishio},
  \citenamefont {Raffray}, \citenamefont {Rao}, \citenamefont {Sagara} \emph
  {et~al.}}]{ihli2008review}%
  \BibitemOpen
  \bibfield  {author} {\bibinfo {author} {\bibfnamefont {T.}~\bibnamefont
  {Ihli}}, \bibinfo {author} {\bibfnamefont {T.}~\bibnamefont {Basu}}, \bibinfo
  {author} {\bibfnamefont {L.}~\bibnamefont {Giancarli}}, \bibinfo {author}
  {\bibfnamefont {S.}~\bibnamefont {Konishi}}, \bibinfo {author} {\bibfnamefont
  {S.}~\bibnamefont {Malang}}, \bibinfo {author} {\bibfnamefont
  {F.}~\bibnamefont {Najmabadi}}, \bibinfo {author} {\bibfnamefont
  {S.}~\bibnamefont {Nishio}}, \bibinfo {author} {\bibfnamefont
  {A.}~\bibnamefont {Raffray}}, \bibinfo {author} {\bibfnamefont
  {C.}~\bibnamefont {Rao}}, \bibinfo {author} {\bibfnamefont {A.}~\bibnamefont
  {Sagara}}, \emph {et~al.},\ }\bibfield  {title} {\bibinfo {title} {Review of
  blanket designs for advanced fusion reactors},\ }\href@noop {} {\bibfield
  {journal} {\bibinfo  {journal} {Fusion Engineering and Design}\ }\textbf
  {\bibinfo {volume} {83}},\ \bibinfo {pages} {912} (\bibinfo {year}
  {2008})}\BibitemShut {NoStop}%
\bibitem [{\citenamefont {Gilbert}\ \emph {et~al.}(2014)\citenamefont
  {Gilbert}, \citenamefont {Zheng}, \citenamefont {Kemp}, \citenamefont
  {Packer}, \citenamefont {Dudarev},\ and\ \citenamefont
  {Sublet}}]{gilbert2014comparative}%
  \BibitemOpen
  \bibfield  {author} {\bibinfo {author} {\bibfnamefont {M.}~\bibnamefont
  {Gilbert}}, \bibinfo {author} {\bibfnamefont {S.}~\bibnamefont {Zheng}},
  \bibinfo {author} {\bibfnamefont {R.}~\bibnamefont {Kemp}}, \bibinfo {author}
  {\bibfnamefont {L.}~\bibnamefont {Packer}}, \bibinfo {author} {\bibfnamefont
  {S.}~\bibnamefont {Dudarev}},\ and\ \bibinfo {author} {\bibfnamefont {J.-C.}\
  \bibnamefont {Sublet}},\ }\bibfield  {title} {\bibinfo {title} {Comparative
  assessment of material performance in demo fusion reactors},\ }\href@noop {}
  {\bibfield  {journal} {\bibinfo  {journal} {Fusion Science and Technology}\
  }\textbf {\bibinfo {volume} {66}},\ \bibinfo {pages} {9} (\bibinfo {year}
  {2014})}\BibitemShut {NoStop}%
\bibitem [{\citenamefont {Gascoine}(2021)}]{gascoine2021towards}%
  \BibitemOpen
  \bibfield  {author} {\bibinfo {author} {\bibfnamefont {M.}~\bibnamefont
  {Gascoine}},\ }\bibfield  {title} {\bibinfo {title} {Towards the fast
  neutron-induced isotope production of 99mtc via the 102ru (n, $\alpha$) 99mo
  reaction.},\ }\href@noop {} {\  (\bibinfo {year} {2021})}\BibitemShut
  {NoStop}%
\bibitem [{\citenamefont {Hartley}\ \emph {et~al.}(1978)\citenamefont
  {Hartley}, \citenamefont {Gore},\ and\ \citenamefont
  {Young}}]{hartley1978potential}%
  \BibitemOpen
  \bibfield  {author} {\bibinfo {author} {\bibfnamefont {J.}~\bibnamefont
  {Hartley}}, \bibinfo {author} {\bibfnamefont {B.}~\bibnamefont {Gore}},\ and\
  \bibinfo {author} {\bibfnamefont {J.}~\bibnamefont {Young}},\ }\bibfield
  {title} {\bibinfo {title} {Potential lithium requirements for fusion power
  plants},\ }\href@noop {} {\bibfield  {journal} {\bibinfo  {journal} {Lithium
  Needs and Resources}\ ,\ \bibinfo {pages} {337}} (\bibinfo {year}
  {1978})}\BibitemShut {NoStop}%
\bibitem [{\citenamefont {Fasel}\ and\ \citenamefont
  {Tran}(2005)}]{fasel2005availability}%
  \BibitemOpen
  \bibfield  {author} {\bibinfo {author} {\bibfnamefont {D.}~\bibnamefont
  {Fasel}}\ and\ \bibinfo {author} {\bibfnamefont {M.}~\bibnamefont {Tran}},\
  }\bibfield  {title} {\bibinfo {title} {Availability of lithium in the context
  of future d--t fusion reactors},\ }\href@noop {} {\bibfield  {journal}
  {\bibinfo  {journal} {Fusion engineering and design}\ }\textbf {\bibinfo
  {volume} {75}},\ \bibinfo {pages} {1163} (\bibinfo {year}
  {2005})}\BibitemShut {NoStop}%
\bibitem [{\citenamefont {De~Les~Valls}\ \emph {et~al.}(2008)\citenamefont
  {De~Les~Valls}, \citenamefont {Sedano}, \citenamefont {Batet}, \citenamefont
  {Ricapito}, \citenamefont {Aiello}, \citenamefont {Gastaldi},\ and\
  \citenamefont {Gabriel}}]{de2008lead}%
  \BibitemOpen
  \bibfield  {author} {\bibinfo {author} {\bibfnamefont {E.~M.}\ \bibnamefont
  {De~Les~Valls}}, \bibinfo {author} {\bibfnamefont {L.}~\bibnamefont
  {Sedano}}, \bibinfo {author} {\bibfnamefont {L.}~\bibnamefont {Batet}},
  \bibinfo {author} {\bibfnamefont {I.}~\bibnamefont {Ricapito}}, \bibinfo
  {author} {\bibfnamefont {A.}~\bibnamefont {Aiello}}, \bibinfo {author}
  {\bibfnamefont {O.}~\bibnamefont {Gastaldi}},\ and\ \bibinfo {author}
  {\bibfnamefont {F.}~\bibnamefont {Gabriel}},\ }\bibfield  {title} {\bibinfo
  {title} {Lead--lithium eutectic material database for nuclear fusion
  technology},\ }\href@noop {} {\bibfield  {journal} {\bibinfo  {journal}
  {Journal of nuclear materials}\ }\textbf {\bibinfo {volume} {376}},\ \bibinfo
  {pages} {353} (\bibinfo {year} {2008})}\BibitemShut {NoStop}%
\bibitem [{\citenamefont {De~Castro}\ \emph {et~al.}(2021)\citenamefont
  {De~Castro}, \citenamefont {Moynihan}, \citenamefont {Stemmley},
  \citenamefont {Szott},\ and\ \citenamefont {Ruzic}}]{de2021lithium}%
  \BibitemOpen
  \bibfield  {author} {\bibinfo {author} {\bibfnamefont {A.}~\bibnamefont
  {De~Castro}}, \bibinfo {author} {\bibfnamefont {C.}~\bibnamefont {Moynihan}},
  \bibinfo {author} {\bibfnamefont {S.}~\bibnamefont {Stemmley}}, \bibinfo
  {author} {\bibfnamefont {M.}~\bibnamefont {Szott}},\ and\ \bibinfo {author}
  {\bibfnamefont {D.}~\bibnamefont {Ruzic}},\ }\bibfield  {title} {\bibinfo
  {title} {Lithium, a path to make fusion energy affordable},\ }\href@noop {}
  {\bibfield  {journal} {\bibinfo  {journal} {Physics of Plasmas}\ }\textbf
  {\bibinfo {volume} {28}} (\bibinfo {year} {2021})}\BibitemShut {NoStop}%
\bibitem [{\citenamefont {Ward}\ \emph {et~al.}(2025)\citenamefont {Ward},
  \citenamefont {Pearson}, \citenamefont {Scott},\ and\ \citenamefont
  {Cardozo}}]{ward2025lithium}%
  \BibitemOpen
  \bibfield  {author} {\bibinfo {author} {\bibfnamefont {S.~H.}\ \bibnamefont
  {Ward}}, \bibinfo {author} {\bibfnamefont {R.~J.}\ \bibnamefont {Pearson}},
  \bibinfo {author} {\bibfnamefont {T.}~\bibnamefont {Scott}},\ and\ \bibinfo
  {author} {\bibfnamefont {N.~J.~L.}\ \bibnamefont {Cardozo}},\ }\bibfield
  {title} {\bibinfo {title} {Lithium enrichment threatens to curb fusion
  deployment},\ }\href@noop {} {\bibfield  {journal} {\bibinfo  {journal}
  {Joule}\ } (\bibinfo {year} {2025})}\BibitemShut {NoStop}%
\bibitem [{\citenamefont {Ueda}\ \emph {et~al.}(2017)\citenamefont {Ueda},
  \citenamefont {Schmid}, \citenamefont {Balden}, \citenamefont {Coenen},
  \citenamefont {Loewenhoff}, \citenamefont {Ito}, \citenamefont {Hasegawa},
  \citenamefont {Hardie}, \citenamefont {Porton},\ and\ \citenamefont
  {Gilbert}}]{ueda2017baseline}%
  \BibitemOpen
  \bibfield  {author} {\bibinfo {author} {\bibfnamefont {Y.}~\bibnamefont
  {Ueda}}, \bibinfo {author} {\bibfnamefont {K.}~\bibnamefont {Schmid}},
  \bibinfo {author} {\bibfnamefont {M.}~\bibnamefont {Balden}}, \bibinfo
  {author} {\bibfnamefont {J.}~\bibnamefont {Coenen}}, \bibinfo {author}
  {\bibfnamefont {T.}~\bibnamefont {Loewenhoff}}, \bibinfo {author}
  {\bibfnamefont {A.}~\bibnamefont {Ito}}, \bibinfo {author} {\bibfnamefont
  {A.}~\bibnamefont {Hasegawa}}, \bibinfo {author} {\bibfnamefont
  {C.}~\bibnamefont {Hardie}}, \bibinfo {author} {\bibfnamefont
  {M.}~\bibnamefont {Porton}},\ and\ \bibinfo {author} {\bibfnamefont
  {M.}~\bibnamefont {Gilbert}},\ }\bibfield  {title} {\bibinfo {title}
  {Baseline high heat flux and plasma facing materials for fusion},\
  }\href@noop {} {\bibfield  {journal} {\bibinfo  {journal} {Nuclear Fusion}\
  }\textbf {\bibinfo {volume} {57}},\ \bibinfo {pages} {092006} (\bibinfo
  {year} {2017})}\BibitemShut {NoStop}%
\bibitem [{\citenamefont {Kirschner}\ \emph {et~al.}(2023)\citenamefont
  {Kirschner}, \citenamefont {Henderson}, \citenamefont {Brezinsek},
  \citenamefont {Romazanov}, \citenamefont {Kovari}, \citenamefont {Baumann},
  \citenamefont {Linsmeier}, \citenamefont {Flynn}, \citenamefont {Hess},
  \citenamefont {Osawa} \emph {et~al.}}]{kirschner2023erosion}%
  \BibitemOpen
  \bibfield  {author} {\bibinfo {author} {\bibfnamefont {A.}~\bibnamefont
  {Kirschner}}, \bibinfo {author} {\bibfnamefont {S.}~\bibnamefont
  {Henderson}}, \bibinfo {author} {\bibfnamefont {S.}~\bibnamefont
  {Brezinsek}}, \bibinfo {author} {\bibfnamefont {J.}~\bibnamefont
  {Romazanov}}, \bibinfo {author} {\bibfnamefont {M.}~\bibnamefont {Kovari}},
  \bibinfo {author} {\bibfnamefont {C.}~\bibnamefont {Baumann}}, \bibinfo
  {author} {\bibfnamefont {C.}~\bibnamefont {Linsmeier}}, \bibinfo {author}
  {\bibfnamefont {E.}~\bibnamefont {Flynn}}, \bibinfo {author} {\bibfnamefont
  {J.}~\bibnamefont {Hess}}, \bibinfo {author} {\bibfnamefont {R.}~\bibnamefont
  {Osawa}}, \emph {et~al.},\ }\bibfield  {title} {\bibinfo {title} {Erosion
  estimates for the divertor and main wall components from step},\ }\href@noop
  {} {\bibfield  {journal} {\bibinfo  {journal} {Nuclear Fusion}\ }\textbf
  {\bibinfo {volume} {63}},\ \bibinfo {pages} {126055} (\bibinfo {year}
  {2023})}\BibitemShut {NoStop}%
\bibitem [{\citenamefont {Fischer}\ \emph {et~al.}(2015)\citenamefont
  {Fischer}, \citenamefont {Bachmann}, \citenamefont {Palermo}, \citenamefont
  {Pereslavtsev},\ and\ \citenamefont {Villari}}]{fischer2015neutronics}%
  \BibitemOpen
  \bibfield  {author} {\bibinfo {author} {\bibfnamefont {U.}~\bibnamefont
  {Fischer}}, \bibinfo {author} {\bibfnamefont {C.}~\bibnamefont {Bachmann}},
  \bibinfo {author} {\bibfnamefont {I.}~\bibnamefont {Palermo}}, \bibinfo
  {author} {\bibfnamefont {P.}~\bibnamefont {Pereslavtsev}},\ and\ \bibinfo
  {author} {\bibfnamefont {R.}~\bibnamefont {Villari}},\ }\bibfield  {title}
  {\bibinfo {title} {Neutronics requirements for a demo fusion power plant},\
  }\href@noop {} {\bibfield  {journal} {\bibinfo  {journal} {Fusion Engineering
  and Design}\ }\textbf {\bibinfo {volume} {98}},\ \bibinfo {pages} {2134}
  (\bibinfo {year} {2015})}\BibitemShut {NoStop}%
\bibitem [{\citenamefont {Kotschenreuther}\ \emph {et~al.}(2007)\citenamefont
  {Kotschenreuther}, \citenamefont {Valanju}, \citenamefont {Mahajan},\ and\
  \citenamefont {Wiley}}]{kotschenreuther2007heat}%
  \BibitemOpen
  \bibfield  {author} {\bibinfo {author} {\bibfnamefont {M.}~\bibnamefont
  {Kotschenreuther}}, \bibinfo {author} {\bibfnamefont {P.}~\bibnamefont
  {Valanju}}, \bibinfo {author} {\bibfnamefont {S.}~\bibnamefont {Mahajan}},\
  and\ \bibinfo {author} {\bibfnamefont {J.}~\bibnamefont {Wiley}},\ }\bibfield
   {title} {\bibinfo {title} {On heat loading, novel divertors, and fusion
  reactors},\ }\href@noop {} {\bibfield  {journal} {\bibinfo  {journal}
  {Physics of plasmas}\ }\textbf {\bibinfo {volume} {14}} (\bibinfo {year}
  {2007})}\BibitemShut {NoStop}%
\bibitem [{\citenamefont {Krasheninnikov}\ \emph {et~al.}(2003)\citenamefont
  {Krasheninnikov}, \citenamefont {Zakharov},\ and\ \citenamefont
  {Pereverzev}}]{Krasheninnikov2003}%
  \BibitemOpen
  \bibfield  {author} {\bibinfo {author} {\bibfnamefont {S.~I.}\ \bibnamefont
  {Krasheninnikov}}, \bibinfo {author} {\bibfnamefont {L.~E.}\ \bibnamefont
  {Zakharov}},\ and\ \bibinfo {author} {\bibfnamefont {G.~V.}\ \bibnamefont
  {Pereverzev}},\ }\bibfield  {title} {\bibinfo {title} {On lithium walls and
  the performance of magnetic fusion devices},\ }\bibfield  {journal} {\bibinfo
   {journal} {Physics of Plasmas}\ }\textbf {\bibinfo {volume} {10}},\ \href
  {https://doi.org/10.1063/1.1558293} {10.1063/1.1558293} (\bibinfo {year}
  {2003})\BibitemShut {NoStop}%
\bibitem [{\citenamefont {Maingi}\ \emph {et~al.}(2012)\citenamefont {Maingi},
  \citenamefont {Boyle}, \citenamefont {Canik}, \citenamefont {Kaye},
  \citenamefont {Skinner}, \citenamefont {Allain}, \citenamefont {Bell},
  \citenamefont {Bell}, \citenamefont {Gerhardt}, \citenamefont {Gray},
  \citenamefont {Jaworski}, \citenamefont {Kaita}, \citenamefont {Kugel},
  \citenamefont {LeBlanc}, \citenamefont {Manickam}, \citenamefont {Mansfield},
  \citenamefont {Menard}, \citenamefont {Osborne}, \citenamefont {Raman},
  \citenamefont {Roquemore}, \citenamefont {Sabbagh}, \citenamefont {Snyder},\
  and\ \citenamefont {Soukhanovskii}}]{Maingi2012}%
  \BibitemOpen
  \bibfield  {author} {\bibinfo {author} {\bibfnamefont {R.}~\bibnamefont
  {Maingi}}, \bibinfo {author} {\bibfnamefont {D.}~\bibnamefont {Boyle}},
  \bibinfo {author} {\bibfnamefont {J.}~\bibnamefont {Canik}}, \bibinfo
  {author} {\bibfnamefont {S.}~\bibnamefont {Kaye}}, \bibinfo {author}
  {\bibfnamefont {C.}~\bibnamefont {Skinner}}, \bibinfo {author} {\bibfnamefont
  {J.}~\bibnamefont {Allain}}, \bibinfo {author} {\bibfnamefont
  {M.}~\bibnamefont {Bell}}, \bibinfo {author} {\bibfnamefont {R.}~\bibnamefont
  {Bell}}, \bibinfo {author} {\bibfnamefont {S.}~\bibnamefont {Gerhardt}},
  \bibinfo {author} {\bibfnamefont {T.}~\bibnamefont {Gray}}, \bibinfo {author}
  {\bibfnamefont {M.}~\bibnamefont {Jaworski}}, \bibinfo {author}
  {\bibfnamefont {R.}~\bibnamefont {Kaita}}, \bibinfo {author} {\bibfnamefont
  {H.}~\bibnamefont {Kugel}}, \bibinfo {author} {\bibfnamefont
  {B.}~\bibnamefont {LeBlanc}}, \bibinfo {author} {\bibfnamefont
  {J.}~\bibnamefont {Manickam}}, \bibinfo {author} {\bibfnamefont
  {D.}~\bibnamefont {Mansfield}}, \bibinfo {author} {\bibfnamefont
  {J.}~\bibnamefont {Menard}}, \bibinfo {author} {\bibfnamefont
  {T.}~\bibnamefont {Osborne}}, \bibinfo {author} {\bibfnamefont
  {R.}~\bibnamefont {Raman}}, \bibinfo {author} {\bibfnamefont
  {A.}~\bibnamefont {Roquemore}}, \bibinfo {author} {\bibfnamefont
  {S.}~\bibnamefont {Sabbagh}}, \bibinfo {author} {\bibfnamefont
  {P.}~\bibnamefont {Snyder}},\ and\ \bibinfo {author} {\bibfnamefont
  {V.}~\bibnamefont {Soukhanovskii}},\ }\bibfield  {title} {\bibinfo {title}
  {The effect of progressively increasing lithium coatings on plasma discharge
  characteristics, transport, edge profiles and elm stability in the national
  spherical torus experiment},\ }\href@noop {} {\bibfield  {journal} {\bibinfo
  {journal} {Nuclear Fusion}\ }\textbf {\bibinfo {volume} {52}},\ \bibinfo
  {pages} {083001} (\bibinfo {year} {2012})}\BibitemShut {NoStop}%
\bibitem [{\citenamefont {Boyle}\ \emph {et~al.}(2023)\citenamefont {Boyle},
  \citenamefont {Anderson}, \citenamefont {Banerjee}, \citenamefont {Bell},
  \citenamefont {Capecchi}, \citenamefont {Elliott}, \citenamefont {Hansen},
  \citenamefont {Kubota}, \citenamefont {LeBlanc}, \citenamefont {Maan},
  \citenamefont {Maingi}, \citenamefont {Majeski}, \citenamefont {Menard},
  \citenamefont {Oliva}, \citenamefont {Rhodes}, \citenamefont
  {Soukhanovskii},\ and\ \citenamefont {Zakharov}}]{Boyle_2023}%
  \BibitemOpen
  \bibfield  {author} {\bibinfo {author} {\bibfnamefont {D.}~\bibnamefont
  {Boyle}}, \bibinfo {author} {\bibfnamefont {J.}~\bibnamefont {Anderson}},
  \bibinfo {author} {\bibfnamefont {S.}~\bibnamefont {Banerjee}}, \bibinfo
  {author} {\bibfnamefont {R.}~\bibnamefont {Bell}}, \bibinfo {author}
  {\bibfnamefont {W.}~\bibnamefont {Capecchi}}, \bibinfo {author}
  {\bibfnamefont {D.}~\bibnamefont {Elliott}}, \bibinfo {author} {\bibfnamefont
  {C.}~\bibnamefont {Hansen}}, \bibinfo {author} {\bibfnamefont
  {S.}~\bibnamefont {Kubota}}, \bibinfo {author} {\bibfnamefont
  {B.}~\bibnamefont {LeBlanc}}, \bibinfo {author} {\bibfnamefont
  {A.}~\bibnamefont {Maan}}, \bibinfo {author} {\bibfnamefont {R.}~\bibnamefont
  {Maingi}}, \bibinfo {author} {\bibfnamefont {R.}~\bibnamefont {Majeski}},
  \bibinfo {author} {\bibfnamefont {J.}~\bibnamefont {Menard}}, \bibinfo
  {author} {\bibfnamefont {S.}~\bibnamefont {Oliva}}, \bibinfo {author}
  {\bibfnamefont {T.}~\bibnamefont {Rhodes}}, \bibinfo {author} {\bibfnamefont
  {V.}~\bibnamefont {Soukhanovskii}},\ and\ \bibinfo {author} {\bibfnamefont
  {L.}~\bibnamefont {Zakharov}},\ }\bibfield  {title} {\bibinfo {title}
  {Extending the low-recycling, flat temperature profile regime in the lithium
  tokamak experiment-$\beta$ (ltx-$\beta$) with ohmic and neutral beam
  heating},\ }\href {https://doi.org/10.1088/1741-4326/acc4da} {\bibfield
  {journal} {\bibinfo  {journal} {Nuclear Fusion}\ }\textbf {\bibinfo {volume}
  {63}},\ \bibinfo {pages} {056020} (\bibinfo {year} {2023})}\BibitemShut
  {NoStop}%
\bibitem [{\citenamefont {Berkery}(2024)}]{Berkery2024}%
  \BibitemOpen
  \bibfield  {author} {\bibinfo {author} {\bibfnamefont {J.~W.}\ \bibnamefont
  {Berkery}},\ }\bibfield  {title} {\bibinfo {title} {Nstx-u research advancing
  the physics of spherical tokamaks (in review)},\ }\href@noop {} {\bibfield
  {journal} {\bibinfo  {journal} {Nuclear Fusion}\ } (\bibinfo {year}
  {2024})}\BibitemShut {NoStop}%
\bibitem [{\citenamefont {Verhaegh}\ \emph {et~al.}(2024)\citenamefont
  {Verhaegh}, \citenamefont {Harrison}, \citenamefont {Lipschultz},
  \citenamefont {Lonigro}, \citenamefont {Kobussen}, \citenamefont {Moulton},
  \citenamefont {Osborne}, \citenamefont {Ryan}, \citenamefont {Theiler},
  \citenamefont {Wijkamp}, \citenamefont {Brida}, \citenamefont {Derks},
  \citenamefont {Doyle}, \citenamefont {Federici}, \citenamefont {Hakola},
  \citenamefont {Henderson}, \citenamefont {Kool}, \citenamefont {Newton},
  \citenamefont {Osawa}, \citenamefont {Pope}, \citenamefont {Reimerdes},
  \citenamefont {Vianello}, \citenamefont {Wischmeier}, \citenamefont {the
  EUROfusion Tokamak Exploitation~Team},\ and\ \citenamefont {the
  MAST-U~Team}}]{Verhaegh_2024}%
  \BibitemOpen
  \bibfield  {author} {\bibinfo {author} {\bibfnamefont {K.}~\bibnamefont
  {Verhaegh}}, \bibinfo {author} {\bibfnamefont {J.}~\bibnamefont {Harrison}},
  \bibinfo {author} {\bibfnamefont {B.}~\bibnamefont {Lipschultz}}, \bibinfo
  {author} {\bibfnamefont {N.}~\bibnamefont {Lonigro}}, \bibinfo {author}
  {\bibfnamefont {S.}~\bibnamefont {Kobussen}}, \bibinfo {author}
  {\bibfnamefont {D.}~\bibnamefont {Moulton}}, \bibinfo {author} {\bibfnamefont
  {N.}~\bibnamefont {Osborne}}, \bibinfo {author} {\bibfnamefont
  {P.}~\bibnamefont {Ryan}}, \bibinfo {author} {\bibfnamefont {C.}~\bibnamefont
  {Theiler}}, \bibinfo {author} {\bibfnamefont {T.}~\bibnamefont {Wijkamp}},
  \bibinfo {author} {\bibfnamefont {D.}~\bibnamefont {Brida}}, \bibinfo
  {author} {\bibfnamefont {G.}~\bibnamefont {Derks}}, \bibinfo {author}
  {\bibfnamefont {R.}~\bibnamefont {Doyle}}, \bibinfo {author} {\bibfnamefont
  {F.}~\bibnamefont {Federici}}, \bibinfo {author} {\bibfnamefont
  {A.}~\bibnamefont {Hakola}}, \bibinfo {author} {\bibfnamefont
  {S.}~\bibnamefont {Henderson}}, \bibinfo {author} {\bibfnamefont
  {B.}~\bibnamefont {Kool}}, \bibinfo {author} {\bibfnamefont {S.}~\bibnamefont
  {Newton}}, \bibinfo {author} {\bibfnamefont {R.}~\bibnamefont {Osawa}},
  \bibinfo {author} {\bibfnamefont {X.}~\bibnamefont {Pope}}, \bibinfo {author}
  {\bibfnamefont {H.}~\bibnamefont {Reimerdes}}, \bibinfo {author}
  {\bibfnamefont {N.}~\bibnamefont {Vianello}}, \bibinfo {author}
  {\bibfnamefont {M.}~\bibnamefont {Wischmeier}}, \bibinfo {author}
  {\bibnamefont {the EUROfusion Tokamak Exploitation~Team}},\ and\ \bibinfo
  {author} {\bibnamefont {the MAST-U~Team}},\ }\bibfield  {title} {\bibinfo
  {title} {Investigations of atomic and molecular processes of nbi-heated
  discharges in the mast upgrade super-x divertor with implications for
  reactors},\ }\href {https://doi.org/10.1088/1741-4326/ad5851} {\bibfield
  {journal} {\bibinfo  {journal} {Nuclear Fusion}\ }\textbf {\bibinfo {volume}
  {64}},\ \bibinfo {pages} {086050} (\bibinfo {year} {2024})}\BibitemShut
  {NoStop}%
\bibitem [{\citenamefont {Sawan}\ and\ \citenamefont
  {Abdou}(2006)}]{sawan_physics_2006}%
  \BibitemOpen
  \bibfield  {author} {\bibinfo {author} {\bibfnamefont {M.}~\bibnamefont
  {Sawan}}\ and\ \bibinfo {author} {\bibfnamefont {M.}~\bibnamefont {Abdou}},\
  }\bibfield  {title} {\bibinfo {title} {Physics and technology conditions for
  attaining tritium self-sufficiency for the dt fuel cycle},\ }\href
  {https://doi.org/10.1016/j.fusengdes.2005.07.035} {\bibfield  {journal}
  {\bibinfo  {journal} {Fusion Engineering and Design}\ }\textbf {\bibinfo
  {volume} {81}},\ \bibinfo {pages} {1131} (\bibinfo {year}
  {2006})}\BibitemShut {NoStop}%
\bibitem [{\citenamefont {Ikeda}(2007)}]{Ikeda2007}%
  \BibitemOpen
  \bibfield  {author} {\bibinfo {author} {\bibfnamefont {K.}~\bibnamefont
  {Ikeda}},\ }\bibfield  {title} {\bibinfo {title} {Progress in the iter
  physics basis},\ }\href@noop {} {\bibfield  {journal} {\bibinfo  {journal}
  {Nuclear Fusion}\ }\textbf {\bibinfo {volume} {47}},\ \bibinfo {pages} {E01}
  (\bibinfo {year} {2007})}\BibitemShut {NoStop}%
\bibitem [{\citenamefont {Kinsey}\ \emph {et~al.}(2011)\citenamefont {Kinsey},
  \citenamefont {Staebler}, \citenamefont {Candy}, \citenamefont {Waltz},\ and\
  \citenamefont {Budny}}]{Kinsey2011}%
  \BibitemOpen
  \bibfield  {author} {\bibinfo {author} {\bibfnamefont {J.~E.}\ \bibnamefont
  {Kinsey}}, \bibinfo {author} {\bibfnamefont {G.~M.}\ \bibnamefont
  {Staebler}}, \bibinfo {author} {\bibfnamefont {J.}~\bibnamefont {Candy}},
  \bibinfo {author} {\bibfnamefont {R.~E.}\ \bibnamefont {Waltz}},\ and\
  \bibinfo {author} {\bibfnamefont {R.~V.}\ \bibnamefont {Budny}},\ }\bibfield
  {title} {\bibinfo {title} {Iter predictions using the gyro verified and
  experimentally validated trapped gyro-landau fluid transport model},\
  }\bibfield  {journal} {\bibinfo  {journal} {Nuclear Fusion}\ }\textbf
  {\bibinfo {volume} {51}},\ \href
  {https://doi.org/10.1088/0029-5515/51/8/083001}
  {10.1088/0029-5515/51/8/083001} (\bibinfo {year} {2011})\BibitemShut
  {NoStop}%
\bibitem [{\citenamefont {Creely}\ \emph {et~al.}(2020)\citenamefont {Creely},
  \citenamefont {Greenwald}, \citenamefont {Ballinger}, \citenamefont
  {Brunner}, \citenamefont {Canik}, \citenamefont {Doody}, \citenamefont
  {F{\"u, T.}l{\"o}p}, \citenamefont {Garnier}, \citenamefont {Granetz},
  \citenamefont {Gray}, \citenamefont {Holland}, \citenamefont {Howard},
  \citenamefont {Hughes}, \citenamefont {Irby}, \citenamefont {Izzo},
  \citenamefont {Kramer}, \citenamefont {Kuang}, \citenamefont {LaBombard},
  \citenamefont {Lin}, \citenamefont {Lipschultz}, \citenamefont {Logan},
  \citenamefont {Lore}, \citenamefont {Marmar}, \citenamefont {Montes},
  \citenamefont {Mumgaard}, \citenamefont {Paz-Soldan}, \citenamefont {Rea},
  \citenamefont {Reinke}, \citenamefont {Rodriguez-Fernandez}, \citenamefont
  {S{\"a}rkim{\"a}ki}, \citenamefont {Sciortino}, \citenamefont {Scott},
  \citenamefont {Snicker}, \citenamefont {Snyder}, \citenamefont {Sorbom},
  \citenamefont {Sweeney}, \citenamefont {Tinguely}, \citenamefont {Tolman},
  \citenamefont {Umansky}, \citenamefont {Vallhagen}, \citenamefont {Varje},
  \citenamefont {Whyte}, \citenamefont {Wright}, \citenamefont {Wukitch},\ and\
  \citenamefont {Zhu}}]{Creely2020}%
  \BibitemOpen
  \bibfield  {author} {\bibinfo {author} {\bibfnamefont {A.~J.}\ \bibnamefont
  {Creely}}, \bibinfo {author} {\bibfnamefont {M.~J.}\ \bibnamefont
  {Greenwald}}, \bibinfo {author} {\bibfnamefont {S.~B.}\ \bibnamefont
  {Ballinger}}, \bibinfo {author} {\bibfnamefont {D.}~\bibnamefont {Brunner}},
  \bibinfo {author} {\bibfnamefont {J.}~\bibnamefont {Canik}}, \bibinfo
  {author} {\bibfnamefont {J.}~\bibnamefont {Doody}}, \bibinfo {author}
  {\bibnamefont {F{\"u, T.}l{\"o}p}}, \bibinfo {author} {\bibfnamefont {D.~T.}\
  \bibnamefont {Garnier}}, \bibinfo {author} {\bibfnamefont {R.}~\bibnamefont
  {Granetz}}, \bibinfo {author} {\bibfnamefont {T.~K.}\ \bibnamefont {Gray}},
  \bibinfo {author} {\bibfnamefont {C.}~\bibnamefont {Holland}}, \bibinfo
  {author} {\bibfnamefont {N.~T.}\ \bibnamefont {Howard}}, \bibinfo {author}
  {\bibfnamefont {J.~W.}\ \bibnamefont {Hughes}}, \bibinfo {author}
  {\bibfnamefont {J.~H.}\ \bibnamefont {Irby}}, \bibinfo {author}
  {\bibfnamefont {V.~A.}\ \bibnamefont {Izzo}}, \bibinfo {author}
  {\bibfnamefont {G.~J.}\ \bibnamefont {Kramer}}, \bibinfo {author}
  {\bibfnamefont {A.~Q.}\ \bibnamefont {Kuang}}, \bibinfo {author}
  {\bibfnamefont {B.}~\bibnamefont {LaBombard}}, \bibinfo {author}
  {\bibfnamefont {Y.}~\bibnamefont {Lin}}, \bibinfo {author} {\bibfnamefont
  {B.}~\bibnamefont {Lipschultz}}, \bibinfo {author} {\bibfnamefont {N.~C.}\
  \bibnamefont {Logan}}, \bibinfo {author} {\bibfnamefont {J.~D.}\ \bibnamefont
  {Lore}}, \bibinfo {author} {\bibfnamefont {E.~S.}\ \bibnamefont {Marmar}},
  \bibinfo {author} {\bibfnamefont {K.}~\bibnamefont {Montes}}, \bibinfo
  {author} {\bibfnamefont {R.~T.}\ \bibnamefont {Mumgaard}}, \bibinfo {author}
  {\bibfnamefont {C.}~\bibnamefont {Paz-Soldan}}, \bibinfo {author}
  {\bibfnamefont {C.}~\bibnamefont {Rea}}, \bibinfo {author} {\bibfnamefont
  {M.~L.}\ \bibnamefont {Reinke}}, \bibinfo {author} {\bibfnamefont
  {P.}~\bibnamefont {Rodriguez-Fernandez}}, \bibinfo {author} {\bibfnamefont
  {K.}~\bibnamefont {S{\"a}rkim{\"a}ki}}, \bibinfo {author} {\bibfnamefont
  {F.}~\bibnamefont {Sciortino}}, \bibinfo {author} {\bibfnamefont {S.~D.}\
  \bibnamefont {Scott}}, \bibinfo {author} {\bibfnamefont {A.}~\bibnamefont
  {Snicker}}, \bibinfo {author} {\bibfnamefont {P.~B.}\ \bibnamefont {Snyder}},
  \bibinfo {author} {\bibfnamefont {B.~N.}\ \bibnamefont {Sorbom}}, \bibinfo
  {author} {\bibfnamefont {R.}~\bibnamefont {Sweeney}}, \bibinfo {author}
  {\bibfnamefont {R.~A.}\ \bibnamefont {Tinguely}}, \bibinfo {author}
  {\bibfnamefont {E.~A.}\ \bibnamefont {Tolman}}, \bibinfo {author}
  {\bibfnamefont {M.}~\bibnamefont {Umansky}}, \bibinfo {author} {\bibfnamefont
  {O.}~\bibnamefont {Vallhagen}}, \bibinfo {author} {\bibfnamefont
  {J.}~\bibnamefont {Varje}}, \bibinfo {author} {\bibfnamefont {D.~G.}\
  \bibnamefont {Whyte}}, \bibinfo {author} {\bibfnamefont {J.~C.}\ \bibnamefont
  {Wright}}, \bibinfo {author} {\bibfnamefont {S.~J.}\ \bibnamefont
  {Wukitch}},\ and\ \bibinfo {author} {\bibfnamefont {J.}~\bibnamefont {Zhu}},\
  }\bibfield  {title} {\bibinfo {title} {Overview of the sparc tokamak},\
  }\href@noop {} {\bibfield  {journal} {\bibinfo  {journal} {Journal of Plasma
  Physics}\ }\textbf {\bibinfo {volume} {86}} (\bibinfo {year}
  {2020})}\BibitemShut {NoStop}%
\bibitem [{shi({\natexlab{b}})}]{shine_flare_facility_2025}%
  \BibitemOpen
  \href {https://www.shinefusion.com/flare} {\bibinfo {title} {{Radiation
  Effects Testing}}} ({\natexlab{b}}),\ \bibinfo {note} {accessed 21 Oct
  2025}\BibitemShut {NoStop}%
\bibitem [{\citenamefont {Kulsrud}\ \emph {et~al.}(1986)\citenamefont
  {Kulsrud}, \citenamefont {Valeo},\ and\ \citenamefont
  {Cowley}}]{Kulsrud1986}%
  \BibitemOpen
  \bibfield  {author} {\bibinfo {author} {\bibfnamefont {R.~M.}\ \bibnamefont
  {Kulsrud}}, \bibinfo {author} {\bibfnamefont {E.~J.}\ \bibnamefont {Valeo}},\
  and\ \bibinfo {author} {\bibfnamefont {S.~C.}\ \bibnamefont {Cowley}},\
  }\bibfield  {title} {\bibinfo {title} {Physics of spin-polarized plasmas},\
  }\bibfield  {journal} {\bibinfo  {journal} {Nuclear Fusion}\ }\textbf
  {\bibinfo {volume} {26}},\ \href
  {https://doi.org/10.1088/0029-5515/26/11/001} {10.1088/0029-5515/26/11/001}
  (\bibinfo {year} {1986})\BibitemShut {NoStop}%
\bibitem [{\citenamefont {Heidbrink}\ \emph {et~al.}(2024)\citenamefont
  {Heidbrink}, \citenamefont {Baylor}, \citenamefont {B{\"u,}scher},
  \citenamefont {Engels}, \citenamefont {Garcia}, \citenamefont {Ghiozzi},
  \citenamefont {Miller}, \citenamefont {Sandorfi},\ and\ \citenamefont
  {Wei}}]{Heidbrink2024}%
  \BibitemOpen
  \bibfield  {author} {\bibinfo {author} {\bibfnamefont {W.~W.}\ \bibnamefont
  {Heidbrink}}, \bibinfo {author} {\bibfnamefont {L.~R.}\ \bibnamefont
  {Baylor}}, \bibinfo {author} {\bibfnamefont {M.}~\bibnamefont
  {B{\"u,}scher}}, \bibinfo {author} {\bibfnamefont {R.~W.}\ \bibnamefont
  {Engels}}, \bibinfo {author} {\bibfnamefont {A.~V.}\ \bibnamefont {Garcia}},
  \bibinfo {author} {\bibfnamefont {A.~G.}\ \bibnamefont {Ghiozzi}}, \bibinfo
  {author} {\bibfnamefont {G.~W.}\ \bibnamefont {Miller}}, \bibinfo {author}
  {\bibfnamefont {A.~M.}\ \bibnamefont {Sandorfi}},\ and\ \bibinfo {author}
  {\bibfnamefont {X.}~\bibnamefont {Wei}},\ }\bibfield  {title} {\bibinfo
  {title} {A research program to measure the lifetime of spin polarized fuel},\
  }\href@noop {} {\bibfield  {journal} {\bibinfo  {journal} {Frontiers in
  Physics}\ } (\bibinfo {year} {2024})}\BibitemShut {NoStop}%
\bibitem [{\citenamefont {Whan~Bae}\ \emph {et~al.}(2025)\citenamefont
  {Whan~Bae}, \citenamefont {Borowiec}, \citenamefont {Badalassi},
  \citenamefont {Parisi}, \citenamefont {Diallo}, \citenamefont {Menard},
  \citenamefont {Khodak},\ and\ \citenamefont {Brown}}]{WhanBae_2025}%
  \BibitemOpen
  \bibfield  {author} {\bibinfo {author} {\bibfnamefont {J.}~\bibnamefont
  {Whan~Bae}}, \bibinfo {author} {\bibfnamefont {K.}~\bibnamefont {Borowiec}},
  \bibinfo {author} {\bibfnamefont {V.}~\bibnamefont {Badalassi}}, \bibinfo
  {author} {\bibfnamefont {J.}~\bibnamefont {Parisi}}, \bibinfo {author}
  {\bibfnamefont {A.}~\bibnamefont {Diallo}}, \bibinfo {author} {\bibfnamefont
  {J.}~\bibnamefont {Menard}}, \bibinfo {author} {\bibfnamefont
  {A.}~\bibnamefont {Khodak}},\ and\ \bibinfo {author} {\bibfnamefont
  {T.}~\bibnamefont {Brown}},\ }\bibfield  {title} {\bibinfo {title}
  {Neutronics analysis of spin-polarized fuel in spherical tokamaks*},\ }\href
  {https://doi.org/10.1088/1741-4326/adf3c6} {\bibfield  {journal} {\bibinfo
  {journal} {Nuclear Fusion}\ }\textbf {\bibinfo {volume} {65}},\ \bibinfo
  {pages} {086051} (\bibinfo {year} {2025})}\BibitemShut {NoStop}%
\bibitem [{\citenamefont {Parisi}\ \emph {et~al.}(2024)\citenamefont {Parisi},
  \citenamefont {Diallo},\ and\ \citenamefont {Schwartz}}]{Parisi_2024_spf}%
  \BibitemOpen
  \bibfield  {author} {\bibinfo {author} {\bibfnamefont {J.}~\bibnamefont
  {Parisi}}, \bibinfo {author} {\bibfnamefont {A.}~\bibnamefont {Diallo}},\
  and\ \bibinfo {author} {\bibfnamefont {J.}~\bibnamefont {Schwartz}},\
  }\bibfield  {title} {\bibinfo {title} {Simultaneous enhancement of tritium
  burn efficiency and fusion power with low-tritium spin-polarized fuel},\
  }\href {https://doi.org/10.1088/1741-4326/ad7da3} {\bibfield  {journal}
  {\bibinfo  {journal} {Nuclear Fusion}\ }\textbf {\bibinfo {volume} {64}},\
  \bibinfo {pages} {126019} (\bibinfo {year} {2024})}\BibitemShut {NoStop}%
\bibitem [{\citenamefont {Schwartz}(2025)}]{Schwartz2025}%
  \BibitemOpen
  \bibfield  {author} {\bibinfo {author} {\bibfnamefont {J.~A.}\ \bibnamefont
  {Schwartz}},\ }\href {https://arxiv.org/abs/2507.11758} {\bibinfo {title}
  {Analytic neutron wall loading from spin-polarized fusion in axisymmetric
  geometries}} (\bibinfo {year} {2025}),\ \Eprint
  {https://arxiv.org/abs/2507.11758} {arXiv:2507.11758 [physics.plasm-ph]}
  \BibitemShut {NoStop}%
\bibitem [{\citenamefont {Panici}\ \emph {et~al.}(2023)\citenamefont {Panici},
  \citenamefont {Conlin}, \citenamefont {Dudt}, \citenamefont {Unalmis},\ and\
  \citenamefont {Kolemen}}]{Panici2023}%
  \BibitemOpen
  \bibfield  {author} {\bibinfo {author} {\bibfnamefont {D.}~\bibnamefont
  {Panici}}, \bibinfo {author} {\bibfnamefont {R.}~\bibnamefont {Conlin}},
  \bibinfo {author} {\bibfnamefont {D.~W.}\ \bibnamefont {Dudt}}, \bibinfo
  {author} {\bibfnamefont {K.}~\bibnamefont {Unalmis}},\ and\ \bibinfo {author}
  {\bibfnamefont {E.}~\bibnamefont {Kolemen}},\ }\bibfield  {title} {\bibinfo
  {title} {The desc stellarator code suite. part 1. quick and accurate
  equilibria computations},\ }\href@noop {} {\bibfield  {journal} {\bibinfo
  {journal} {Journal of Plasma Physics}\ }\textbf {\bibinfo {volume} {89}},\
  \bibinfo {pages} {955890303} (\bibinfo {year} {2023})}\BibitemShut {NoStop}%
\bibitem [{Note1()}]{Note1}%
  \BibitemOpen
  \bibinfo {note} {A fully polarized perpendicular mode increases the D-T
  reaction cross section by 50\% \cite {Kulsrud1982} and can increase the
  overall fusion power by $\sim 80-90\%$ \cite
  {Smith2018,Heidbrink2024}}\BibitemShut {NoStop}%
\bibitem [{\citenamefont {El-Guebaly}(1995)}]{el1995neutronics}%
  \BibitemOpen
  \bibfield  {author} {\bibinfo {author} {\bibfnamefont {L.}~\bibnamefont
  {El-Guebaly}},\ }\bibfield  {title} {\bibinfo {title} {Neutronics analysis
  for the stellarator power plant study spps},\ }in\ \href@noop {} {\emph
  {\bibinfo {booktitle} {Proceedings of 16th International Symposium on Fusion
  Engineering}}},\ Vol.~\bibinfo {volume} {2}\ (\bibinfo {organization}
  {IEEE},\ \bibinfo {year} {1995})\ pp.\ \bibinfo {pages}
  {1162--1165}\BibitemShut {NoStop}%
\bibitem [{\citenamefont {El-Guebaly}\ \emph {et~al.}(2008)\citenamefont
  {El-Guebaly}, \citenamefont {Wilson}, \citenamefont {Henderson},
  \citenamefont {Sawan}, \citenamefont {Sviatoslavsky}, \citenamefont
  {Tautges}, \citenamefont {Slaybaugh}, \citenamefont {Kiedrowski},
  \citenamefont {Ibrahim}, \citenamefont {Team} \emph
  {et~al.}}]{el2008designing}%
  \BibitemOpen
  \bibfield  {author} {\bibinfo {author} {\bibfnamefont {L.}~\bibnamefont
  {El-Guebaly}}, \bibinfo {author} {\bibfnamefont {P.}~\bibnamefont {Wilson}},
  \bibinfo {author} {\bibfnamefont {D.}~\bibnamefont {Henderson}}, \bibinfo
  {author} {\bibfnamefont {M.}~\bibnamefont {Sawan}}, \bibinfo {author}
  {\bibfnamefont {G.}~\bibnamefont {Sviatoslavsky}}, \bibinfo {author}
  {\bibfnamefont {T.}~\bibnamefont {Tautges}}, \bibinfo {author} {\bibfnamefont
  {R.}~\bibnamefont {Slaybaugh}}, \bibinfo {author} {\bibfnamefont
  {B.}~\bibnamefont {Kiedrowski}}, \bibinfo {author} {\bibfnamefont
  {A.}~\bibnamefont {Ibrahim}}, \bibinfo {author} {\bibfnamefont
  {A.}~\bibnamefont {Team}}, \emph {et~al.},\ }\bibfield  {title} {\bibinfo
  {title} {Nuclear challenges and progress in designing stellarator fusion
  power plants},\ }\href@noop {} {\bibfield  {journal} {\bibinfo  {journal}
  {Energy conversion and management}\ }\textbf {\bibinfo {volume} {49}},\
  \bibinfo {pages} {1859} (\bibinfo {year} {2008})}\BibitemShut {NoStop}%
\bibitem [{\citenamefont {Slaybaugh}\ \emph {et~al.}(2009)\citenamefont
  {Slaybaugh}, \citenamefont {Wilson}, \citenamefont {El-Guebaly},\ and\
  \citenamefont {Marriott}}]{slaybaugh2009three}%
  \BibitemOpen
  \bibfield  {author} {\bibinfo {author} {\bibfnamefont {R.}~\bibnamefont
  {Slaybaugh}}, \bibinfo {author} {\bibfnamefont {P.}~\bibnamefont {Wilson}},
  \bibinfo {author} {\bibfnamefont {L.}~\bibnamefont {El-Guebaly}},\ and\
  \bibinfo {author} {\bibfnamefont {E.}~\bibnamefont {Marriott}},\ }\bibfield
  {title} {\bibinfo {title} {Three-dimensional neutron source models for
  toroidal fusion energy systems},\ }\href@noop {} {\bibfield  {journal}
  {\bibinfo  {journal} {Fusion engineering and design}\ }\textbf {\bibinfo
  {volume} {84}},\ \bibinfo {pages} {1774} (\bibinfo {year}
  {2009})}\BibitemShut {NoStop}%
\bibitem [{\citenamefont {Lion}\ \emph {et~al.}(2022)\citenamefont {Lion},
  \citenamefont {Warmer},\ and\ \citenamefont {Wang}}]{lion2022deterministic}%
  \BibitemOpen
  \bibfield  {author} {\bibinfo {author} {\bibfnamefont {J.}~\bibnamefont
  {Lion}}, \bibinfo {author} {\bibfnamefont {F.}~\bibnamefont {Warmer}},\ and\
  \bibinfo {author} {\bibfnamefont {H.}~\bibnamefont {Wang}},\ }\bibfield
  {title} {\bibinfo {title} {A deterministic method for the fast evaluation and
  optimisation of the 3d neutron wall load for generic stellarator
  configurations},\ }\href@noop {} {\bibfield  {journal} {\bibinfo  {journal}
  {Nuclear Fusion}\ }\textbf {\bibinfo {volume} {62}},\ \bibinfo {pages}
  {076040} (\bibinfo {year} {2022})}\BibitemShut {NoStop}%
\bibitem [{\citenamefont {Lion}\ \emph {et~al.}(2025)\citenamefont {Lion},
  \citenamefont {Angl{\`e}s}, \citenamefont {Bonauer}, \citenamefont {Navarro},
  \citenamefont {Ceron}, \citenamefont {Davies}, \citenamefont {Drevlak},
  \citenamefont {Foppiani}, \citenamefont {Geiger}, \citenamefont {Goodman}
  \emph {et~al.}}]{lion2025stellaris}%
  \BibitemOpen
  \bibfield  {author} {\bibinfo {author} {\bibfnamefont {J.}~\bibnamefont
  {Lion}}, \bibinfo {author} {\bibfnamefont {J.-C.}\ \bibnamefont
  {Angl{\`e}s}}, \bibinfo {author} {\bibfnamefont {L.}~\bibnamefont {Bonauer}},
  \bibinfo {author} {\bibfnamefont {A.~B.}\ \bibnamefont {Navarro}}, \bibinfo
  {author} {\bibfnamefont {S.~C.}\ \bibnamefont {Ceron}}, \bibinfo {author}
  {\bibfnamefont {R.}~\bibnamefont {Davies}}, \bibinfo {author} {\bibfnamefont
  {M.}~\bibnamefont {Drevlak}}, \bibinfo {author} {\bibfnamefont
  {N.}~\bibnamefont {Foppiani}}, \bibinfo {author} {\bibfnamefont
  {J.}~\bibnamefont {Geiger}}, \bibinfo {author} {\bibfnamefont
  {A.}~\bibnamefont {Goodman}}, \emph {et~al.},\ }\bibfield  {title} {\bibinfo
  {title} {Stellaris: A high-field quasi-isodynamic stellarator for a
  prototypical fusion power plant},\ }\href@noop {} {\bibfield  {journal}
  {\bibinfo  {journal} {Fusion Engineering and Design}\ }\textbf {\bibinfo
  {volume} {214}},\ \bibinfo {pages} {114868} (\bibinfo {year}
  {2025})}\BibitemShut {NoStop}%
\bibitem [{\citenamefont {Clark}\ \emph {et~al.}(2025)\citenamefont {Clark},
  \citenamefont {Goh}, \citenamefont {Ramirez}, \citenamefont {Pflug},
  \citenamefont {Smandych}, \citenamefont {Kessing}, \citenamefont {Moreno},
  \citenamefont {Bohm}, \citenamefont {Wilson}, \citenamefont {Singh} \emph
  {et~al.}}]{clark2025breeder}%
  \BibitemOpen
  \bibfield  {author} {\bibinfo {author} {\bibfnamefont {D.}~\bibnamefont
  {Clark}}, \bibinfo {author} {\bibfnamefont {B.}~\bibnamefont {Goh}}, \bibinfo
  {author} {\bibfnamefont {S.}~\bibnamefont {Ramirez}}, \bibinfo {author}
  {\bibfnamefont {E.}~\bibnamefont {Pflug}}, \bibinfo {author} {\bibfnamefont
  {J.}~\bibnamefont {Smandych}}, \bibinfo {author} {\bibfnamefont
  {J.}~\bibnamefont {Kessing}}, \bibinfo {author} {\bibfnamefont
  {C.}~\bibnamefont {Moreno}}, \bibinfo {author} {\bibfnamefont
  {T.}~\bibnamefont {Bohm}}, \bibinfo {author} {\bibfnamefont {P.}~\bibnamefont
  {Wilson}}, \bibinfo {author} {\bibfnamefont {L.}~\bibnamefont {Singh}}, \emph
  {et~al.},\ }\bibfield  {title} {\bibinfo {title} {Breeder blanket and tritium
  fuel cycle feasibility of the infinity two fusion pilot plant},\ }\href@noop
  {} {\bibfield  {journal} {\bibinfo  {journal} {Journal of Plasma Physics}\
  }\textbf {\bibinfo {volume} {91}},\ \bibinfo {pages} {E86} (\bibinfo {year}
  {2025})}\BibitemShut {NoStop}%
\bibitem [{\citenamefont {Ketring}\ \emph {et~al.}(2002)\citenamefont
  {Ketring}, \citenamefont {Embree}, \citenamefont {Bailey}, \citenamefont
  {Tyler}, \citenamefont {Gawenis}, \citenamefont {Cutler}, \citenamefont
  {Jurisson},\ and\ \citenamefont {Engelbrecht}}]{ketring2002production}%
  \BibitemOpen
  \bibfield  {author} {\bibinfo {author} {\bibfnamefont {A.}~\bibnamefont
  {Ketring}}, \bibinfo {author} {\bibfnamefont {M.}~\bibnamefont {Embree}},
  \bibinfo {author} {\bibfnamefont {K.}~\bibnamefont {Bailey}}, \bibinfo
  {author} {\bibfnamefont {T.}~\bibnamefont {Tyler}}, \bibinfo {author}
  {\bibfnamefont {J.}~\bibnamefont {Gawenis}}, \bibinfo {author} {\bibfnamefont
  {C.}~\bibnamefont {Cutler}}, \bibinfo {author} {\bibfnamefont
  {S.}~\bibnamefont {Jurisson}},\ and\ \bibinfo {author} {\bibfnamefont
  {H.}~\bibnamefont {Engelbrecht}},\ }\bibfield  {title} {\bibinfo {title}
  {Production and supply of high specific activity radioisotopes for
  radiotherapy applications},\ }\href@noop {} {\bibfield  {journal} {\bibinfo
  {journal} {World Journal of Nuclear Medicine}\ }\textbf {\bibinfo {volume}
  {1}} (\bibinfo {year} {2002})}\BibitemShut {NoStop}%
\bibitem [{\citenamefont {Swanson}\ \emph {et~al.}(2025)\citenamefont
  {Swanson}, \citenamefont {Gates}, \citenamefont {Kumar}, \citenamefont
  {Martin}, \citenamefont {Kruger}, \citenamefont {Dudt}, \citenamefont
  {Bonofiglo},\ and\ \citenamefont {team}}]{swanson2025scoping}%
  \BibitemOpen
  \bibfield  {author} {\bibinfo {author} {\bibfnamefont {C.}~\bibnamefont
  {Swanson}}, \bibinfo {author} {\bibfnamefont {D.}~\bibnamefont {Gates}},
  \bibinfo {author} {\bibfnamefont {S.}~\bibnamefont {Kumar}}, \bibinfo
  {author} {\bibfnamefont {M.}~\bibnamefont {Martin}}, \bibinfo {author}
  {\bibfnamefont {T.}~\bibnamefont {Kruger}}, \bibinfo {author} {\bibfnamefont
  {D.}~\bibnamefont {Dudt}}, \bibinfo {author} {\bibfnamefont {P.}~\bibnamefont
  {Bonofiglo}},\ and\ \bibinfo {author} {\bibfnamefont {T.~E.}\ \bibnamefont
  {team}},\ }\bibfield  {title} {\bibinfo {title} {The scoping, design, and
  plasma physics optimization of the eos neutron source stellarator},\
  }\href@noop {} {\bibfield  {journal} {\bibinfo  {journal} {Nuclear Fusion}\
  }\textbf {\bibinfo {volume} {65}},\ \bibinfo {pages} {026053} (\bibinfo
  {year} {2025})}\BibitemShut {NoStop}%
\bibitem [{\citenamefont {Menard}\ \emph {et~al.}(2016)\citenamefont {Menard},
  \citenamefont {Brown}, \citenamefont {El-Guebaly}, \citenamefont {Boyer},
  \citenamefont {Canik}, \citenamefont {Colling}, \citenamefont {Raman},
  \citenamefont {Wang}, \citenamefont {Zhai}, \citenamefont {Buxton},
  \citenamefont {Covele}, \citenamefont {D'Angelo}, \citenamefont {Davis},
  \citenamefont {Gerhardt}, \citenamefont {Gryaznevich}, \citenamefont {Harb},
  \citenamefont {Hender}, \citenamefont {Kaye}, \citenamefont {Kingham},
  \citenamefont {Kotschenreuther}, \citenamefont {Mahajan}, \citenamefont
  {Maingi}, \citenamefont {Marriott}, \citenamefont {Meier}, \citenamefont
  {Mynsberge}, \citenamefont {Neumeyer}, \citenamefont {Ono}, \citenamefont
  {Park}, \citenamefont {Sabbagh}, \citenamefont {Soukhanovskii}, \citenamefont
  {Valanju},\ and\ \citenamefont {Woolley}}]{Menard2016}%
  \BibitemOpen
  \bibfield  {author} {\bibinfo {author} {\bibfnamefont {J.~E.}\ \bibnamefont
  {Menard}}, \bibinfo {author} {\bibfnamefont {T.}~\bibnamefont {Brown}},
  \bibinfo {author} {\bibfnamefont {L.}~\bibnamefont {El-Guebaly}}, \bibinfo
  {author} {\bibfnamefont {M.}~\bibnamefont {Boyer}}, \bibinfo {author}
  {\bibfnamefont {J.}~\bibnamefont {Canik}}, \bibinfo {author} {\bibfnamefont
  {B.}~\bibnamefont {Colling}}, \bibinfo {author} {\bibfnamefont
  {R.}~\bibnamefont {Raman}}, \bibinfo {author} {\bibfnamefont
  {Z.}~\bibnamefont {Wang}}, \bibinfo {author} {\bibfnamefont {Y.}~\bibnamefont
  {Zhai}}, \bibinfo {author} {\bibfnamefont {P.}~\bibnamefont {Buxton}},
  \bibinfo {author} {\bibfnamefont {B.}~\bibnamefont {Covele}}, \bibinfo
  {author} {\bibfnamefont {C.}~\bibnamefont {D'Angelo}}, \bibinfo {author}
  {\bibfnamefont {A.}~\bibnamefont {Davis}}, \bibinfo {author} {\bibfnamefont
  {S.}~\bibnamefont {Gerhardt}}, \bibinfo {author} {\bibfnamefont
  {M.}~\bibnamefont {Gryaznevich}}, \bibinfo {author} {\bibfnamefont
  {M.}~\bibnamefont {Harb}}, \bibinfo {author} {\bibfnamefont {T.~C.}\
  \bibnamefont {Hender}}, \bibinfo {author} {\bibfnamefont {S.}~\bibnamefont
  {Kaye}}, \bibinfo {author} {\bibfnamefont {D.}~\bibnamefont {Kingham}},
  \bibinfo {author} {\bibfnamefont {M.}~\bibnamefont {Kotschenreuther}},
  \bibinfo {author} {\bibfnamefont {S.}~\bibnamefont {Mahajan}}, \bibinfo
  {author} {\bibfnamefont {R.}~\bibnamefont {Maingi}}, \bibinfo {author}
  {\bibfnamefont {E.}~\bibnamefont {Marriott}}, \bibinfo {author}
  {\bibfnamefont {E.~T.}\ \bibnamefont {Meier}}, \bibinfo {author}
  {\bibfnamefont {L.}~\bibnamefont {Mynsberge}}, \bibinfo {author}
  {\bibfnamefont {C.}~\bibnamefont {Neumeyer}}, \bibinfo {author}
  {\bibfnamefont {M.}~\bibnamefont {Ono}}, \bibinfo {author} {\bibfnamefont
  {J.~K.}\ \bibnamefont {Park}}, \bibinfo {author} {\bibfnamefont {S.~A.}\
  \bibnamefont {Sabbagh}}, \bibinfo {author} {\bibfnamefont {V.}~\bibnamefont
  {Soukhanovskii}}, \bibinfo {author} {\bibfnamefont {P.}~\bibnamefont
  {Valanju}},\ and\ \bibinfo {author} {\bibfnamefont {R.}~\bibnamefont
  {Woolley}},\ }\bibfield  {title} {\bibinfo {title} {Fusion nuclear science
  facilities and pilot plants based on the spherical tokamak},\ }\href@noop {}
  {\bibfield  {journal} {\bibinfo  {journal} {Nuclear Fusion}\ }\textbf
  {\bibinfo {volume} {56}},\ \bibinfo {pages} {106023} (\bibinfo {year}
  {2016})}\BibitemShut {NoStop}%
\bibitem [{\citenamefont {Hartwig}\ \emph {et~al.}(2012)\citenamefont
  {Hartwig}, \citenamefont {Haakonsen}, \citenamefont {Mumgaard},\ and\
  \citenamefont {Bromberg}}]{Hartwig2012}%
  \BibitemOpen
  \bibfield  {author} {\bibinfo {author} {\bibfnamefont {Z.}~\bibnamefont
  {Hartwig}}, \bibinfo {author} {\bibfnamefont {C.}~\bibnamefont {Haakonsen}},
  \bibinfo {author} {\bibfnamefont {R.}~\bibnamefont {Mumgaard}},\ and\
  \bibinfo {author} {\bibfnamefont {L.}~\bibnamefont {Bromberg}},\ }\bibfield
  {title} {\bibinfo {title} {An initial study of demountable high-temperature
  superconducting toroidal field magnets for the vulcan tokamak conceptual
  design},\ }\href@noop {} {\bibfield  {journal} {\bibinfo  {journal} {Fusion
  Engineering and Design}\ }\textbf {\bibinfo {volume} {87}},\ \bibinfo {pages}
  {201} (\bibinfo {year} {2012})}\BibitemShut {NoStop}%
\bibitem [{\citenamefont {Sorbom}\ \emph {et~al.}(2015)\citenamefont {Sorbom},
  \citenamefont {Ball}, \citenamefont {Palmer}, \citenamefont {Mangiarotti},
  \citenamefont {Sierchio}, \citenamefont {Bonoli}, \citenamefont {Kasten},
  \citenamefont {Sutherland}, \citenamefont {Barnard}, \citenamefont
  {Haakonsen}, \citenamefont {Goh}, \citenamefont {Sung},\ and\ \citenamefont
  {Whyte}}]{Sorbom2015}%
  \BibitemOpen
  \bibfield  {author} {\bibinfo {author} {\bibfnamefont {B.~N.}\ \bibnamefont
  {Sorbom}}, \bibinfo {author} {\bibfnamefont {J.}~\bibnamefont {Ball}},
  \bibinfo {author} {\bibfnamefont {T.~R.}\ \bibnamefont {Palmer}}, \bibinfo
  {author} {\bibfnamefont {F.~J.}\ \bibnamefont {Mangiarotti}}, \bibinfo
  {author} {\bibfnamefont {J.~M.}\ \bibnamefont {Sierchio}}, \bibinfo {author}
  {\bibfnamefont {P.}~\bibnamefont {Bonoli}}, \bibinfo {author} {\bibfnamefont
  {C.}~\bibnamefont {Kasten}}, \bibinfo {author} {\bibfnamefont {D.~A.}\
  \bibnamefont {Sutherland}}, \bibinfo {author} {\bibfnamefont {H.~S.}\
  \bibnamefont {Barnard}}, \bibinfo {author} {\bibfnamefont {C.~B.}\
  \bibnamefont {Haakonsen}}, \bibinfo {author} {\bibfnamefont {J.}~\bibnamefont
  {Goh}}, \bibinfo {author} {\bibfnamefont {C.}~\bibnamefont {Sung}},\ and\
  \bibinfo {author} {\bibfnamefont {D.~G.}\ \bibnamefont {Whyte}},\ }\bibfield
  {title} {\bibinfo {title} {{ARC: A compact, high-field, fusion nuclear
  science facility and demonstration power plant with demountable magnets}},\
  }\href@noop {} {\bibfield  {journal} {\bibinfo  {journal} {Fusion Engineering
  and Design}\ }\textbf {\bibinfo {volume} {100}},\ \bibinfo {pages} {378}
  (\bibinfo {year} {2015})}\BibitemShut {NoStop}%
\bibitem [{\citenamefont {Wesson}(2012)}]{Wesson2012}%
  \BibitemOpen
  \bibfield  {author} {\bibinfo {author} {\bibfnamefont {J.}~\bibnamefont
  {Wesson}},\ }\href@noop {} {\emph {\bibinfo {title} {{Tokamaks}}}},\ \bibinfo
  {edition} {4th}\ ed.\ (\bibinfo  {publisher} {Oxford University Press},\
  \bibinfo {address} {Oxford},\ \bibinfo {year} {2012})\BibitemShut {NoStop}%
\bibitem [{\citenamefont {Dimov}\ \emph {et~al.}(1976)\citenamefont {Dimov},
  \citenamefont {Zadkaidakov},\ and\ \citenamefont
  {Kishinevsky}}]{Dimov1976AmbipolarMirrors}%
  \BibitemOpen
  \bibfield  {author} {\bibinfo {author} {\bibfnamefont {G.~I.}\ \bibnamefont
  {Dimov}}, \bibinfo {author} {\bibfnamefont {V.~V.}\ \bibnamefont
  {Zadkaidakov}},\ and\ \bibinfo {author} {\bibfnamefont {M.~E.}\ \bibnamefont
  {Kishinevsky}},\ }\bibfield  {title} {\bibinfo {title} {Open trap with
  ambipolar mirrors},\ }in\ \href@noop {} {\emph {\bibinfo {booktitle}
  {Proceedings of the 6th International Conference on Plasma Physics and
  Controlled Nuclear Fusion Research}}}\ (\bibinfo  {publisher} {IAEA},\
  \bibinfo {year} {1976})\ pp.\ \bibinfo {pages} {CN--35/C4}\BibitemShut
  {NoStop}%
\bibitem [{\citenamefont {Baldwin}\ and\ \citenamefont
  {Logan}(1979)}]{Baldwin1979}%
  \BibitemOpen
  \bibfield  {author} {\bibinfo {author} {\bibfnamefont {D.~E.}\ \bibnamefont
  {Baldwin}}\ and\ \bibinfo {author} {\bibfnamefont {B.~G.}\ \bibnamefont
  {Logan}},\ }\bibfield  {title} {\bibinfo {title} {Improved tandem mirror
  fusion reactor},\ }\href {https://doi.org/10.1103/PhysRevLett.43.1318}
  {\bibfield  {journal} {\bibinfo  {journal} {Phys. Rev. Lett.}\ }\textbf
  {\bibinfo {volume} {43}},\ \bibinfo {pages} {1318} (\bibinfo {year}
  {1979})}\BibitemShut {NoStop}%
\bibitem [{\citenamefont {Post}(1987)}]{post1987magnetic}%
  \BibitemOpen
  \bibfield  {author} {\bibinfo {author} {\bibfnamefont {R.}~\bibnamefont
  {Post}},\ }\bibfield  {title} {\bibinfo {title} {The magnetic mirror approach
  to fusion},\ }\href@noop {} {\bibfield  {journal} {\bibinfo  {journal}
  {Nuclear Fusion}\ }\textbf {\bibinfo {volume} {27}},\ \bibinfo {pages} {1579}
  (\bibinfo {year} {1987})}\BibitemShut {NoStop}%
\bibitem [{\citenamefont {Frank}\ \emph {et~al.}(2025)\citenamefont {Frank},
  \citenamefont {Viola}, \citenamefont {Petrov}, \citenamefont {Anderson},
  \citenamefont {Bindl}, \citenamefont {Biswas}, \citenamefont {Caneses-Marin},
  \citenamefont {Endrizzi}, \citenamefont {Furlong}, \citenamefont {Harvey}
  \emph {et~al.}}]{frank2025confinement}%
  \BibitemOpen
  \bibfield  {author} {\bibinfo {author} {\bibfnamefont {S.}~\bibnamefont
  {Frank}}, \bibinfo {author} {\bibfnamefont {J.}~\bibnamefont {Viola}},
  \bibinfo {author} {\bibfnamefont {Y.}~\bibnamefont {Petrov}}, \bibinfo
  {author} {\bibfnamefont {J.}~\bibnamefont {Anderson}}, \bibinfo {author}
  {\bibfnamefont {D.}~\bibnamefont {Bindl}}, \bibinfo {author} {\bibfnamefont
  {B.}~\bibnamefont {Biswas}}, \bibinfo {author} {\bibfnamefont {J.~F.}\
  \bibnamefont {Caneses-Marin}}, \bibinfo {author} {\bibfnamefont {D.~A.}\
  \bibnamefont {Endrizzi}}, \bibinfo {author} {\bibfnamefont {K.}~\bibnamefont
  {Furlong}}, \bibinfo {author} {\bibfnamefont {R.}~\bibnamefont {Harvey}},
  \emph {et~al.},\ }\bibfield  {title} {\bibinfo {title} {Confinement
  performance predictions for a high field axisymmetric tandem mirror},\
  }\href@noop {} {\bibfield  {journal} {\bibinfo  {journal} {Journal of Plasma
  Physics}\ }\textbf {\bibinfo {volume} {91}},\ \bibinfo {pages} {E110}
  (\bibinfo {year} {2025})}\BibitemShut {NoStop}%
\bibitem [{\citenamefont {Kesner}(1980)}]{kesner1980axisymmetric}%
  \BibitemOpen
  \bibfield  {author} {\bibinfo {author} {\bibfnamefont {J.}~\bibnamefont
  {Kesner}},\ }\bibfield  {title} {\bibinfo {title} {Axisymmetric sloshing-ion
  tandem-mirror plugs},\ }\href@noop {} {\bibfield  {journal} {\bibinfo
  {journal} {Nuclear Fusion}\ }\textbf {\bibinfo {volume} {20}},\ \bibinfo
  {pages} {557} (\bibinfo {year} {1980})}\BibitemShut {NoStop}%
\bibitem [{\citenamefont {Wurzel}\ and\ \citenamefont
  {Hsu}(2022)}]{Wurzel2022}%
  \BibitemOpen
  \bibfield  {author} {\bibinfo {author} {\bibfnamefont {S.~E.}\ \bibnamefont
  {Wurzel}}\ and\ \bibinfo {author} {\bibfnamefont {S.~C.}\ \bibnamefont
  {Hsu}},\ }\bibfield  {title} {\bibinfo {title} {Progress toward fusion energy
  breakeven and gain as measured against the lawson criterion},\ }\bibfield
  {journal} {\bibinfo  {journal} {Physics of Plasmas}\ }\textbf {\bibinfo
  {volume} {29}},\ \href {https://doi.org/10.1063/5.0083990}
  {10.1063/5.0083990} (\bibinfo {year} {2022})\BibitemShut {NoStop}%
\end{thebibliography}%

\end{document}